\documentclass[onecolumn]{emulateapj}
\usepackage{amsmath,amssymb,graphicx,natbib,subfigure}

\newcommand{\mh}{{M_\bullet}}
\newcommand{\msun}{{M_\odot}}

\newcommand{\beq}{\begin{equation}}
\newcommand{\eeq}{\end{equation}}
\newcommand{\sbh}{SBH}
\newcommand{\sbhs}{SBHs}

\begin{document}

\title{Gravitational Encounters and the Evolution of Galactic Nuclei. III. Anomalous Relaxation}
\author{David Merritt}
\affil{Department of Physics and Center for Computational Relativity
and Gravitation, Rochester Institute of Technology, Rochester, NY 14623}

\begin{abstract}
This paper is the third in a series presenting the results of 
direct numerical integrations of the Fokker-Planck equation 
for stars orbiting a supermassive black hole (\sbh) at the center of a galaxy.
The algorithm of Paper II included diffusion coefficients that described the 
effects of random (``classical'') and correlated (``resonant'') relaxation.
In this paper, the diffusion coefficients of Paper II have been generalized to account for the
effects of ``anomalous relaxation,'' the qualitatively different way in which eccentric orbits
evolve in the regime of rapid relativistic precession.
Two functional forms for the anomalous diffusion coefficients are investigated, based on
 power-law or exponential modifications of the resonant diffusion coefficients.
The parameters defining the modified coefficients are first constrained by comparing the 
results of Fokker-Planck integrations with previously-published $N$-body integrations.
Steady-state solutions are then obtained via the Fokker-Planck equation for models
with properties similar to those of the Milky Way nucleus.
Inclusion of anomalous relaxation leads to the formation of less prominent cores than in the case
of resonant relaxation alone, due to the lengthening of diffusion timescales for eccentric orbits.
Steady-state capture rates of stars by the \sbh\  are found to always be less, by as much as an
order of magnitude, than capture rates in the presence of resonant relaxation alone.
\end{abstract}


\section{Introduction}

Paper I of this series \citep{Paper1} described a numerical algorithm for integrating the Fokker-Planck
equation for $f(E,L,t)$, the phase-space density of stars orbiting a supermassive black hole 
(SBH) at the center of a galaxy; $E$ and $L$ are respectively the orbital energy and angular
momentum per unit mass of a star.
Paper II \citep{Paper2} presented steady-state and time-dependent solutions for $f$
based on diffusion coefficients that describe the effects of
both ``classical'' (random) and ``resonant'' (correlated) relaxation; the latter becomes progressively
more important relative to the former as one moves inside the gravitational influence sphere of the \sbh.
A single value for the stellar mass, $m_\star$, was assumed in both papers.

This paper extends the treatment of Paper II to include the qualitatively different sort of evolution
experienced by eccentric orbits near a \sbh\ \citep{MAMW2011}.
Such orbits undergo rapid apsidal precession due to the effects of general relativity (GR), 
at an orbit-averaged rate
\beq\label{Equation:dwdtGR}
\left|\frac{d\omega}{dt}\right| = 
\frac{3\left(G\mh\right)^{3/2}}{a^{5/2}\left(1-e^2\right)c^2}
\eeq
\citep[``Schwarzschild precession'';][Equation~4.205]{DEGN}.
Here $a$ and $e$ are the orbital semimajor axis and eccentricity
respectively and $\omega$ is the argument of periapsis.
Precession described by Equation~(\ref{Equation:dwdtGR}) affects the collective evolution
in two, distinct ways. (i) The ``coherence time'' is defined as the mean precession time for orbits
at a given energy \citep{RauchTremaine1996}. 
Near the \sbh\, coherence times become progressively shorter due to Schwarzschild precession. 
This effect was correctly accounted for in Papers I and II and in many earlier treatments of
resonant relaxation.
(ii) At any energy, sufficiently eccentric orbits undergo apsidal precession on a shorter timescale
than other orbits of the same energy, due to the strong eccentricity dependence of Equation~(\ref{Equation:dwdtGR}).
Such high-eccentricity orbits might be expected to evolve in a manner qualitatively different than 
described by the equations of resonant relaxation, since their orientation with respect to the torquing potential
changes in a time short compared with the coherence time.

Since it is the high-eccentricity orbits that are most amenable to capture by the \sbh,
Schwarzschild precession was recognized early on as a potentially important mediating factor with regard to 
rates of capture from tightly-bound orbits around a \sbh\
 \citep{HopmanAlexander2006,Madigan2011}.

The first, fully self-consistent investigation of orbital evolution in this regime
\citep{MAMW2011} revealed a new phenomenon.
Stars near the \sbh\ undergo random walks in angular momentum due to resonant relaxation, but when their
eccentricities reach a certain maximum value (depending on $a$), their trajectories ``bounce,'' returning
after roughly one coherence time to lower values of $e$, where they continue to evolve under the
influence of resonant relaxation.
The locus of reflection in the ($a,e$) plane was termed the ``Schwarzschild barrier'' (SB) and an
approximate analytic expression for its location, $L=L_\mathrm{SB}(E)$, was derived.
Subsequent  studies have confirmed this phenomenon using different integration schemes for
the $N$-body equations of motion \citep{Brem2014,Hamers2014}.

A characteristic of motion near and below the SB ($L\lesssim L_\mathrm{SB}(E)$) is that the
apsidal precession time is short compared with the coherence time, and  with the
time over which resonant relaxation would be able to change $L$ in the absence of the rapid precession.
Angular momentum evolution in the region below the SB was called ``anomalous relaxation''
by \citet{Hamers2014}.
This name reflects the fact the the evolution in this regime is qualitatively different than the evolution
described by the equations of either classical or resonant relaxation.
For instance: diffusion rates in this regime drop rapidly with decreasing $L$, and there is 
a net drift in the direction of increasing $L$.

The direct $N$-body integrations of \citet{MAMW2011} showed that the SB is not completely impermeable,
although captures by the \sbh\ were found to occur at a rate that was about an order of magnitude lower than 
 in simulations that omitted the first post-Newtonian (1PN) terms from the equations of motion; that is, 
the terms that generate Schwarzschild precession.
Extrapolating the capture rates in those small-$N$ simulations to real galaxies is not straightforward.
One reason is the absence, in the $N$-body models, of stars initially distant from the \sbh\ 
that would diffuse inward and replace those lost to the \sbh, thus establishing a steady state.
Another reason was pointed out by Hamers et al. (2014).
At sufficiently low $L$, anomalous diffusion rates can become so low that
{\it classical} relaxation  once again sets the timescale for angular momentum evolution.
Using an approximate test-particle algorithm, Hamers et al. were able to simulate systems of
much larger $N$ and to cleanly delineate three regimes of angular momentum evolution, 
at energies for which the SB exists:
\begin{enumerate}
\item $L_\mathrm{SB}(E) \lesssim L \le 1$ (resonant relaxation)
\item $L_\mathrm{NR}(E) \lesssim L \lesssim L_\mathrm{SB}(E)$ (anomalous relaxation)
\item $L_\mathrm{lc}(E) \le L\lesssim L_\mathrm{NR}( E)$ (classical relaxation)
\end{enumerate}
(see their Figure 1). Here $L_\mathrm{NR}$ is the angular momentum at which Schwarzschild
precession is so rapid that the torques driving resonant relaxation are almost completely ineffective
 at changing $L$, so that classical relaxation dominates the evolution once more.
 $L_\mathrm{lc}$ is the angular momentum at the edge of the loss cone.
\citet{Hamers2014} derived an approximate expression for $L_\mathrm{NR}(E)$ and showed that
in the simulations of \citet{MAMW2011}, 
classical relaxation dominated the evolution in $L$ over much of the $(a,e)$ plane,
including even some regions with $L>L_\mathrm{SB}(E)$. 
They argued that this fact would complicate the extrapolation of the  $N$-body results to 
real galaxies.

This paper, the third in a series, addresses these issues by incorporating into the Fokker-Planck algorithm expressions for the diffusion coefficients that account for anomalous relaxation.
By integrating $f(E,L)$ forward in time using these new diffusion coefficients,
steady-state solutions are constructed that are valid fully into 
the ``Schwarzschild'' regime defined in Paper II -- roughly an order of magnitude nearer to the
\sbh\ than the solutions of Paper II, or indeed any other published simulation.

Section~\ref{Section:Method} reviews the numerical algorithm used here; 
further details are given in Papers I and II.
Section~\ref{Section:AR} presents the functional forms adopted for the anomalous diffusion coefficients.
Since there does not yet exist a good theory for orbital evolution in this regime, different parametrized
forms for the  diffusion coefficients are considered and constrained by comparison
with previously-published simulations.
Section~\ref{Section:Results} presents steady-state solutions for $f(E,L)$ with parameters
chosen to describe the nuclear
cluster of the Milky Way; the results are compared with those of Paper II that did not 
incorporate anomalous relaxation.
Section~\ref{Section:Discussion} discusses some implications of the results obtained here 
and \S \ref{Section:Summary} sums up.

\section{Method}
\label{Section:Method}

As in Papers I and II, stars are assumed to have a single mass, $m_\star$, 
and to be close enough to the black hole 
(SBH) that the gravitational potential defining their unperturbed orbits is
\beq
\Phi(r) = -\frac{G\mh}{r} \equiv -\psi(r) 
\eeq
with $\mh$ the \sbh\ mass, assumed constant in time.
Unperturbed orbits respect the two isolating integrals $E$, the energy per unit mass,
and $L$, the angular momentum per unit mass.
Following \citet{CohnKulsrud1978} these are replaced by ${\cal E}$ and ${\cal R}$ where
\begin{eqnarray}
{\cal E} \equiv -E = -\frac{v^2}{2} +\psi(r), \ \ 
{\cal R} \equiv \frac{L^2}{L_c^2} ;
\end{eqnarray}
$L_c({\cal E})$ is the angular momentum of a circular orbit of energy ${\cal E}$
so that  $0\le {\cal R}\le 1$.
${\cal E}$ and ${\cal R}$ are related to the semimajor axis $a$ and eccentricity $e$ 
of the Kepler orbit via
\beq\label{Equation:semivsE}
a = \frac{G\mh}{2{\cal E}},\ \ \ \ e^2 = 1 - {\cal R} .
\eeq
Spin of the \sbh\ is ignored.

The time dependence of the phase-space number density of stars,
$f({\cal E},{\cal R})$, is described by the orbit-averaged Fokker-Planck equation
\begin{eqnarray}
{\cal J}\frac{\partial f}{\partial t} &=& 
-\frac{\partial}{\partial{\cal E}}\left({\cal J}\phi_{\cal E}\right) 
- {\cal J}\frac{\partial}{\partial {\cal R}}\phi_{\cal R}, \nonumber \\
-\phi_{\cal E} &=& D_{\cal E\cal E}\frac{\partial f}{\partial {\cal E}} + 
D_{\cal E\cal R} \frac{\partial f}{\partial {\cal R}} +
D_{\cal E} f,\ \ 
-\phi_{\cal R} = D_{\cal R\cal E}\frac{\partial f}{\partial {\cal E}} + 
D_{\cal R\cal R} \frac{\partial f}{\partial {\cal R}} + D_{\cal R} f
\label{Equation:FPFluxConserve}
\end{eqnarray}
with flux coefficients
\begin{eqnarray}\label{Equation:DefineFluxCoefs}
D_{\cal E} &=& -\langle\Delta{\cal E}\rangle 
- \frac{5}{4{\cal E}}\langle\left(\Delta{\cal E}\right)^2\rangle
+ \frac12\frac{\partial}{\partial{\cal E}}\langle\left(\Delta{\cal E}\right)^2\rangle
+ \frac12\frac{\partial}{\partial{\cal R}}\langle\Delta{\cal E}\Delta {\cal R}\rangle \;,
\nonumber \\
D_{\cal R} &=& -\langle\Delta{\cal R}\rangle
- \frac{5}{4{\cal E}}\langle\Delta{\cal E}\Delta{\cal R}\rangle 
+ \frac12\frac{\partial}{\partial{\cal E}}\langle\Delta{\cal E}\Delta{\cal R}\rangle 
+ \frac12\frac{\partial}{\partial{\cal R}}\langle\left(\Delta {\cal R}\right)^2\rangle \;,
\nonumber \\
D_{\cal E\cal E} &=& \frac12 \langle\left(\Delta{\cal E}\right)^2\rangle \;,
D_{\cal E\cal R} = D_{\cal R\cal E} = \frac12\langle\Delta{\cal E}\Delta{\cal R}\rangle\; ,
D_{\cal R\cal R} = \frac12\langle\left(\Delta{\cal R}\right)^2\rangle 
\end{eqnarray}
and ${\cal J}\equiv \sqrt{2}\pi^3G^3\mh^3{\cal E}^{-5/2}$
\citep[][5.5.1]{DEGN}.
Quantities in $\langle \; \rangle$ are orbit-averaged diffusion coefficients.
The functional forms of the diffusion coefficients are discussed below.

Loss of stars into the \sbh\ is controlled by the choice of $r_\mathrm{lc}$, the radius of the physical
loss sphere around the \sbh, and by the conditions imposed on $f$ at the
loss-cone boundary, ${\cal R} = {\cal R}_\mathrm{lc}({\cal E})$, defined as
\begin{eqnarray}\label{Equation:CaptureCond}
{\cal R}_\mathrm{lc} ({\cal E})&=& 
2\frac{\cal E}{{\cal E}_\mathrm{lc}} \left(1-\frac12\frac{\cal E}{{\cal E}_\mathrm{lc}}\right),\ \ 
{\cal E} \le {\cal E}_\mathrm{lc}, \ \ \ \ 
{\cal E}_\mathrm{lc} \equiv \frac{G\mh}{2r_\mathrm{lc}} \;.
\end{eqnarray}
${\cal R}_\mathrm{lc}$ is the normalized angular momentum of an orbit with (Newtonian)
periapsis at $r_\mathrm{lc}$.
The ${\cal R}$-directed flux of stars across the loss-cone boundary is
\beq\label{Equation:DefineFofE}
F({\cal E}) \; d{\cal E}= -{\cal J}({\cal E}) \phi_{{\cal R}}({\cal R}_\mathrm{lc}) \; d{\cal E} 
\equiv -{\cal J}({\cal E})\; \phi_{{\cal R},\mathrm{lc}}({\cal E}) \; d{\cal E} .
\eeq
Two quantities that play important roles in angular momentum diffusion near the loss-cone boundary 
are ${\cal D}$,
\begin{equation}\label{Equation:calDofElc}
{\cal D}({\cal E}) \equiv
\frac{\langle\left(\Delta{\cal R}\right)^2\rangle_t}{2{\cal R}}\bigg|_{{\cal R}={\cal R}_\mathrm{lc}}
= \frac{D_{\cal R R} ({\cal E}, {\cal R}_\mathrm{lc})}{{\cal R}_\mathrm{lc}}
\end{equation}
and $q_\mathrm{lc}$,
\begin{eqnarray}\label{Equation:Defineqlc}
q_\mathrm{lc}({\cal E}) \equiv  \frac{P({\cal E}) {\cal D}({\cal E})}{{\cal R}_\mathrm{lc}({\cal E})}.
\end{eqnarray}
${\cal D}^{-1}$ is effectively an orbit-averaged, angular momentum relaxation time at energy ${\cal E}$.
The quantity $q_\mathrm {lc}$
measures the change in angular momentum per orbital period, compared with the
size of the loss cone.
The loss-cone boundary conditions adopted in all the integrations presented here were the ``Cohn-Kulsrud boundary conditions'' defined in Paper I.
No attempt is made to solve for $f$ inside the loss cone, 
i.e. at ${\cal R}< {\cal R}_\mathrm{lc}$, since $f$ does not satisfy Jeans's theorem in this region.

Solutions are obtained on a ($N_x \times N_z$) grid in $(X,Z)$, where
\begin{eqnarray}
X &\equiv & \ln R = \ln \left[\frac{L}{L_c({\cal E})}\right]^2  , \nonumber \\
Z &\equiv & \ln\left(1 + \beta {\cal E}^*\right) = \ln \left(1+\beta {\cal E}/c^2\right) .
\end{eqnarray}
Integrations presented here used $N_x=N_z=64$ grid points.
The code adopts units such that
\beq\label{Equation:Units}
G=\mh=c=1 
\eeq
allowing the results to be scaled to different masses of the \sbh.
Dimensionless parameters that must be specified before the start of an integration include
$m_\star/\mh$, $\ln\Lambda$ and $\Theta_\mathrm{lc}\equiv r_\mathrm{lc}/r_g$.

In Paper II, the diffusion coefficients had the forms
\begin{eqnarray}\label{Equation:CombinedDiffCoefs}
\langle\Delta{\cal E}\rangle &=&  \langle\Delta{\cal E}\rangle_\mathrm{CK},\ \ \ \ 
\langle\left(\Delta{\cal E}\right)^2\rangle = \langle\left(\Delta{\cal E}\right)^2\rangle_\mathrm{CK}, 
\ \ \ \ 
\langle\Delta{\cal E}\Delta{\cal R}\rangle =  
\langle\Delta{\cal E}\Delta{\cal R}\rangle_\mathrm{CK} ,
\nonumber \\
\langle\Delta{\cal R}\rangle &=& \langle\Delta{\cal R}\rangle_\mathrm{CK}
+ \langle\Delta{\cal R}\rangle_\mathrm{RR}, \ \ \ \ 
\langle\left(\Delta{\cal R}\right)^2\rangle =
\langle\left(\Delta{\cal R}\right)^2\rangle_\mathrm{CK} + 
 \langle\left(\Delta{\cal R}\right)^2\rangle_\mathrm{RR} \;.
\end{eqnarray}
The subscript CK indicates that the diffusion coefficient is computed as in \citet{CohnKulsrud1978};
their derivation was based on standard assumptions about randomness of encounters
\citep{Rosenbluth1957}.
The subscript RR refers to ``resonant relaxation'' \citep{RauchTremaine1996}.
The resonant diffusion coefficients were expressed as
\begin{eqnarray}
\langle \Delta{\cal R}\rangle_\mathrm{RR} = 2 A({\cal E}) \left(1 - 2{\cal R}\right), \ \ \ \ 
\langle\left(\Delta {\cal R}\right)^2\rangle_\mathrm{RR} = 4 A({\cal E}) {\cal R}\left(1-{\cal R}\right) .
\label{Equation:RRDiffCoef}
\end{eqnarray}
The term containing the ${\cal E}$ dependence is
\begin{eqnarray}
A(a) =  \alpha_s^2\left[\frac{M_\star}{\mh}\right]^2 \frac{1}{N} 
\frac{t_\mathrm{coh}}{P^2} , \ \ \ \ \alpha_s = 1.6, \ \ \ \ a = \frac{G\mh}{2{\cal E}} .
\label{Equation:RRDiffCoef2}
\end{eqnarray}
Here $N\equiv N(r<a)$ is the number of stars instantaneously at radii than $a$,
$M_\star = m_\star N$, $P$ is the Kepler (radial) period, and $t_\mathrm{coh}$ 
is the coherence time, defined as
\begin{eqnarray}\label{Equation:Definetcoh}
t_\mathrm{coh}^{-1} &\equiv& t_\mathrm{coh,M}^{-1} + t_\mathrm{coh,S}^{-1} \nonumber \\
t_\mathrm{coh,M}(a) &=& \frac{\mh}{Nm_\star}P\;, \ \ \ \ 
t_\mathrm{coh,S}(a) = \frac{1}{12} \frac{a}{r_g} P .
\end{eqnarray}
$t_\mathrm{coh,M}$ is the mean precession time for stars of semimajor axis $a$ due to
the distributed mass around the \sbh\ (``mass precession''), and
$t_\mathrm{coh,S}$ is the mean precession time due to the 1PN corrections to the
Newtonian equations of motion (``Schwarzschild precession'').

\section{Anomalous diffusion coefficients}
\label{Section:AR}

The diffusion coefficients (\ref{Equation:CombinedDiffCoefs}) 
are affected by general relativity (GR) to the extent that 
GR determines the coherence time via equation (\ref{Equation:Definetcoh}).
Another GR-related phenomenon is the Schwarzschild barrier (SB), 
the tendency of orbits near the \sbh\ to avoid high eccentricities.
The SB was first observed in $N$-body simulations \citep{MAMW2011},
as a locus in the ($E,L$) plane where trajectories ``bounced'' during the course
of their random walks in $L$.
At energies where the angular momentum associated with the bounce,
$L_\mathrm{SB} (E)$, exceeds $L_\mathrm{lc} (E)$, 
far fewer stars are captured by the \sbh\ than in simulations
that neglect the effects of GR.
The \citet{MAMW2011} study revealed that orbits experiencing the ``bounce'' were 
of such high eccentricity that their GR precession times were short compared with those
of typical (i.e., less eccentric) stars at the same $a$.

\citet{Hamers2014} coined the term ``anomalous relaxation'' to describe the
behavior of orbits in this high-eccentricity regime, $L\lesssim L_\mathrm{SB}(E)$.
Those authors verified the existence of the SB  via an independent set of $N$-body 
integrations, and also carried out test-particle integrations, using a much larger number of
stars, from which they numerically evaluated the rates of diffusion in the anomalous regime.

Based on these, and other, studies, two analytic expressions have been proposed for the location of the SB.
The first compares the GR precession time with the time for the
$\sqrt{N}$ torques to change $L$ \citep{MAMW2011}:
\beq\label{Equation:SB}
{\cal R}_\mathrm{SB}^{(i)} (a) \approx \left(\frac{r_g}{a}\right)^2
\left[\frac{\mh}{M_\star(a)}\right]^2 N(a) \;.
\eeq
The second \citep{Hamers2014,BarorAlexander2014} 
compares the GR precession time with the coherence time:
\beq\label{Equation:SB2}
{\cal R}_\mathrm{SB}^{(ii)} (a) \approx 4 \frac{r_g}{a} \frac{t_\mathrm{coh}(a)}{P(a)}\;.
\eeq
In spite of their disparate functional forms, the two expressions can yield numerically
similar relations for ${\cal R}_\mathrm{SB}(a)$, as illustrated below.
The former relation appears to more accurately reproduce the barrier location in numerical studies to date;
while the latter relation arises naturally when matching diffusion coefficients in the resonant
and anomalous regimes \citep{Hamers2014}.

Evaluating the former expression in the case of an unmodified Bahcall-Wolf cusp, 
$n(r)\propto r^{-7/4}$, yields
\beq\label{Equation:SBBW}
\frac{a_\mathrm{SB}}{r_g} \approx 
\left(\frac{\mh}{2m_\star}\right)^{4/13}
\left(\frac{r_m}{r_g}\right)^{5/13}
\left(1-e^2\right)^{-4/13}
\eeq
where $r_m$ is the radius containing a mass in stars of $2\mh$.
The barrier as given by Equation~(\ref{Equation:SBBW}) extends between the radii
$a_\mathrm{min}$ and $a_\mathrm{max}$, where
\begin{subequations}\label{Equation:SBlimits}
\begin{eqnarray}
\frac{a_\mathrm{min}}{r_g} &\approx& \left(\frac{\mh}{2m_\star}\right)^{4/13}
\left(\frac{r_m}{r_g}\right)^{5/13},
\\
\frac{a_\mathrm{max}}{r_g} &\approx& \left(4\Theta\right)^{-4/9}
\left(\frac{\mh}{m_\star}\right)^{4/9}
\left(\frac{r_m}{r_g}\right)^{5/9} .
\end{eqnarray}
\end{subequations}
The first relation follows from setting $e=0$.
The second relation is the intersection of Equation~(\ref{Equation:SBBW}) with the curve $a(1-e)=\Theta r_g$,
the periapsis of an orbit that intersects the loss sphere of radius $r_\mathrm{lc}=\Theta r_g$.
Taking parameter values appropriate for the Milky Way:
\beq
\mh = 4\times 10^6\msun,\ \ \ \ m_\star = 1, \ \ \ \ \Theta = 15 \nonumber
\eeq
yields
\begin{eqnarray}\label{Equation:SBBWMW}
a_\mathrm{min} \approx 6 \left(\frac{r_m}{\mathrm{pc}}\right)^{5/13} \mathrm{mpc} ,
\ \ \ \ 
a_\mathrm{max}\approx 140 \left(\frac{r_m}{\mathrm{pc}}\right)^{5/9} \mathrm{mpc} \;.
\end{eqnarray}

``Anomalous relaxation'' is defined as angular-momentum diffusion of orbits with
$L\lesssim L_\mathrm{SB}(E)$.
Paper I presented a derivation, based on a simple Hamiltonian model,
 of the diffusion coefficients in the  anomalous regime:
\begin{eqnarray}
\langle\Delta{\cal R}\rangle_\mathrm{AR} =
\frac{5}{\tau}  {\cal R}^2, \ \ \ \ 
\langle(\Delta{\cal R})^2\rangle_\mathrm{AR} 
= \frac{4}{\tau} {\cal R}^3 
\label{Equation:DiffAR12}
\end{eqnarray}
where $\tau(a) \equiv t_\mathrm{coh}(a)/(A_{\sqrt{N}})^2$
and
\beq\label{Equation:DefineAD}
A_{\sqrt{N}} \equiv \frac{1}{2\sqrt{N(a)}}
\frac{M_\star(a)}{\mh} \frac{a}{r_g} .
\eeq
The rapid, power-law drop predicted in the diffusion rates for ${\cal R}<{\cal R}_\mathrm{SB}$
is due to the adiabatic invariance of $L$ under the effects of rapid precession.

The derivation leading to equations (\ref{Equation:DiffAR12})-(\ref{Equation:DefineAD}) was
very approximate and one would like to verify those functional forms 
by  comparison with $N$-body integrations.
\citet{Hamers2014} attempted to do this.
However, it was found that the value of $N$ accessible to high-accuracy simulations ($N\lesssim 100$)
was so small that the effects of anomalous relaxation could not be cleanly differentiated from the 
effects of classical relaxation at small $L$.
Application of a more approximate, test-particle approach allowed Hamers et al. to increase
the effective value of $N$ by two orders of magnitude.
The diffusion coefficients extracted from these experiments 
were found to be reasonably well described by equations  (\ref{Equation:DiffAR12})-(\ref{Equation:DefineAD}).

In the present treatment,
we account for the effects of anomalous relaxation by modifying the 
angular momentum diffusion coefficients (\ref{Equation:CombinedDiffCoefs}).
We consider two sorts of modification with  different functional forms: 
a power-law modification, which reproduces equations~(\ref{Equation:DiffAR12}) at small ${\cal R}$;
and an exponential modification, which implies a much more rapid decrease in the diffusion rate toward
small ${\cal R}$.

\subsection{Power-law modification}
\label{Section:AR_PL}

To account for anomalous relaxation, the angular-momentum diffusion coefficients of
Equation~(\ref{Equation:CombinedDiffCoefs}) are modified as follows:
\begin{eqnarray}\label{Equation:ARDiffCoefsPL}
\langle \Delta{\cal R}\rangle = \langle\Delta {\cal R}\rangle_\mathrm{CK} +
g_1({\cal E},{\cal R}) \langle\Delta{\cal R}\rangle_\mathrm{RR},
\ \ \ \ 
\langle (\Delta{\cal R})^2\rangle = \langle(\Delta {\cal R})^2\rangle_\mathrm{CK} +
g_2({\cal E},{\cal R}) \langle(\Delta{\cal R})^2\rangle_\mathrm{RR}.
\end{eqnarray}
The functions $g_1({\cal R})$ and $g_2({\cal R})$ should have certain properties.
Both $g_1$ and $g_2$ should tend to one as ${\cal R}\rightarrow 1$.
{\bf Since}
\begin{eqnarray}
\langle\Delta{\cal R}\rangle_\mathrm{RR} \rightarrow 2A({\cal E}),\ \ \ \ 
\langle(\Delta{\cal R})^2\rangle_\mathrm{RR} \rightarrow 4A({\cal E}) {\cal R}  \nonumber
\end{eqnarray}
{\bf as ${\cal R}\rightarrow 0$ (Equation~\ref{Equation:RRDiffCoef}),}
we require $g_1\rightarrow {\cal R}^2$ and $g_2\rightarrow {\cal R}^2$ for small ${\cal R}$
so that the small-${\cal R}$ behavior of Equation (\ref{Equation:DiffAR12}) is reproduced.
The transition between the two regimes should occur at ${\cal R}\sim {\cal R}_\mathrm{SB}$ for
both functions.

An ad hoc functional form for $g_2$ that satisfies these requirements is
\beq
g_2({\cal E},{\cal R}) = \left\{1+\left[\frac{R_2({\cal E})}{\cal R}\right]^n\right\}^{-2/n} 
\label{Equation:Defineg2}
\eeq
where ${\cal R}_2\approx {\cal R}_\mathrm{SB}({\cal E})$.
The parameter $n$ determines the rapidity of transition between the large-${\cal R}$ and
small-${\cal R}$ regimes.

The same functional form might be adopted for $g_1$. 
Rather than make that choice, we first consider another possible constraint on $g_1$ and $g_2$.

The ${\cal R}$-directed flux coefficients that appear in the Fokker-Planck equation are
given by Equations~(\ref{Equation:DefineFluxCoefs}):
\begin{subequations}\label{Equation:RFluxCoefs}
\begin{eqnarray}
D_{\cal R} &=& -\langle\Delta {\cal R}\rangle -\frac{5}{4{\cal E}}\langle\Delta{\cal E}\Delta{\cal R}\rangle
+ \frac12\frac{\partial}{\partial{\cal E}}\langle\Delta{\cal E}\Delta{\cal R}\rangle
+ \frac12\frac{\partial}{\partial {\cal R}} \langle(\Delta {\cal R})^2\rangle \\
&\approx &  -\langle\Delta {\cal R}\rangle
+ \frac12\frac{\partial}{\partial {\cal R}} \langle(\Delta {\cal R})^2\rangle, 
\label{Equation:RFluxCoefsb}\\
D_{\cal RR} &=& \frac12\langle(\Delta {\cal R})^2\rangle .
\end{eqnarray}
\end{subequations}
Since the ${\cal R}$-directed {\it flux} is
\begin{eqnarray}
\phi_{\cal R} = 
-D_{{\cal R}{\cal E}} \frac{\partial f}{\partial {\cal E}} - D_{\cal RR}\frac{\partial f}{\partial {\cal R}}
- D_{\cal R} f \approx 
 - D_{\cal RR}\frac{\partial f}{\partial {\cal R}} - D_{\cal R} f,
\end{eqnarray}
it is reasonable to require that the diffusion coefficients in ${\cal R}$ satisfy
\beq\label{Equation:DRDRRCond}
D_{\cal R}\rightarrow 0,\ \ \ \ D_{\cal RR}\rightarrow 0
\eeq
at the boundaries ${\cal R}= \{0,1\}$;
in other words, that 
\begin{subequations}\label{Equation:DRDRRconds}
\begin{eqnarray}
\langle\Delta{\cal R}\rangle = \frac12\frac{\partial}{\partial {\cal R}} \langle(\Delta{\cal R})^2\rangle, 
\label{Equation:DRDRRcondsa} \ \ \ \ 
\langle(\Delta{\cal R})^2\rangle = 0
\label{Equation:DRDRRcondsb}
\end{eqnarray}
\end{subequations}
at ${\cal R} = \{0,1\}$.
Both the classical (Cohn-Kulsrud), and the resonant diffusion coefficients adopted here and in Papers I 
and II satisfy these conditions.

Suppose that a stronger condition is imposed: $D_{\cal R}\equiv 0$, $0\le {\cal R}\le 1$.
In this case, the Fokker-Planck equation describing diffusion in angular momentum 
reduces to
\begin{eqnarray}
\frac{\partial f}{\partial t} = -\frac{\partial\phi_{\cal R}}{\partial {\cal R}}  = 
\frac{\partial}{\partial {\cal R}}\left(D_{\cal RR}\frac{\partial f}{\partial {\cal R}}\right)
= \frac12 \frac{\partial}{\partial {\cal R}}\left(\langle(\Delta{\cal R})^2\rangle \frac{\partial f}{\partial {\cal R}}\right)
\end{eqnarray}
which has a steady-state (zero-flux) solution $
f({\cal R}) = \mathrm{constant}$
regardless of the functional form of $\langle(\Delta{\cal R})^2\rangle$.
It could be argued that $f({\cal R})=$ constant is a reasonable form for a time-independent $f$,
since it corresponds to an isotropic, or ``maximum entropy,'' state.
The resonant diffusion coefficients adopted in Papers I and II satisfy this stronger condition.
Setting $D_{\cal R}\equiv 0$ also implies zero drift, i.e., zero flux in ${\cal R}$ in the absence of a gradient.
For this reason, $D_{\cal R}\equiv 0$ will henceforth be called a ``zero-drift'' condition.

Returning now to the anomalous diffusion coefficients, we ask: what functional form for $g_1$ 
is required for zero drift?
That is:
\begin{eqnarray}\label{Equation:DRequalszero}
0 &=& D_{\cal R} = -g_1({\cal R})\langle\Delta {\cal R}\rangle_\mathrm{RR} +  \frac12\frac{\partial}{\partial {\cal R}} 
\left[g_2({\cal R})  \langle(\Delta {\cal R})^2\rangle_\mathrm{RR}\right] .
\end{eqnarray}
Assuming that $g_2$ is given by Equation (\ref{Equation:Defineg2}), the result is
\begin{eqnarray}\label{Equation:Defineg10}
g_1({\cal R}) =  \left\{1+\left[\frac{R_2({\cal E})}{\cal R}\right]^n\right\}^{-2/n} 
+ \frac {2(1-{\cal R})}{1-2{\cal R}} \left(\frac{{\cal R}_2}{{\cal R}}\right)^n
 \left\{1+\left[\frac{R_2({\cal E})}{\cal R}\right]^n\right\}^{-2/n-1} .
\end{eqnarray}
Since the zero-drift argument is one of plausibility only,
we will consider a slightly more general expression for $g_1$ that includes
``zero drift'' as a special case:
\begin{eqnarray}\label{Equation:Defineg1}
g_1({\cal R}) =  \left\{1+\left[\frac{R_1({\cal E})}{\cal R}\right]^n\right\}^{-2/n} 
+ \frac {2(1-{\cal R})}{1-2{\cal R}} \left(\frac{{\cal R}_1}{{\cal R}}\right)^n
 \left\{1+\left[\frac{R_1({\cal E})}{\cal R}\right]^n\right\}^{-2/n-1} .
\end{eqnarray}
The only difference between Equations~(\ref{Equation:Defineg10}) and (\ref{Equation:Defineg1}) 
is the introduction of a second parameter, ${\cal R}_1$, in place of ${\cal R}_2$.
This generalization still satisfies the zero-flux condition (\ref{Equation:DRDRRCond}) at ${\cal R}=0$,
regardless of  ${\cal R}_1/{\cal R}_2$.
At ${\cal R}=1$, $g_1$ and $g_2$ are both very close to one (especially since $n$ will be chosen
to be large) so that the resonant diffusion coefficients are recovered and the zero-flux condition is satisfied, 
again for any choice of ${\cal R}_1/{\cal R}_2$.

Correspondence of these expressions
with the  diffusion coefficients of Equation (\ref{Equation:DiffAR12}) at small ${\cal R}$ would require
\begin{eqnarray}
{\cal R}_1^2({\cal E}) &=& (6/5) \tau A({\cal E}) , \ \ \ \ 
{\cal R}_2^2({\cal E}) = \tau A({\cal E}) ,
\end{eqnarray}
where
\beq
\tau A({\cal E}) = 4\alpha_s^2 \left(\frac{t_\mathrm{coh}}{P}\right)^2 \left(\frac{r_g}{a}\right)^2 .
\eeq
To within factors of order unity, these relations imply
${\cal R}_1\approx {\cal R}_2\approx {\cal R}^{(ii)}_\mathrm{SB}$ (equation~\ref{Equation:SB2}).

It is shown in the Appendix that at small ${\cal R}$, these choices for $g_1$ and $g_2$ imply
\begin{eqnarray}
D_{\cal R} \rightarrow 6\lambda A({\cal E}) \left(\frac{{\cal R}}{{\cal R}_2}\right)^2 ,
\ \ \ \ \lambda \equiv 1 - \frac{{\cal R}_2^2}{{\cal R}_1^2}
\end{eqnarray}
so that the direction of the drift is determined by the relative sizes of ${\cal R}_1$ and ${\cal R}_2$, 
as follows:
\begin{subequations}\label{Equation:FluxDirection}
\begin{eqnarray}
{\cal R}_1 &<& {\cal R}_2\; \rightarrow\; D_{\cal R} < 0\; \rightarrow\; \phi_{\cal R}>0 \\
{\cal R}_1 &>& {\cal R}_2\; \rightarrow\; D_{\cal R} > 0\; \rightarrow\;  \phi_{\cal R}<0 
\end{eqnarray}
\end{subequations}
where the expressions for $\phi_{\cal R}$ assume $f({\cal R}) = \mathrm{constant}$.
Furthermore both the form of the steady-state $f({\cal R})$, and the steady-state flux (assuming the presence
of a sink, i.e. that $f({\cal R})=0$ at ${\cal R}\le {\cal R}_0$), depend sensitively on $\lambda$, as shown in Figures~\ref{Figure:fofRAppend} and \ref{Figure:beta} from the Appendix.
In the zero-drift ($\lambda=0$) case, the steady-state flux is reduced by a factor
\beq
\eta \approx 2 \left(\frac{{\cal R}_0}{{\cal R}_2}\right)^2 \log\left(\frac{{\cal R}_2}{{\cal R}_0}\right)
\eeq
compared with the flux that would obtain in the absence of anomalous relaxation
(note that an empty loss cone has been assumed).
For nonzero $\lambda$, the reduction factor can be greater or smaller than this, tending
toward a maximum value of one for large ${\cal R}_1/{\cal R}_2$ (Figure~\ref{Figure:beta}).

\subsection{Exponential modification}
\label{Section:AR_EXP}

\citet{BarorAlexander2014} suggested a different functional form for $\langle(\Delta{\cal R})^2\rangle$
in the anomalous regime:
\beq
\langle(\Delta {\cal R})^2\rangle \propto \exp\left(-\frac{1}{\pi}\frac{{\cal R}_\mathrm{SB}^2}{{\cal R}^2}\right),
\eeq
an exponential cut-off toward small ${\cal R}$. 
While there does not seem to be strong support for this functional form in any published numerical
simulations, we consider it here for completeness, and because it serves to illustrate how sensitively
the evolution of $f$ depends on the form assumed for the anomalous diffusion coefficients.

Proceeding as above, we write
\begin{eqnarray}\label{Equation:ARDiffCoefsEXP}
\langle \Delta{\cal R}\rangle = \langle\Delta {\cal R}\rangle_\mathrm{CK} +
h_1({\cal E},{\cal R}) \langle\Delta{\cal R}\rangle_\mathrm{RR},
\ \ \ \ 
\langle (\Delta{\cal R})^2\rangle = \langle(\Delta {\cal R})^2\rangle_\mathrm{CK} +
h_2({\cal E},{\cal R}) \langle(\Delta{\cal R})^2\rangle_\mathrm{RR}.
\end{eqnarray}
Suppose
\beq\label{Equation:Defineh2}
h_2({\cal E},{\cal R}) = \exp\left(-\frac{{\cal R}_4^2}{{\cal R}^2}\right)
\eeq
where ${\cal R}_4 ({\cal E})\approx {\cal R}_\mathrm{SB}({\cal E})$.
Since ${\cal R}_4\ll 1$, $h_2({\cal E}, 1)\approx 1$.
The zero-drift condition would imply
\beq
\langle\Delta{\cal R}\rangle = 2A({\cal E})
\left[1-2{\cal R} + 2(1-{\cal R})\left(\frac{{\cal R}_4}{\cal R}\right)^2\right]
\exp\left(-\frac{{\cal R}_4^2}{{\cal R}^2}\right) . \nonumber
\eeq
We  again generalize this expression by defining a second parameter, ${\cal R}_3$, and writing
\beq\label{Equation:Defineh1}
h_1({\cal R}) = \left[1 + \frac{2(1-{\cal R})}{1-2{\cal R}}\left(\frac{{\cal R}_3}{{\cal R}}\right)^2\right]
\exp\left(-\frac{{\cal R}_3^2}{{\cal R}^2}\right) .
\eeq
At small ${\cal R}$, these expressions imply a flux coefficient
\begin{eqnarray}
D_{\cal R} \rightarrow 4A({\cal E}) \left(\frac{{\cal R}_4}{\cal R}\right)^2
\exp\left(-\frac{{\cal R}_4^2}{{\cal R}^2}\right)
\left[1 - \frac{{\cal R}_3^2}{{\cal R}_4^2}\exp{\left(-\frac{{\cal R}_3^2 - {\cal R}_4^2}{{\cal R}^2}\right)}\right]
\end{eqnarray}
showing that in this case, as in the power-law case, 
the sign of $D_{\cal R}$ is determined by the relative sizes
of ${\cal R}_3$ and ${\cal R}_4$.
Figures~\ref{Figure:fofRAppend} and \ref{Figure:beta} illustrate the properties of the steady-state solutions
$f({\cal R})$.
The behavior of $f$ at small ${\cal R}$ depends very sensitively on  ${\cal R}_3/{\cal R}_4$.
The reduction in the steady-state flux, for ${\cal R}_3={\cal R}_4$, is
\beq
\eta \approx 2\frac{{\cal R}_4}{{\cal R}_0} 
\log \left(\frac{{\cal R}_4}{{\cal R}_0}\right)
e^{-{\cal R}_4/{\cal R}_0} .
\eeq

\subsection{Constraining the functional forms of the diffusion coefficients in the anomalous regime}
\label{Section:AR_Constrain}

Based on the analysis presented above and in the Appendix, many properties of the steady-state
solutions are expected to depend sensitively on the functional forms of the diffusion coefficients 
in the anomalous regime, and in particular on the ratios ${\cal R}_1/{\cal R}_2$ or ${\cal R}_3/{\cal R}_4$.

In Paper I, the functional forms of the diffusion coefficients in the {\it resonant} regime were successfully 
constrained by comparison with the numerical results of \citet{Hamers2014}.
Those authors used a test-particle integrator to extract values of the angular momentum diffusion
coefficients for stars orbiting near a \sbh\ in nuclei with $n(r) \propto r^{-2}$ and $r^{-1}$.

The Hamers et al. integrations included post-Newtonian terms in the equations of motion,
both for the field and test stars, and the suppression of angular momentum diffusion below
the Schwarzschild barrier was clearly seen.
Figure~\ref{Figure:dfit} provides an illustration: it shows the first- and second-order diffusion 
coefficients for stars in a single energy bin, in integrations of a model with 
$n(r) \propto r^{-2}$.\footnote{These data were kindly provided by A. Hamers.}
Overplotted are analytic diffusion coefficients from the two families considered above:
power-law (Equations~\ref{Equation:DiffCoefsAppendix}, {\ref{Equation:g1g2Appendix}) 
and exponential (Equations~\ref{Equation:DiffCoefsAppendix}, \ref{Equation:h1h2Appendix}).

These figures, and similar ones made for stars at other energies, motivate the following conclusions
(some of which were presented already by Hamers et al.):
\begin{enumerate}
\item An exponential dependence of the diffusion coefficients on ${\cal R}$ in the anomalous regime
\citep{BarorAlexander2014} is ruled out.
\item The power-law dependence of $\langle\Delta{\cal R}\rangle$ and $\langle(\Delta{\cal R})^2\rangle$
on ${\cal R}$ predicted in Paper I, and reproduced here in Equations (\ref{Equation:DiffAR12}),
is confirmed, particularly in the case of the second-order coefficient for which the noise is smallest.
\item The value of ${\cal R}$ that defines the transition between the resonant and anomalous regimes,
called here ${\cal R}_2$, is well predicted by Equation (\ref{Equation:SB2}).
\end{enumerate}

\begin{figure}[h!]
\centering
\mbox{\subfigure{\includegraphics[angle=0.,width=3.in]{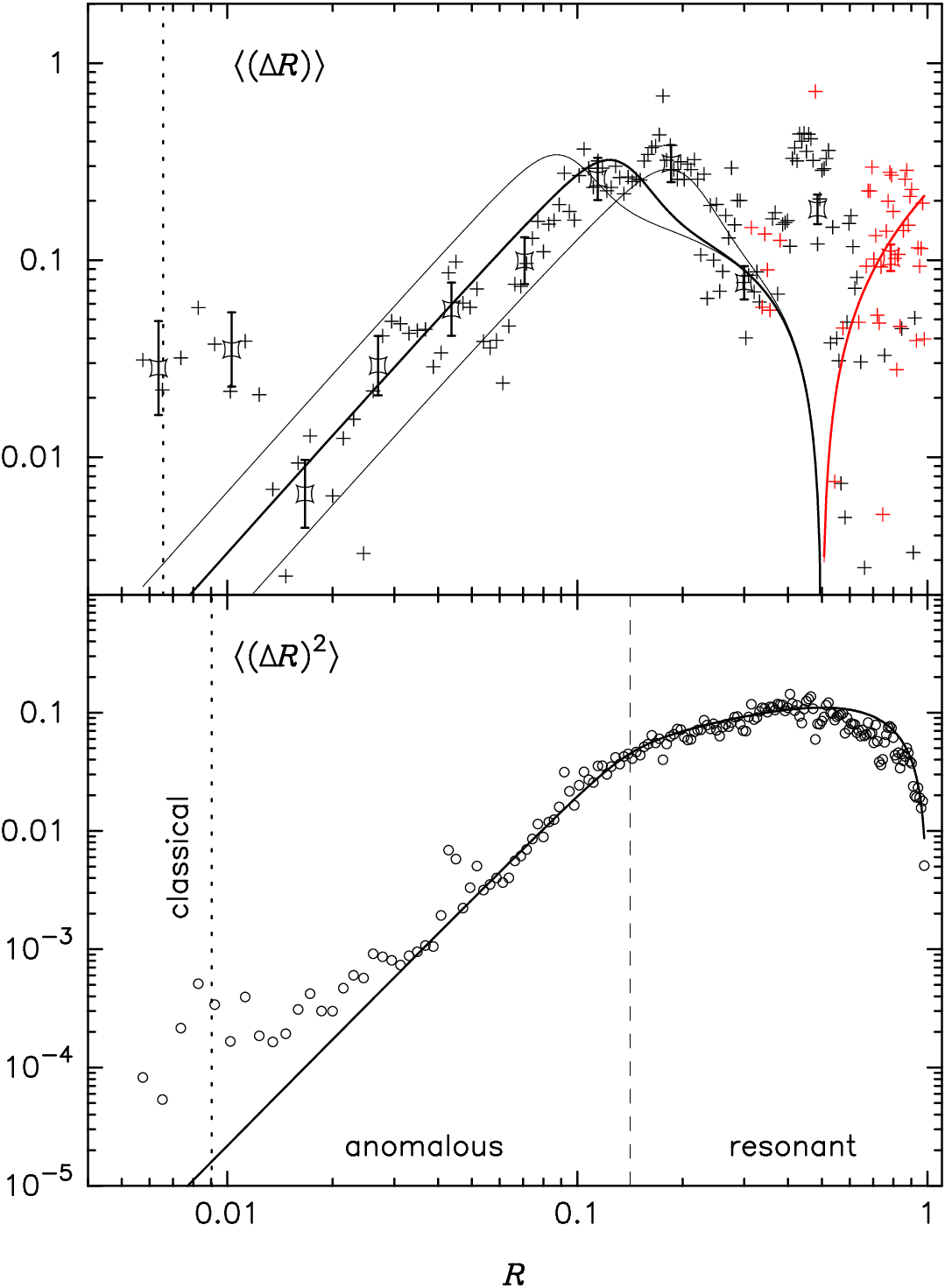}}\quad
\subfigure{\includegraphics[angle=0.,width=3.in]{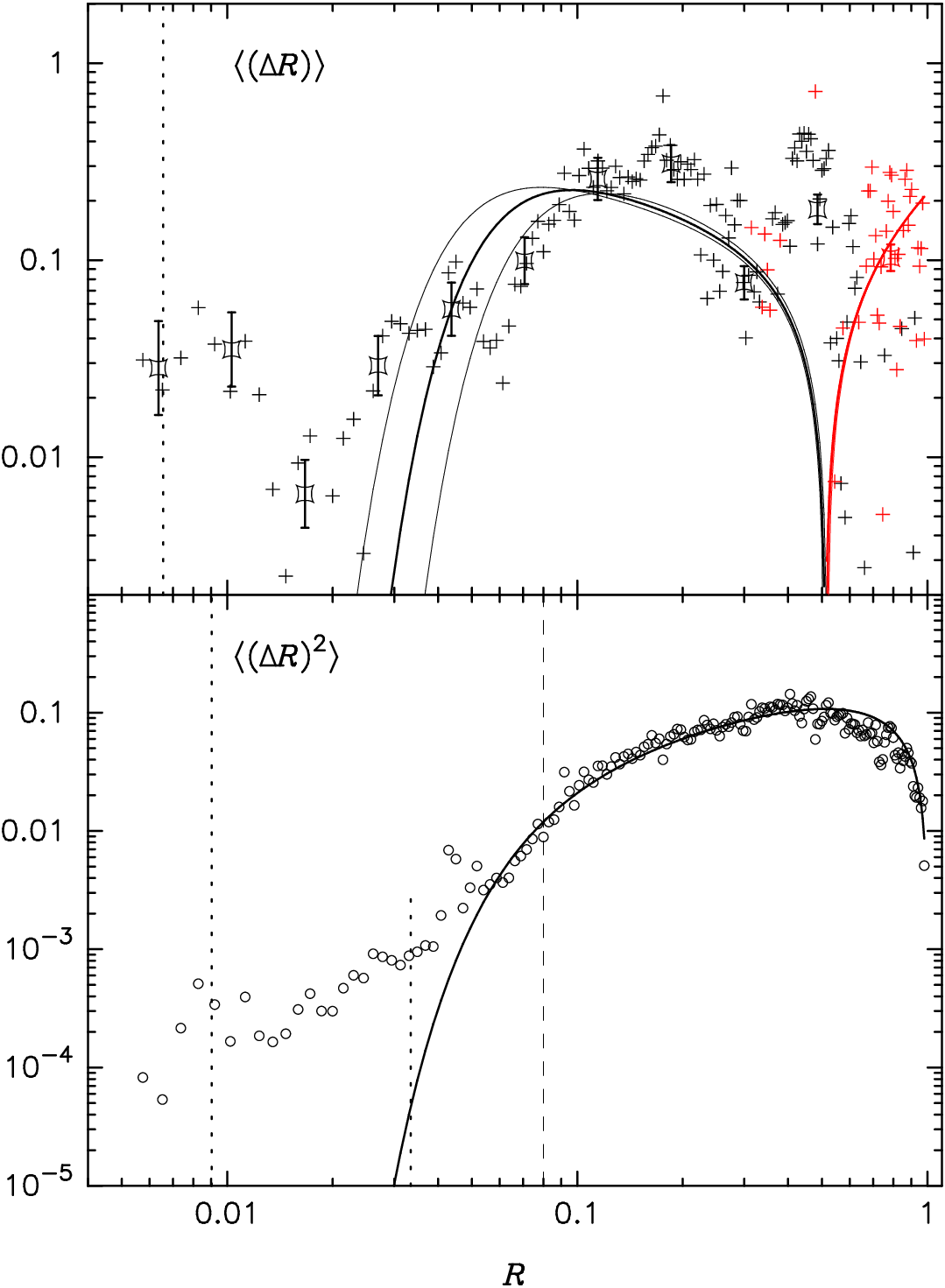} }}
\caption{Fits to diffusion coefficients extracted from the test-particle integrations
of  \citet{Hamers2014}, for the radial bin $\langle a \rangle = 11.9$ mpc.
In the upper panels, red symbols are $-\langle\Delta{\cal R}\rangle$.
{\bf Diamond symbols are binned data, with errors, to guide the eye; fits were based on the
unbinned data.}
{\bf Left:} analytic curves are Equations (\ref{Equation:DiffCoefsAppendix}), (\ref{Equation:g1g2Appendix}),
the power-law model for anomalous relaxation.
${\cal R}_2 = 0.110$ (shown by the dashed line in the bottom panel);
in the upper panel, ${\cal R}_1/{\cal R}_2 = \{1,1.5,0.7\}$.
{\bf Right:} fits of the exponential model for anomalous relaxation,
Equations  (\ref{Equation:DiffCoefsAppendix}), (\ref{Equation:h1h2Appendix}).
${\cal R}_4 = 0.080$, ${\cal R}_3/{\cal R}_4 = \{1,1.25,0.8\}$.
Vertical dotted lines are estimates of the value of ${\cal R}$ at which classical relaxation
dominates anomalous relaxation (Eqs.~\ref{Equation:RNR1}), assuming that the power-law
forms of the anomalous diffusion coefficients are correct;
the analytic forms would not be expected to describe the data below these lines.
In the lower-right panel, the additional dotted line at ${\cal R}\approx 0.33$
is Eq.~(\ref{Equation:RNR4}), an estimate of where classical relaxation would begin to dominate
if the exponential forms of the anomalous diffusion coefficients were correct.
}
\label{Figure:dfit}
\end{figure}

An attempt was made to find the best-fit value(s) of ${\cal R}_1/{\cal R}_2$
 by searching for parameter values that optimized the fits to data
like those in Figure~\ref{Figure:dfit}.
Unfortunately, the results so obtained were found not to be robust.
This was due in part to the greater noise associated with data below the Schwarzschild barrier,
particularly in the case of the first-order coefficient.
In addition, the much greater variation in the amplitudes of $\langle\Delta{\cal R}\rangle$ and $\langle(\Delta{\cal R})^2\rangle$ at a single energy meant that the best-fit solution depended sensitively on the
relative weighting of the data at different values of ${\cal R}$.
In the end, all that could be concluded was that the first-order diffusion coefficients
are consistent with ${\cal R}_1={\cal R}_2$, the ``zero-drift'' condition; but that 
values of ${\cal R}_1/{\cal R}_2$ moderately greater or less than one could not be excluded using
these data.

The fact that the steady-state form of $f({\cal R})$, and the loss-cone flux,
depend sensitively on ${\cal R}_1/{\cal R}_2$ suggests 
a second way to constrain the anomalous diffusion coefficients:
insert them into the Fokker-Planck equation and integrate forward from initial conditions
like the ones used by Merritt et al. (2011) in their small-$N$ simulations.
As discussed in Hamers et al., the value of $N$ in those simulations was too small to allow
direct extraction of the diffusion coefficients.
However, the time-averaged, or integrated, properties of the $N$-body models were reasonably well
determined, particularly given that  multiple ($\sim 8$) realizations of the same initial conditions
were integrated, allowing the  variance in the results to be  reduced by averaging.

The \citet{MAMW2011} initial conditions consisted of 50 stars, of mass $m_\star=50\msun$,  distributed as
$n(r) \propto r^{-2}$ around a \sbh\ of mass $\mh=10^6\msun$.
The initial distribution was truncated for orbits with semi-major axes above 10 mpc and below
0.1 mpc.
Integrations were carried out both with, and without, post-Newtonian terms in the 
equations of motion, up to order 2.5 PN.
Capture of stars by the \sbh\ was allowed to occur whenever the orbital periapsis fell below
$8r_g= 8 G\mh/c^2$.
Each realization of the initial conditions was integrated for a time corresponding to $2\times 10^6$ yr
(with PN terms) and $10^7$ yr (without PN terms).
The average capture rate in the relativistic integrations was about one event per $10^6$ yr;
in the absence of the PN terms, mean capture rates were about a factor 20 higher.

There is no ambiguity in representing the Merritt et al. $N$-body
initial conditions as a smooth $f({\cal E}, {\cal R})$.
However, the Fokker-Planck algorithm has a number of parameters  that must be specified,
in addition to those that define the anomalous diffusion coefficients ($n$ and ${\cal R}_1/{\cal R}_2$,
or ${\cal R}_3/{\cal R}_4$).
Those  parameters include
\beq
{\cal E}_\mathrm{min}, \ \ \ \ 
\ln\Lambda,\ \ \ \ 
\frac{r_\mathrm{lc}}{r_g} \ \ \ \ 
\Delta t,\ \ \ \ 
N_X,\ \ \ \  N_Z .
\eeq
${\cal E}_\mathrm{min}$ is the binding energy at the edge of the (${\cal E}, {\cal R}$) grid;
it should be small enough that few stars diffuse to ${\cal E}<{\cal E}_\mathrm{min}$
during the course of the integration.
The value $10^{-8} c^2$ was chosen, which is the energy of an orbit with semimajor axis
$a = 5\times 10^{7} r_g \approx$ 2.5 pc, or $\sim 250$ times the maximum $a$-value of
the initial conditions.
The number of grid points was $N_X=N_Z=64$.
The quantity $\ln\Lambda$ was set to $15$ in most of the integrations, except for one set
in which smaller and larger values (from 11 to 19) were tried.
The Coulomb logarithm only appears in the expressions for the classical diffusion coefficients; 
since evolution of these models is dominated by resonant relaxation, the results are expected to be
weakly dependent on $\ln\Lambda$ and this was found to be the case.
The integration time step, $\Delta t$, was set to 2000 yr, i.e. $10^3$ steps per integration.

A natural choice for the parameter $r_\mathrm{lc}/r_g$ in the Fokker-Planck integrations would
be $8$,  the same value assumed in the $N$-body integrations.
In the case of ``plunges'' -- captures that occur without significant energy loss due to
gravitational radiation --  this would be the correct choice. 
However, some of the $N$-body capture events were ``EMRIs,'' for which capture was preceded
by significant energy loss due to the 2.5PN terms.
No such loss terms are included in the Fokker-Planck integrations described here.
Roughly speaking, the effect of the 2.5PN terms is to shift the location of the loss cone toward
larger ${\cal R}$ (i.e. larger orbital periapsis) at each ${\cal E}$ (see e.g. Figure 5 of Merritt et al. 2011).
To evaluate the effect on the Fokker-Planck results of ignoring the 2.5PN terms, a set of
integrations was carried out setting $r_\mathrm{lc}/r_g=32$, four times its value in the
$N$-body integrations.

\begin{figure}[h!]
\centering
\mbox{\subfigure{\includegraphics[angle=0.,width=3.in]{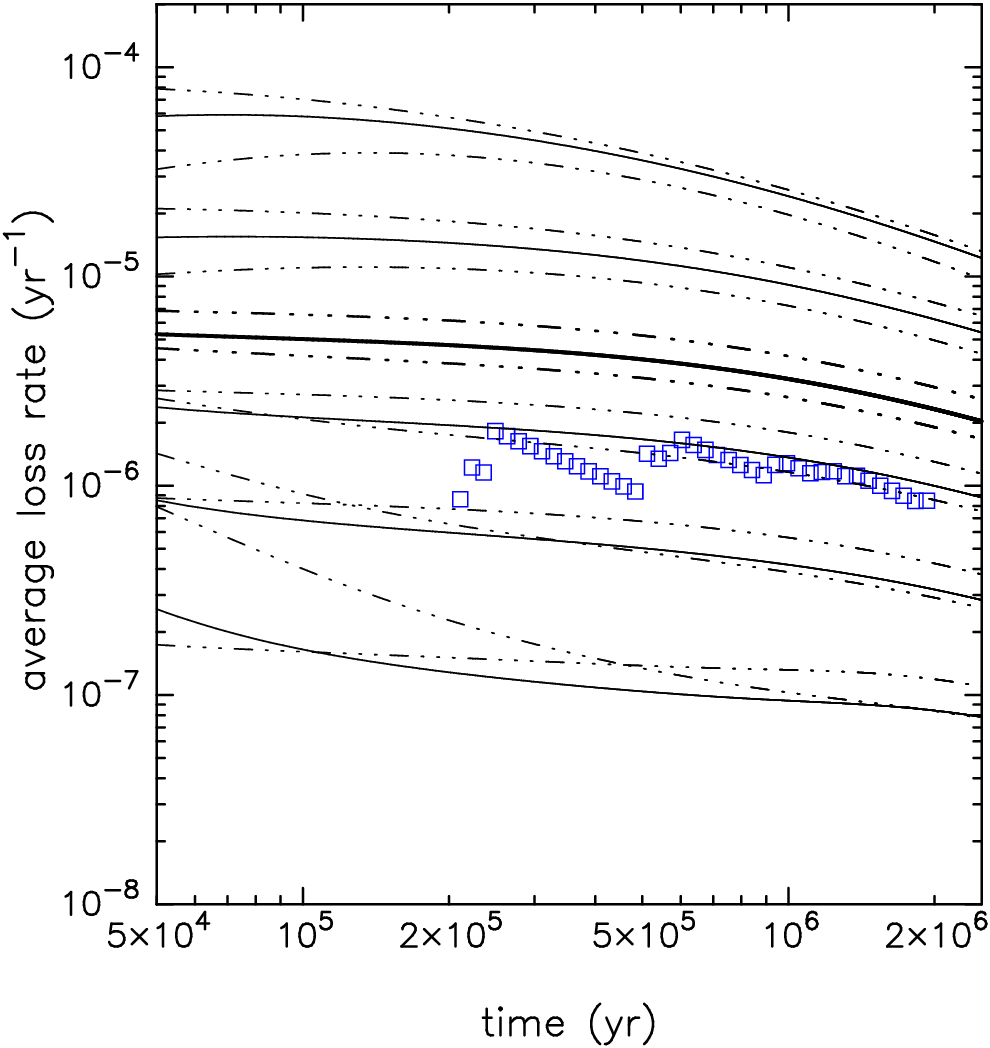}}\quad
\subfigure{\includegraphics[angle=0.,width=3.in]{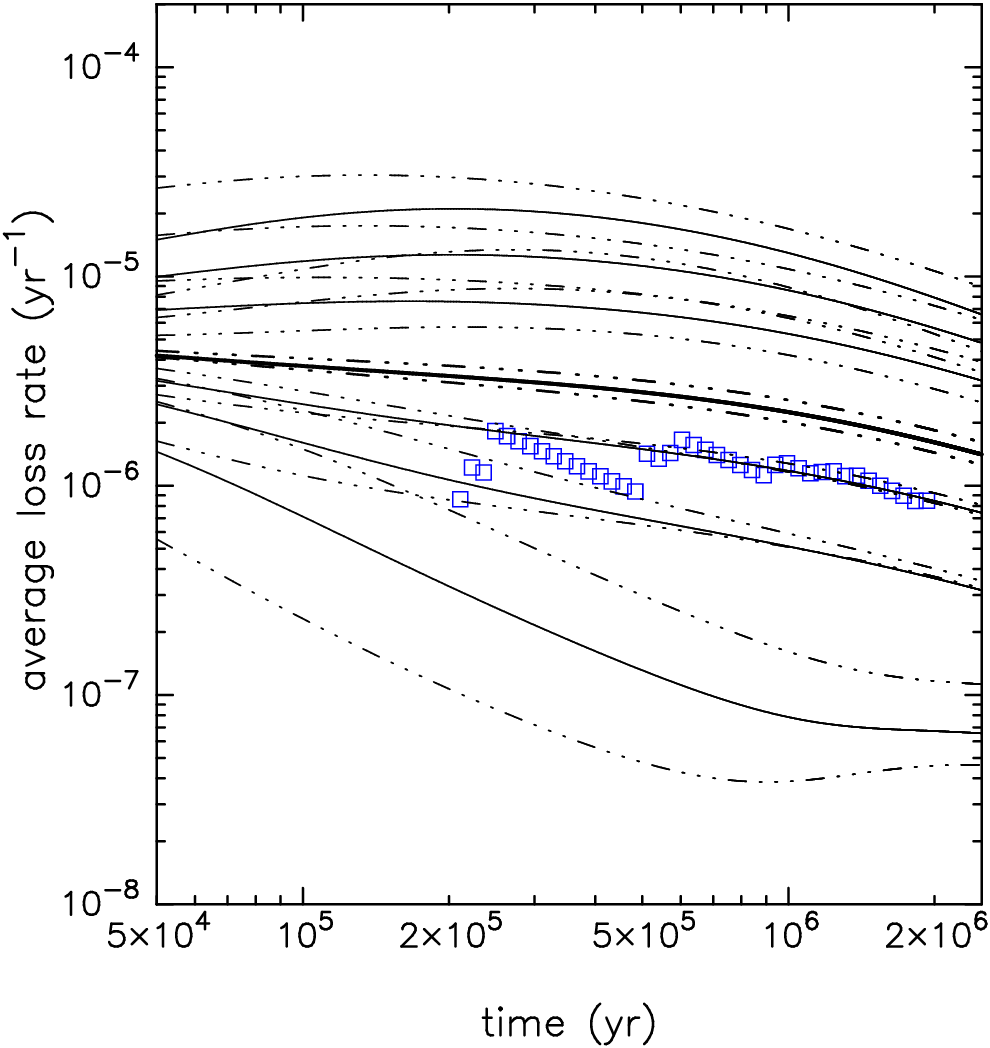} }}
\caption{Time-averaged loss rates, defined as the total number of stars lost until time $t$,
divided by $t$. (Blue) squares are from the $N$-body integrations of \citet{MAMW2011}
(Figure 2c of that paper); curves are from the Fokker-Planck integrations described in 
\S\ref{Section:AR_Constrain}.
{\bf The sudden jumps in the $N$-body loss rates at early times correspond to single capture events.}
Left (right) panel shows results using the power-law (exponential) forms of the anomalous 
diffusion coefficients in the Fokker-Planck code.
{\bf Left:} the six sets of curves are for ${\cal R}_1/{\cal R}_2 = \{2,1.2,1,0.9,0.8,0.7\}$,
from top to bottom.
The different line styles are explained in the text.
{\bf Right:} the six sets of curves are for ${\cal R}_3/{\cal R}_4 = \{1.4,1.2,1.1,1,0.95,0.9,0.7\}$,
from top to bottom.
Curves for ${\cal R}_1/{\cal R}_2=1$ or ${\cal R}_3/{\cal R}_4=1$ (``zero drift'') are thicker.
}
\label{Figure:lossrate}
\end{figure}

Figure~\ref{Figure:lossrate} compares time-averaged loss rates in the $N$-body and
Fokker-Planck models, defined as the total number of stars lost until time $t$ divided by $t$.
The left panel shows results using the power-law forms of the anomalous diffusion coefficients,
Equations (\ref{Equation:DiffCoefsAppendix}), (\ref{Equation:g1g2Appendix}), with $n=8$;
the right panel shows results using the exponential forms,
Equations (\ref{Equation:DiffCoefsAppendix}), (\ref{Equation:h1h2Appendix}).
Each panel shows results for several values of ${\cal R}_1/{\cal R}_2$ (power-law)
or ${\cal R}_3/{\cal R}_4$ (exponential), as specified in the caption.
For each value of this ratio, three integrations were carried out, varying the way in which 
${\cal R}_2$ or ${\cal R}_4$ were related to ${\cal R}_\mathrm{SB}$.
One of the three integrations (shown by the solid curves) equated ${\cal R}_2$ or
${\cal R}_4$ with ${\cal R}_\mathrm{SB}^{(ii)}$, Equation (\ref{Equation:SB2}).
The other two integrations adopted larger or smaller values: by a factor two or one-half
(in the power-law models), or by a factor $\sqrt{\pi}$ or $1/\sqrt{\pi}$ (exponential).
These additional integrations are shown by the dash-dotted curves in Figure~\ref{Figure:lossrate}. 
Larger (smaller) values of ${\cal R}_2$ or ${\cal R}_4$ generally resulted in smaller (larger)
loss rates, at least at early times.

Figure~\ref{Figure:lossrate} suggests that both functional forms of the anomalous diffusion
coefficients are able to reproduce the $N$-body capture rates, as long as
${\cal R}_1/{\cal R}_2$ or ${\cal R}_3/{\cal R}_4$ are not too different from one.
The best correspondence is achieved, in both cases, when this ratio is slightly less than one,
and this result remains unchanged even when the values of ${\cal R}_2({\cal E})$ or
${\cal R}_4({\cal E})$ are substantially modified.
Recall that ${\cal R}_1<{\cal R}_2$ or ${\cal R}_3<{\cal R}_4$ imply $D_{\cal R}<0$,
i.e. $\phi_{\cal R}>0$, i.e. a drift toward larger ${\cal R}$ (Equation~\ref{Equation:FluxDirection}).

Changing the parameter $n$ in the power-law diffusion coefficients from $n=8$ to $n=32$ had almost
no discernible effect on the loss rates.
Varying $\ln\Lambda$ or $r_\mathrm{lc}/r_g$ in the amounts described above did result
in noticeable changes, but by amounts comparable with the ranges shown in Figure~\ref{Figure:lossrate}
due to variations in the definition of ${\cal R}_2$ or ${\cal R}_4$.
In every set of integrations, correspondence with the $N$-body loss rates was best for 
${\cal R}_1/{\cal R}_2\lesssim 1$ or 
${\cal R}_3/{\cal R}_4\lesssim 1$.

\begin{figure}[h!]
\centering
\mbox{\subfigure{\includegraphics[angle=0.,width=2.2in]{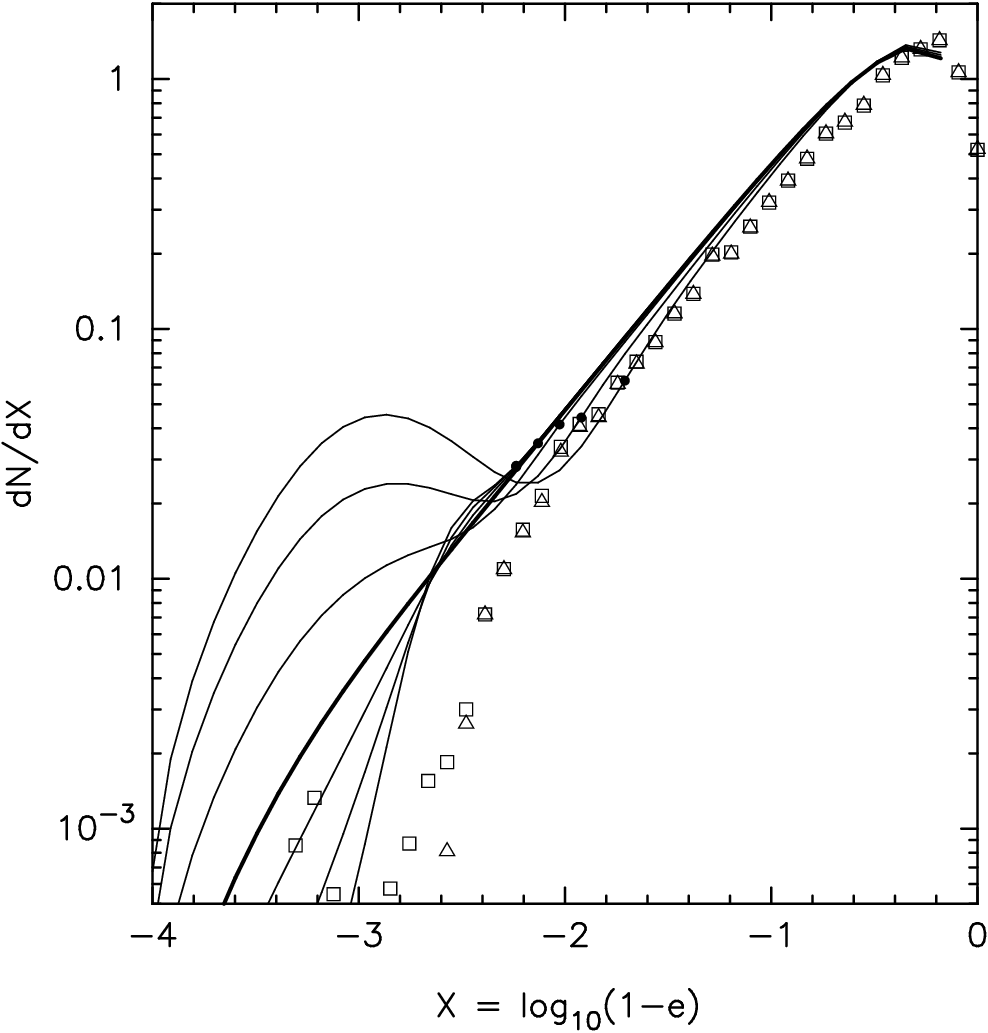}}\quad
\subfigure{\includegraphics[angle=0.,width=2.2in]{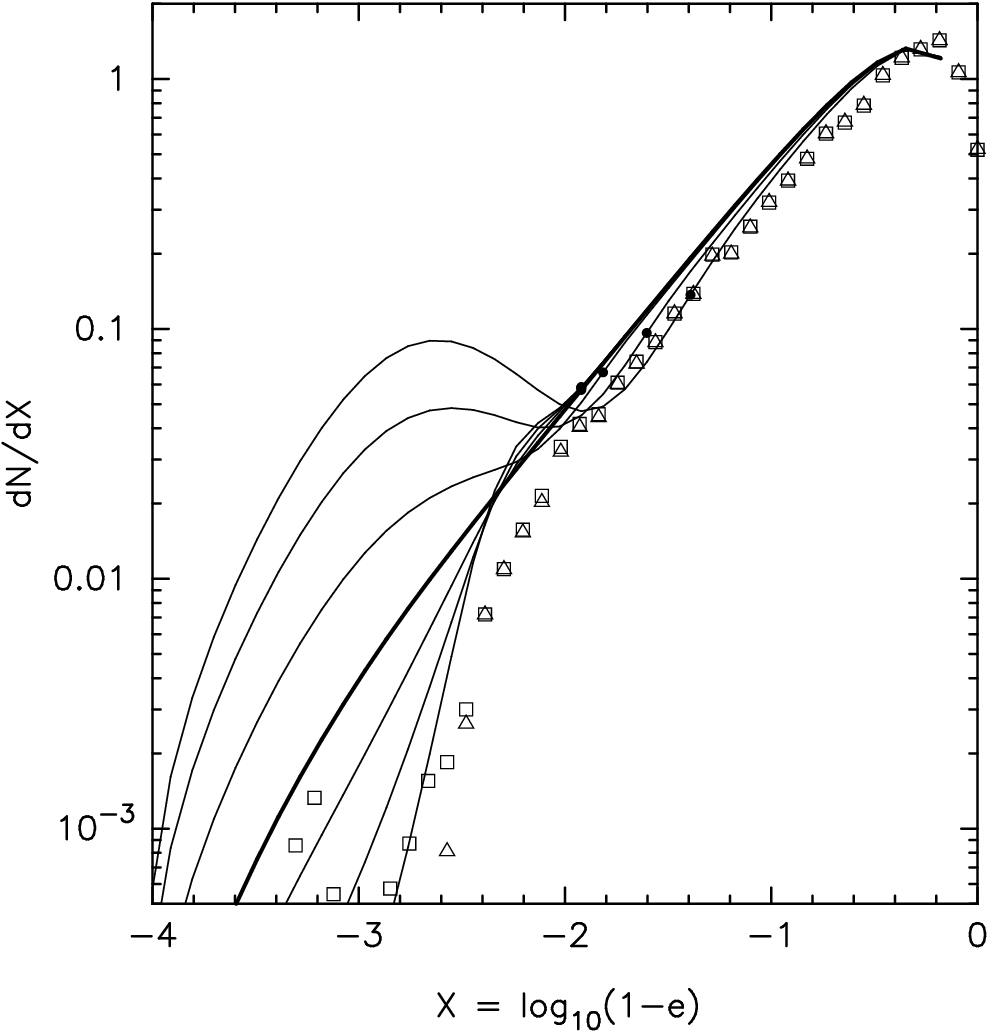}}\quad
\subfigure{\includegraphics[angle=0.,width=2.2in]{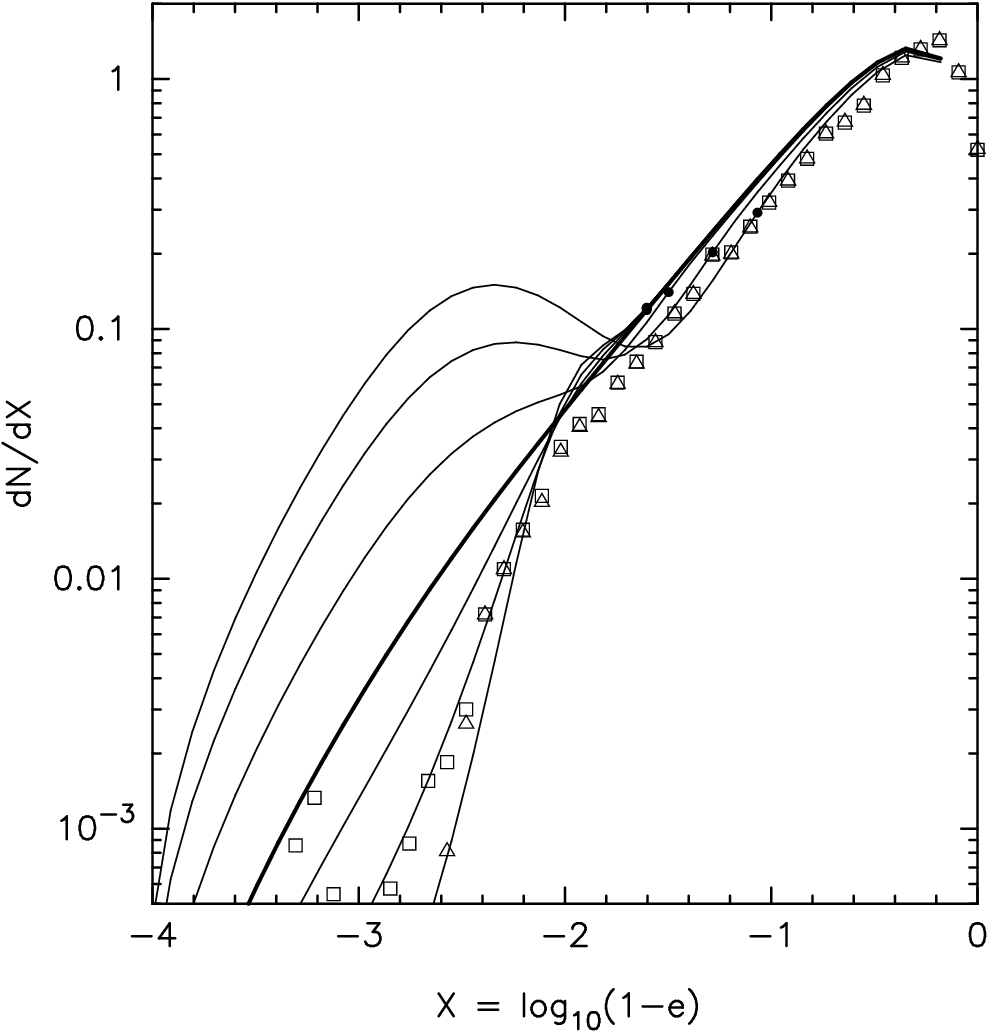}}}
\mbox{\subfigure{\includegraphics[angle=0.,width=2.2in]{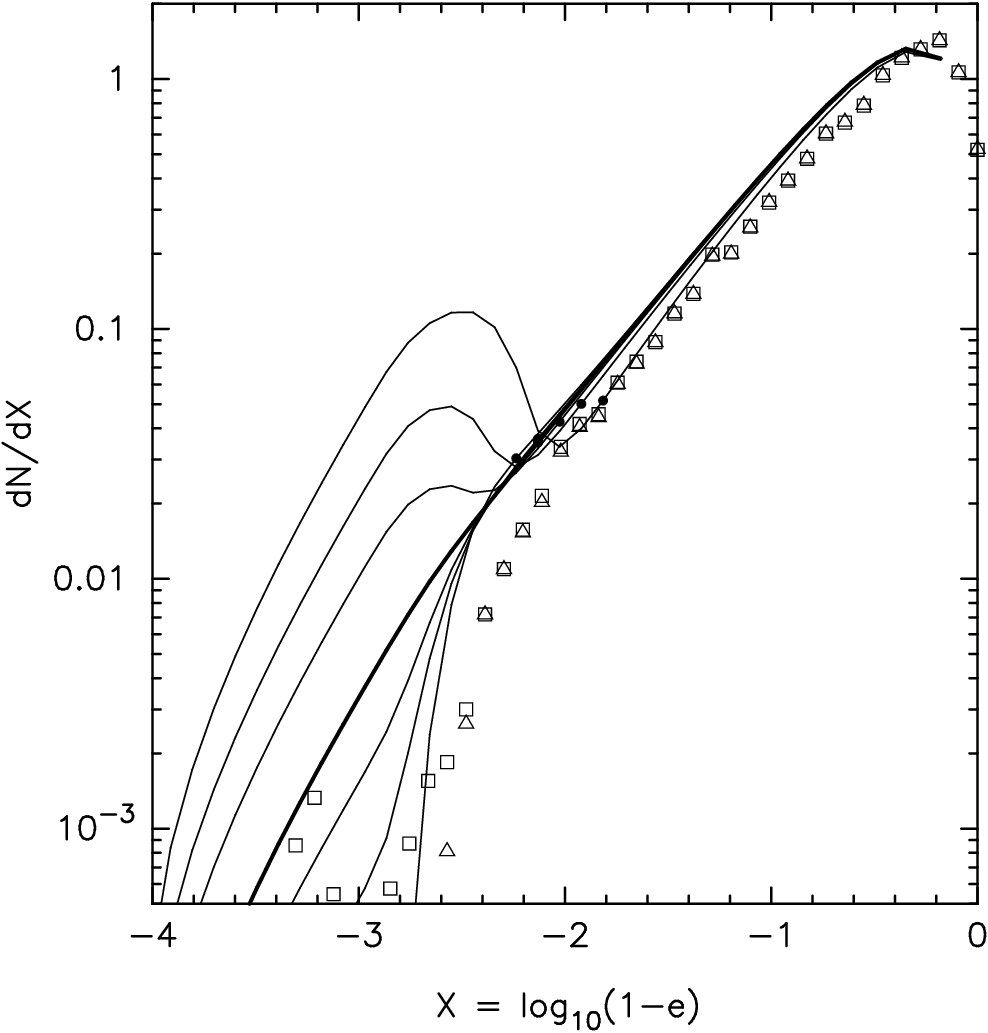}}\quad
\subfigure{\includegraphics[angle=0.,width=2.2in]{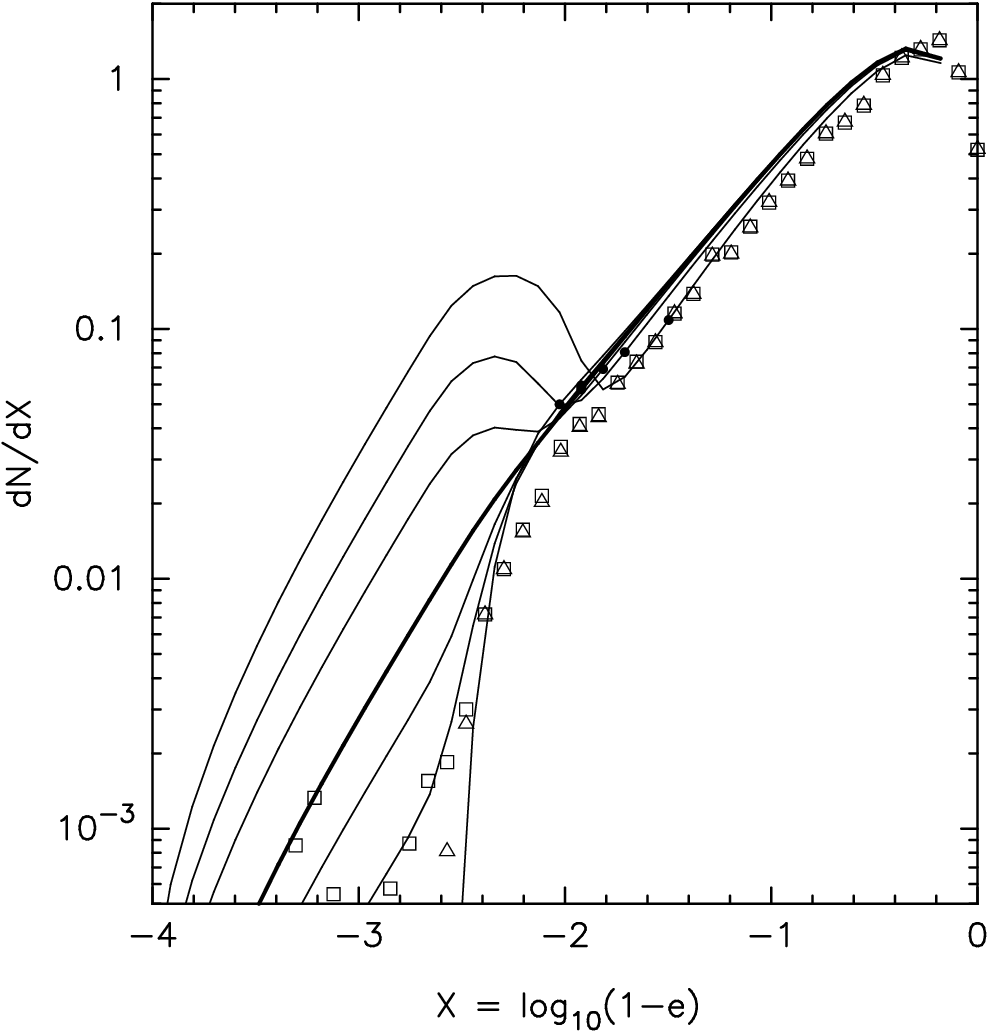}}\quad
\subfigure{\includegraphics[angle=0.,width=2.2in]{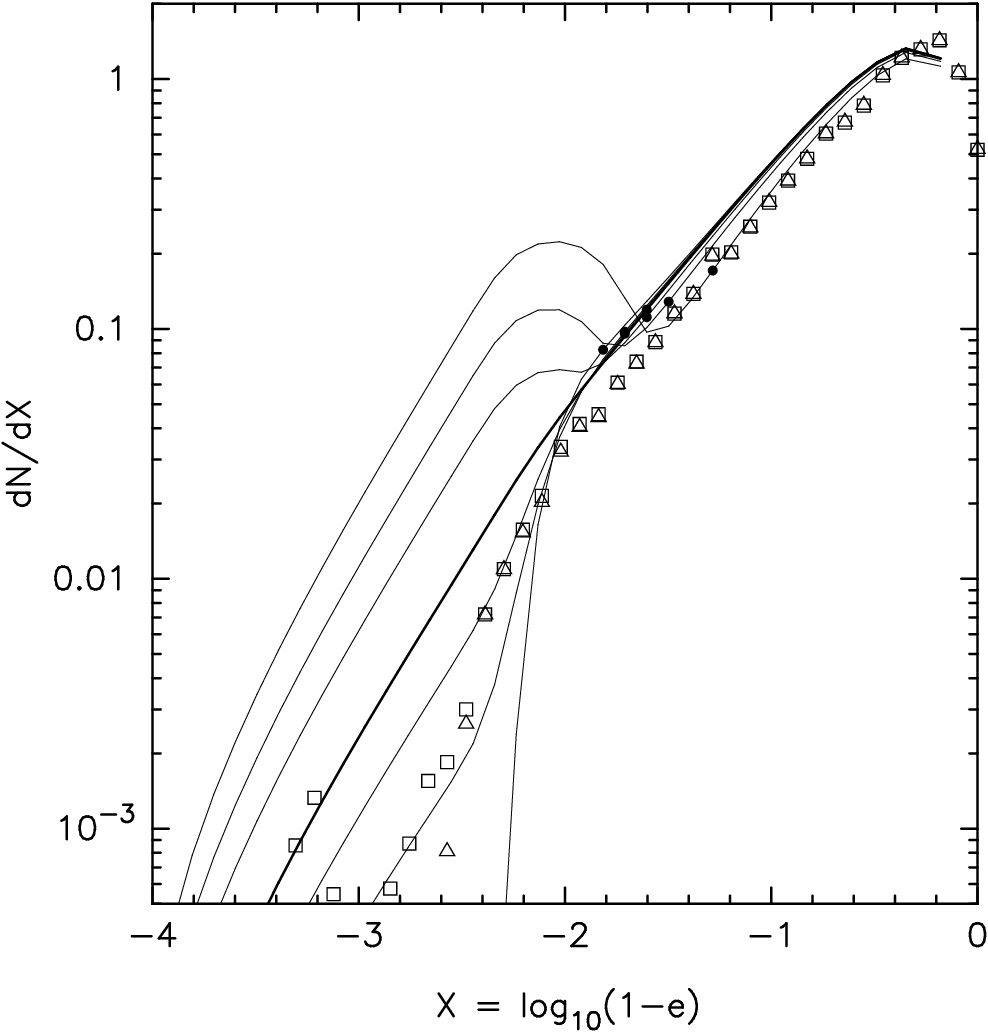}}}
\caption{Time-averaged angular momentum distributions for stars at a single energy.
The symbols are from the $N$-body integrations of \citet{MAMW2011} (Figure 11 from that paper);
the triangles exclude stars that eventually became EMRIs while the squares include those stars.
Curves were extracted from Fokker-Planck integrations that used the power-law
(top) or exponential (bottom) forms of the anomalous diffusion coefficients; the curves show
$f({\cal R}; {\cal E}_4)$ where ${\cal E}_4$
is the energy corresponding to orbits of semimajor axis 4 mpc.
The $N$-body data were computed using stars with instantaneous $a$ values in a range
$\Delta \log_{10} a = \pm 0.05$ centered on $a=4$ mpc. 
Curves are distinguished by the value of 
${\cal R}_1/{\cal R}_2 = \{0.7,0.8,0.9,1,1.2,1.5,2\}$ (top) 
or ${\cal R}_3/{\cal R}_4 = \{0.7,0.9,0.95,1,1.1,1.2,1.4\}$ (bottom);
the larger this ratio, the larger the value of $f$ at small $X$.
The middle panel used ${\cal R}_2={\cal R}_\mathrm{SB}^{(ii)}$ or
 ${\cal R}_4={\cal R}_\mathrm{SB}^{(ii)}$.
 Left and right panels used
 ${\cal R}_2={\cal R}_\mathrm{SB}^{(ii)}/2$ and
 ${\cal R}_2={\cal R}_\mathrm{SB}^{(ii)}\times 2$ respectively (top) or
  ${\cal R}_4={\cal R}_\mathrm{SB}^{(ii)}/\sqrt{\pi}$ and
 ${\cal R}_4={\cal R}_\mathrm{SB}^{(ii)}\times \sqrt{\pi}$ (bottom)
 Filled circles mark ${\cal R} = {\cal R}_2$ (top) or ${\cal R} = {\cal R}_4$ (bottom).
\label{Figure:dNdXNB}}
\end{figure}

Figure~\ref{Figure:dNdXNB} makes another comparison between 
Fokker-Planck and $N$-body models.
Plotted there are time-averaged angular momentum distributions at a single energy.
These are displayed as $dN/dX$, where $X$ is defined as in Figure 11 of \citet{MAMW2011}:
\beq
X \equiv \log_{10}\left(1-e\right)
\eeq
with $e=\sqrt{1-{\cal R}}$ the orbital eccentricity.
Figure~\ref{Figure:dNdXNB} implies that values of ${\cal R}_1/{\cal R}_2$ or ${\cal R}_3/{\cal R}_4$ 
of unity (``zero drift'') or greater can be securely ruled out -- consistent with Figure~\ref{Figure:lossrate}.
In the case of the power-law forms of the diffusion coefficients (upper panels),
the best correspondence with the $N$-body results seems to occur for 
\beq
{\cal R}_2 \approx 2 {\cal R}_\mathrm{SB}^{(ii)},  \ \ \ \ 
0.8 \lesssim {\cal R}_1/{\cal R}_2 \lesssim 0.9.
\eeq
In the case of the exponential forms of the diffusion coefficients (lower panels),
correspondence with the $N$-body results seems to require
\beq
{\cal R}_2 \approx \sqrt{\pi}\;  {\cal R}_\mathrm{SB}^{(ii)},  \ \ \ \ 
0.9 \lesssim {\cal R}_1/{\cal R}_2 \lesssim 0.95.
\eeq
Once again, the best correspondence is achieved with diffusion coefficients that imply a non-zero drift,
in the direction of increasing ${\cal R}$, in the anomalous regime.

\begin{figure}[h!]
\centering
\mbox{\subfigure{\includegraphics[angle=0.,width=3.in]{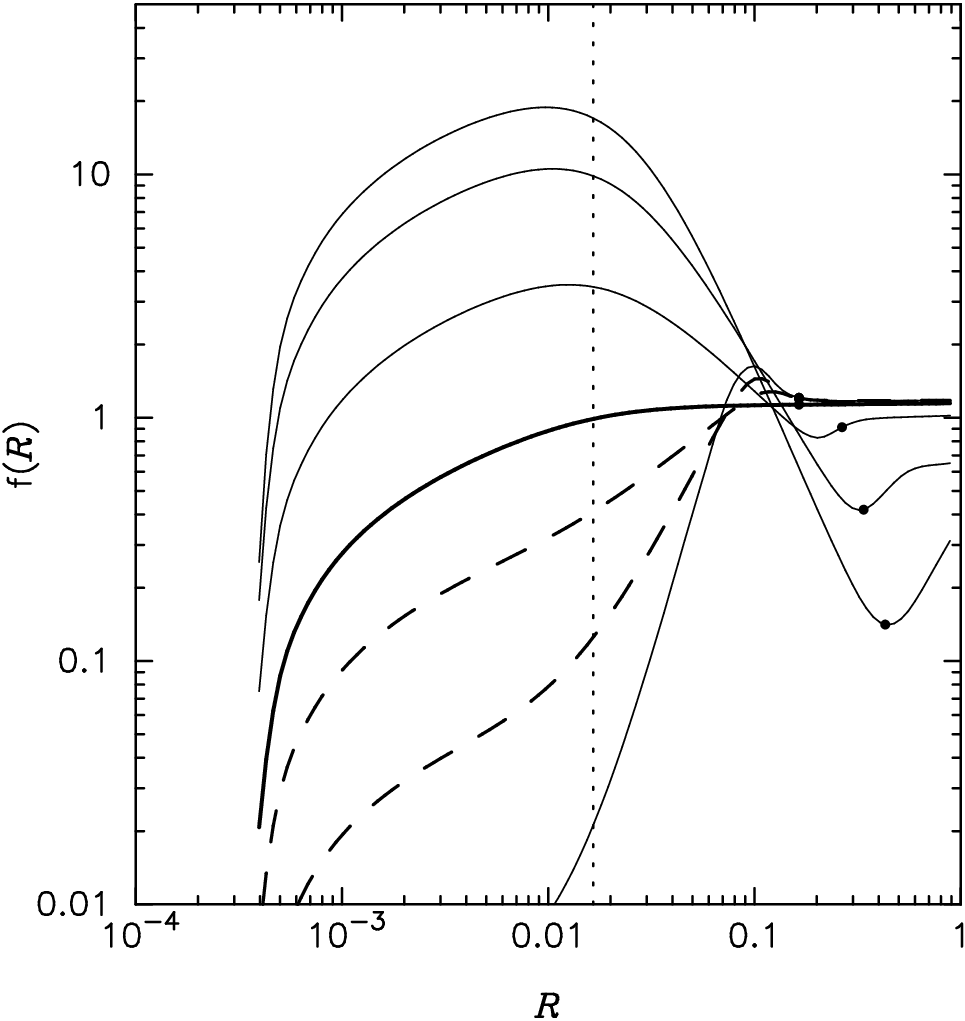}}\quad
\subfigure{\includegraphics[angle=0.,width=3.in]{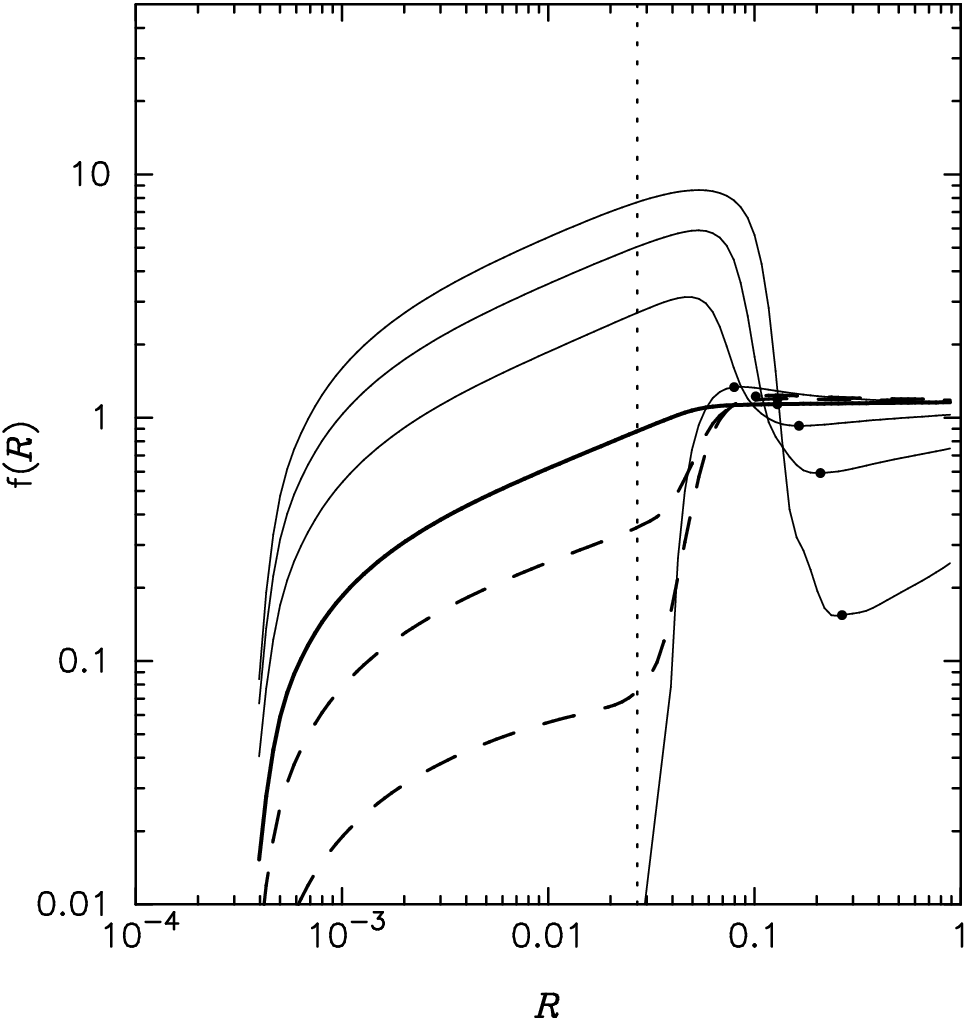} }}
\caption{Plots of $f({\cal R})$ at the final time step of the Fokker-Planck integrations described in 
\S\ref{Section:AR_Constrain}, for stars at a single energy, corresponding
to orbits with semimajor axes of $4$ mpc. 
The left panel shows integrations based on the power-law form of the anomalous diffusion 
coefficients; the right panel is based on the exponential form.
On the left, the parameter  ${\cal R}_2=2{\cal R}_\mathrm{SB}^{(ii)}$, and the 
different curves correspond to  ${\cal R}_1/{\cal R}_2 = \{0.7,0.8,0.9,1,1.2,1.5,2\}$;
the larger this ratio, the larger the value of $f$ at small ${\cal R}$.
On the right,  ${\cal R}_4=\sqrt{\pi} {\cal R}_\mathrm{SB}^{(ii)}$ and
${\cal R}_3/{\cal R}_4 = \{0.7,0.9,0.95,1,1.1,1.2,1.4\}$.
The thick solid curves in both panels are for  ${\cal R}_1/{\cal R}_2 = 1$.
The dashed curves show models that were judged to best reproduce the $N$-body data,
based on results like those in Figures~\ref{Figure:lossrate} and \ref{Figure:dNdXNB}.
Filled circles mark ${\cal R} = {\cal R}_\mathrm{SB}$; vertical dotted lines show the values
of ${\cal R}$ below which classical relaxation is expected to dominate anomalous relaxation
at this energy (Eqs.~\ref{Equation:RNR1} and~\ref{Equation:RNR4}).
}
\label{Figure:fofRlastNB}
\end{figure}

Figure~\ref{Figure:fofRlastNB} shows representative plots of $f({\cal R}; {\cal E})$ at one ${\cal E}$
from the Fokker-Planck models at the final time step (roughly $2.5\times 10^6$ yr).
Dashed curves show models having parameters similar to those found to correspond best to the
$N$-body results.
These solutions always exhibit a rapid drop in the steady-state $f({\cal R})$ below the Schwarzschild barrier.
That drop would be even steeper in the absence of classical relaxation, which dominates the
evolution in ${\cal R}$ at small ${\cal R}$ (the region below the vertical dotted lines in the figure), 
thus maintaining a relatively high diffusion rate at small ${\cal R}$.

One final argument  can be made in support of diffusion coefficients that satisfy  $D_{\cal R}<0$.
As described in \cite{AntoniniMerritt2013}, stars orbiting near the Schwarzschild barrier are observed
to exhibit a ``buoyancy'' phenomenon: should they cross the barrier from below 
(${\cal R}<{\cal R}_\mathrm{SB}$) 
to above (${\cal R}>{\cal R}_\mathrm{SB}$), they tend to remain above, and vice-versa.
This behavior is consistent with a drift toward larger ${\cal R}$, as implied by $D_{\cal R}<0$.

\subsection{Dominance of classical relaxation at small ${\cal R}$}
\label{Section:AR_NR}

In the regime of anomalous relaxation (${\cal R}<{\cal R}_\mathrm{SB}({\cal E})$), timescales for
angular momentum diffusion become  long for small ${\cal R}$.
One consequence is that {\it classical} relaxation, which (by assumption) is not affected by precession,
can once again dominate the evolution in angular momentum \citep{Hamers2014}.

We estimate the value of ${\cal R}$ at which this occurs by equating the first and second terms on the
right hand sides of Equation (\ref{Equation:ARDiffCoefsPL}).
We simplify the expressions by using the limiting forms of the diffusion coefficients as ${\cal R}\rightarrow 0$.

For the classical diffusion coefficients, Equations~(11) and (12) from \citet{Hamers2014},
together with the transformation Equations~(32) from Paper I, yield
\begin{subequations}
\begin{eqnarray}
\langle\Delta {\cal R}\rangle_\mathrm{CK} &\rightarrow& A^H({\cal E}),\ \ \ \ 
\langle(\Delta {\cal R})^2\rangle_\mathrm{CK} \rightarrow 2 {\cal R} A^H({\cal E}), \\
A^H(a) &=& \frac{\ln\Lambda}{C_\mathrm{NRR}(\gamma)} \left(\frac{m_\star}{\mh}\right)^2 \frac{N(a)}{P(a)}
\end{eqnarray}
\end{subequations}
where the symbol ``$\rightarrow$'' denotes the limit of small ${\cal R}$ and the function
$C_\mathrm{NRR}(\gamma)$ is given in Appendix B of Hamers et al. (2014); 
their calculation assumes $\rho(r) \propto r^{-\gamma}$.

For the anomalous diffusion coefficients,
Equations (\ref{Equation:ARDiffCoefsPL}), (\ref{Equation:Defineg2})  and (\ref{Equation:RRDiffCoef2})
give the limiting forms
\begin{eqnarray}
\langle\Delta {\cal R}\rangle_\mathrm{AR} &\rightarrow& 6A({\cal E})
\left(\frac{{\cal R}}{{\cal R}_1}\right)^2,\ \ \ \ 
\langle(\Delta {\cal R})^2\rangle_\mathrm{AR} \rightarrow 4A({\cal E}){\cal R} 
\left(\frac{{\cal R}}{{\cal R}_2}\right)^2
\end{eqnarray}
in the power-law case, with $A(a)$ defined in Equation~(\ref{Equation:RRDiffCoef2}). 
Equating $\langle\Delta{\cal R}\rangle_\mathrm{CK}$ with $\langle\Delta{\cal R}\rangle_\mathrm{AR}$ or 
$\langle(\Delta{\cal R})^2\rangle_\mathrm{CK}$ with $\langle(\Delta{\cal R})^2\rangle_\mathrm{AR}$ yields
\begin{subequations}\label{Equation:RNR1}
\begin{eqnarray}
\frac{{\cal R}^2}{{\cal R}_1^2} &=& \frac{\ln\Lambda}{6\, C_\mathrm{NR}\alpha^2} \frac{P}{t_\mathrm{coh}}
\ \ \ \ (\mathrm{1st}\; \mathrm{order}), \\
\frac{{\cal R}^2}{{\cal R}_2^2} &=& \frac{\ln\Lambda}{2\, C_\mathrm{NR}\alpha^2} \frac{P}{t_\mathrm{coh}}
\ \ \ \ (\mathrm{2nd}\; \mathrm{order}) .
\end{eqnarray}
\end{subequations}
Replacing ${\cal R}_1$ and ${\cal R}_2$ by ${\cal R}_\mathrm{SB}^{(ii)}$
in Eqs.~(\ref{Equation:RNR1}) and setting $\alpha=1.6$ yields
\begin{eqnarray} \label{Equation:RNR3}
{\cal R}^2 &=& (1.0,3.1)\; \frac{\ln\Lambda}{C_\mathrm{NR}} \left(\frac{r_g}{a}\right)^2
\frac{t_\mathrm{coh}}{P} 
\end{eqnarray}
where the constants in parentheses refer to first- and second-order diffusion coefficients respectively.
Equation~(\ref{Equation:RNR3}) is similar to Equation~(23a) of \citet{Hamers2014}.

Equations~(\ref{Equation:RNR1}) were used to plot the vertical dotted lines in the left-hand panels of
Figures~\ref{Figure:dfit} and~\ref{Figure:fofRlastNB} (note that these two figures refer to different
mass models).
In Figure~\ref{Figure:dfit}, the line predicts reasonably well the value of ${\cal R}$ at which
the data begin to deviate from the fitted curves.
In the case of Figure~\ref{Figure:fofRlastNB}, the effects of classical relaxation can be seen
in the steady-state $f({\cal R})$, which drops more gradually to zero below the SB than it
would if only anomalous relaxation were acting (Figure~\ref{Figure:fofRAppend}).

If the exponential forms of the anomalous diffusion coefficients are correct, then
the value of ${\cal R}$ at which $\langle(\Delta{\cal R})^2\rangle_\mathrm{CK}=
\langle(\Delta{\cal R})^2\rangle_\mathrm{AR}$ becomes
\begin{eqnarray} \label{Equation:RNR4}
\frac{{\cal R}^2}{{\cal R}_4^2} = \left[\log\left(\frac{2\alpha^2C_\mathrm{NR}}{\ln\Lambda}
\frac{t_\mathrm{coh}}{P}\right)\right]^{-1} .
\end{eqnarray}
This value for ${\cal R}$ is plotted, in addition to the value given by 
Equation~(\ref{Equation:RNR3}), on the lower right-hand panel of
Figure~\ref{Figure:dfit}.
Note that under this hypothesis, the range in ${\cal R}$ over which anomalous relaxation
would be relevant would become very small; in effect, classical relaxation would dominate the evolution
everywhere below (and sometimes even above) the Schwarzschild barrier.
We reiterate that there is no support for the exponential form of the anomalous diffusion coefficients
in any published numerical simulations.

\section{Steady-state solutions}
\label{Section:Results}

Paper II presented steady-state solutions for $f({\cal E}, {\cal R})$
obtained via integrations of the Fokker-Planck equation with diffusion coefficients given by
Equation~(\ref{Equation:CombinedDiffCoefs}) (no anomalous relaxation) and 
outer boundary condition
\beq\label{Equation:OuterBoundary}
f^*({\cal E}^*_\mathrm{min}, {\cal R}^*,t^*) = f^*({\cal E}^*_\mathrm{min},{\cal R}^*,0)
\eeq
with $f$ the phase-space density and
${\cal E}_\mathrm{min}$ the minimum value of ${\cal E}$ on the energy grid.
(Asterisks  denote dimensionless quantities; see Equation~\ref{Equation:Units} and Paper I.)
Initial conditions for $f$ were based on an isotropic power-law model, 
$n\propto r^{-7/4}$, $f\propto {\cal E}^{1/4}$,
 but with a simple modification to account for the presence of the loss cone:
 \beq
 f({\cal E}, {\cal R}, t=0) = 0,\ \ \ \ {\cal R}\le {\cal R}_\mathrm{lc}({\cal E}).
 \eeq
Among the parameters that were varied in the integrations of Paper II
were $m_\star/\mh$ and the initial density at large radii;
the latter was chosen to have one of three values, bracketing the estimated value for the 
nucleus of the Milky Way.
The physical radius of the loss sphere, $r_\mathrm{lc}$, was fixed at $15 r_g=15 G\mh/c^2$,
roughly the tidal disruption radius of a solar-type star.
\begin{figure}[h]
\centering
\includegraphics[angle=-90.,width=4.5in]{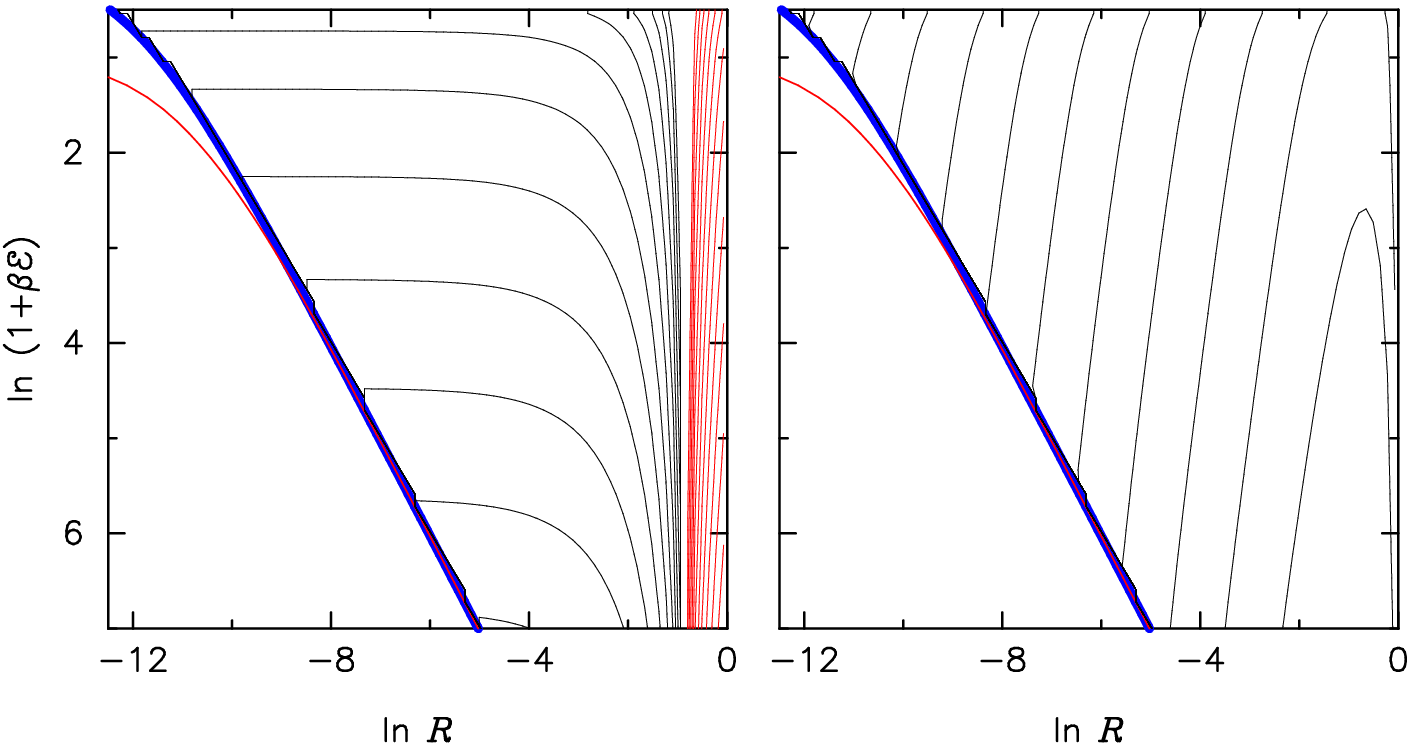} 
\includegraphics[angle=-90.,width=4.5in]{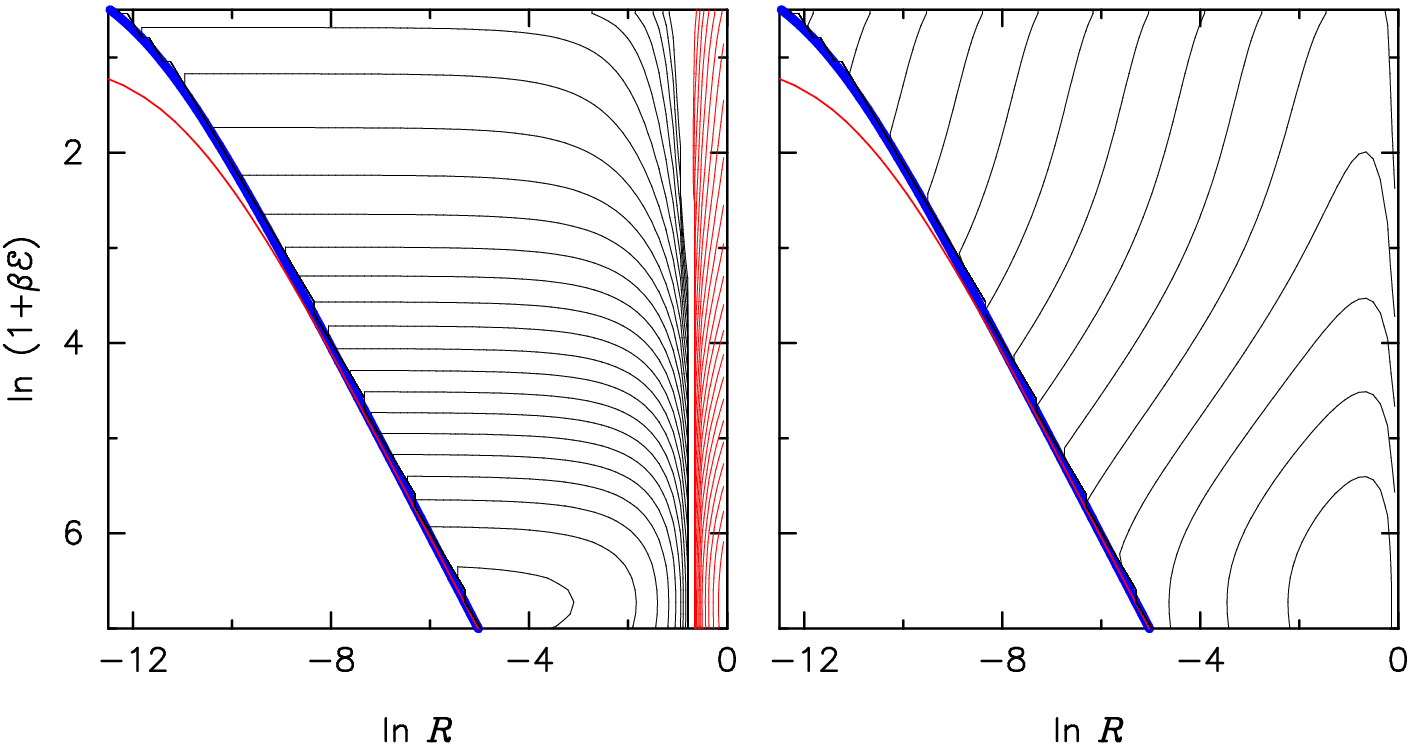}  
\includegraphics[angle=-90.,width=4.5in]{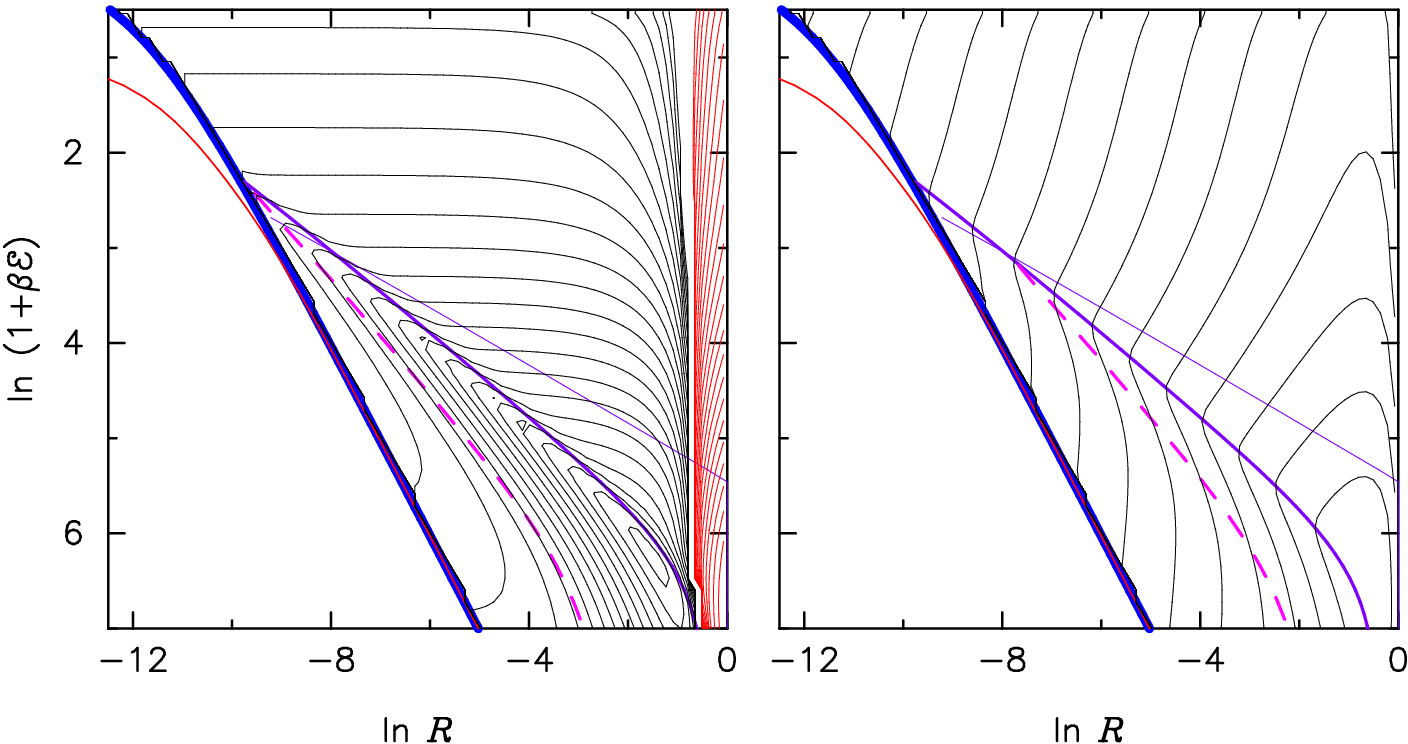} 
\caption{Angular momentum diffusion coefficients at $t=0$ in the
models of \S\ref{Section:Results}.
{\bf Top:}  $\langle\Delta{\cal R}\rangle_\mathrm{CK}$ (left),
$\langle\left(\Delta{\cal R}\right)^2\rangle_\mathrm{CK}$ (right).
{\bf Middle:}  $\langle\Delta{\cal R}\rangle_\mathrm{CK} + \langle\Delta{\cal R}\rangle_\mathrm{RR}$  (left),
$\langle\left(\Delta{\cal R}\right)^2\rangle_\mathrm{CK} + 
\langle\left(\Delta{\cal R}\right)^2\rangle_\mathrm{RR}$ (right).
{\bf Bottom:} Modified forms of the diffusion coefficients that account for anomalous relaxation,
Equations (\ref{Equation:ARDiffCoefsPL}), with parameters as given in the text.
The physical loss cone radius is shown as the thick (blue) curve in each panel; 
$f=0$ is assumed inside this curve at $t=0$ (``empty loss cone'').
The thin (red) curve that lies inside the loss cone is the quantity ${\cal R}_0({\cal E})$ defined
in Paper I.
In the panels showing $\langle\Delta{\cal R}\rangle$, the red contours indicate 
$-\langle\Delta{\cal R}\rangle$, i.e. $\langle\Delta{\cal R}\rangle<0$ in these regions.
In the lower panels, two expressions for the location of
the Schwarzschild barrier, ${\cal R}_\mathrm{SB}^{(i)}({\cal E})$
and ${\cal R}_\mathrm{SB}^{(ii)}({\cal E})$, are shown respectively as the thin and thick
magenta curves.
The dashed curves in the lower panels indicate where the timescales for classical
and anomalous relaxation are equal.
Contour values are the same in all frames.
\label{Figure:diffR}}
\end{figure}

We repeated a subset of those integrations, now using the modified expressions 
for the angular-momentum diffusion coefficients that account for anomalous relaxation: 
either power-law (Equation~\ref{Equation:ARDiffCoefsPL}) or exponential (Equation~\ref{Equation:ARDiffCoefsEXP}).

The dimensionless parameter $m_\star/\mh$ was set to $2.5\times 10^{-7}$.
Assuming $\mh=4\times 10^6\msun$, the (dimensional) stellar mass becomes
$m_\star = 1.0\msun$.
The outer boundary condition was chosen, as in Paper II, to give one of the following three values 
for the mass density at one parsec:
\beq\label{Equation:rhoatonepc}
\{1.9\times 10^4, 3.5\times 10^5,   6.1\times 10^6\} \msun \mathrm{pc}^{-3} 
\eeq
where again $\mh=4\times 10^6\msun$ has been assumed.
For each of these parameter choices, two other parameters that appear in the
expressions for the anomalous diffusion coefficients were varied:
\begin{enumerate}
\item ${\cal R}_1/{\cal R}_2$ (power-law) or ${\cal R}_3/{\cal R}_4$ (exponential)  ;
\item The ratio ${\cal R}_2/{\cal R}_\mathrm{SB}$
or ${\cal R}_4/{\cal R}_\mathrm{SB}$.
\end{enumerate}
Based on the results described in the previous section, the following parameter values
were considered:
\begin{itemize}
\item ${\cal R}_1/{\cal R}_2 = \{0.8,1.0,1.2\}$; \ \ ${\cal R}_3/{\cal R}_4 = \{0.9,1.0,1.1\}$
\item ${\cal R}_2/{\cal R}_\mathrm{SB} = \{0.5,1.0,2.0\}$;\ \ 
${\cal R}_4/{\cal R}_\mathrm{SB} = \{0.5,1.0,2.0\}$ .
\end{itemize}
For ${\cal R}_\mathrm{SB}({\cal E})$, the expression (\ref{Equation:SB2}) was used.
The parameter $n$ that appears in the power-law form of the anomalous diffusion coefficients
was set to 8 in all integrations.

Figure~\ref{Figure:diffR} plots the diffusion coefficients $\langle\Delta{\cal R}\rangle$,
$\langle(\Delta{\cal R})^2\rangle$ as computed by the code at $t=0$,
in models having the middle of the three values for the mass density at 
one parsec (Equation~\ref{Equation:rhoatonepc}).
The top two frames plot the classical diffusion coefficients:
\beq
\langle\Delta{\cal R}\rangle_\mathrm{CK}, \ \ \ \ 
\langle\left(\Delta{\cal R}\right)^2\rangle_\mathrm{CK}.\nonumber
\eeq
The middle two frames plot diffusion coefficients that account for resonant relaxation:
\beq
\langle\Delta{\cal R}\rangle_\mathrm{CK} + \langle\Delta{\cal R}\rangle_\mathrm{RR}, \ \ \ \ 
\langle\left(\Delta{\cal R}\right)^2\rangle_\mathrm{CK} + 
\langle\left(\Delta{\cal R}\right)^2\rangle_\mathrm{RR} . \nonumber
\eeq
These are the same expressions adopted in the integrations of Paper I.
The lower set of frames show the diffusion coefficients of Equation~(\ref{Equation:ARDiffCoefsPL}), which
account for anomalous relaxation (power-law modification, with $n=8$,
${\cal R}_2 = 2 {\cal R}_\mathrm{SB}^{(ii)}$, and ${\cal R}_1 = {\cal R}_2$).
In the case of the exponential modification (not shown in this figure), the diffusion coefficients
drop off more steeply below the Schwarzschild barrier.
However, this drop is mediated, in all models, by the presence of classical diffusion, 
which dominates again at sufficiently small ${\cal R}$, as discussed above.

\begin{figure}[h!]
\centering
\mbox{
\subfigure{\includegraphics[angle=0.,width=2.25in]{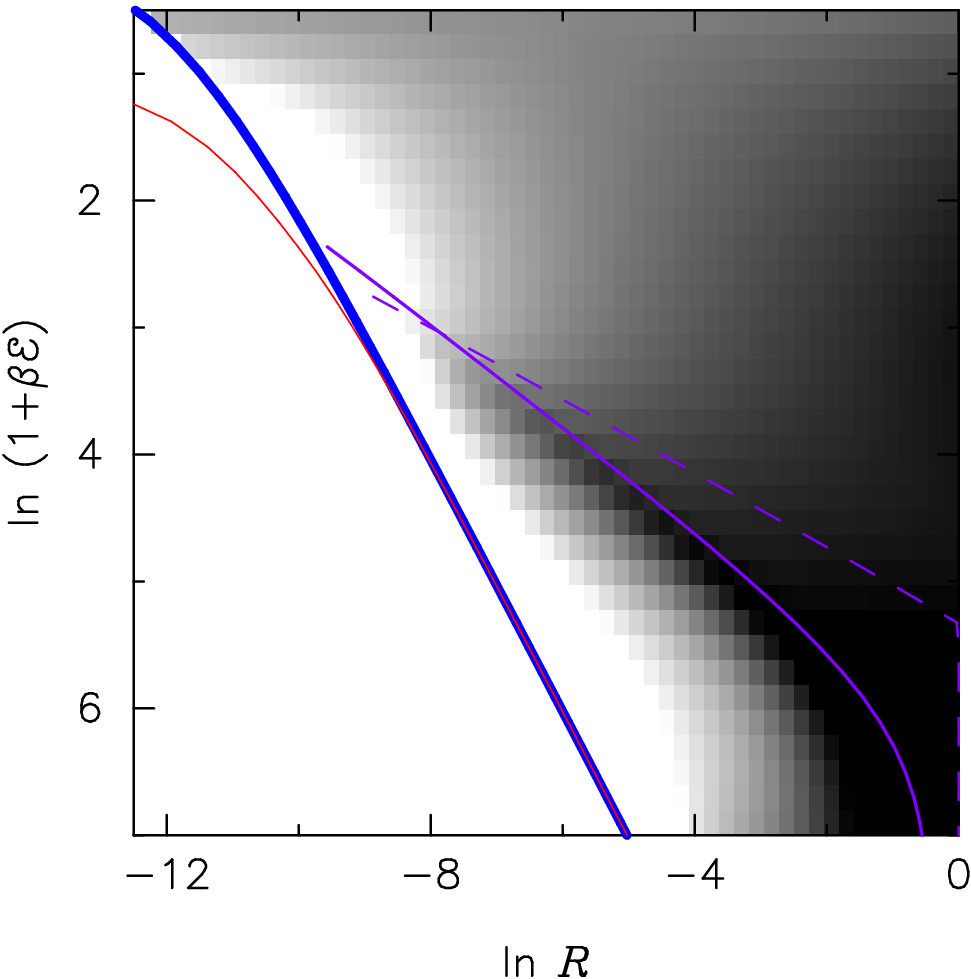}}\quad
\subfigure{\includegraphics[angle=0.,width=2.25in]{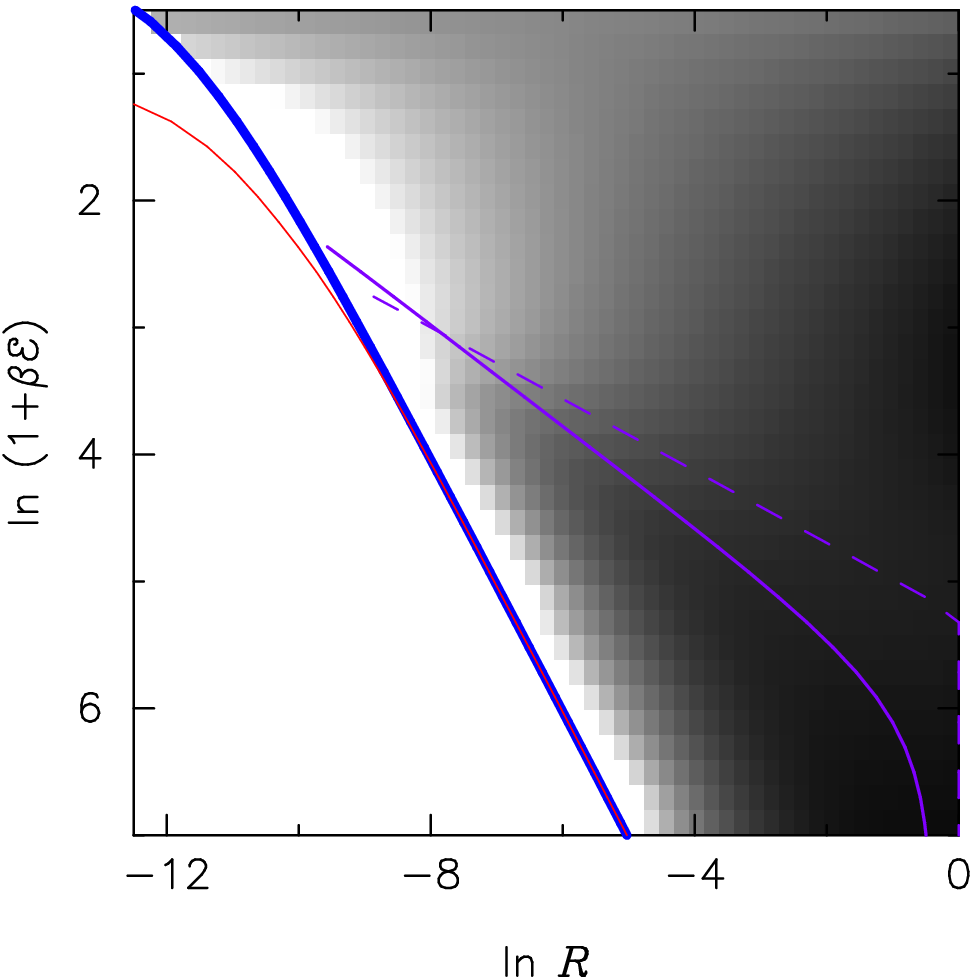}}\quad
\subfigure{\includegraphics[angle=0.,width=2.25in]{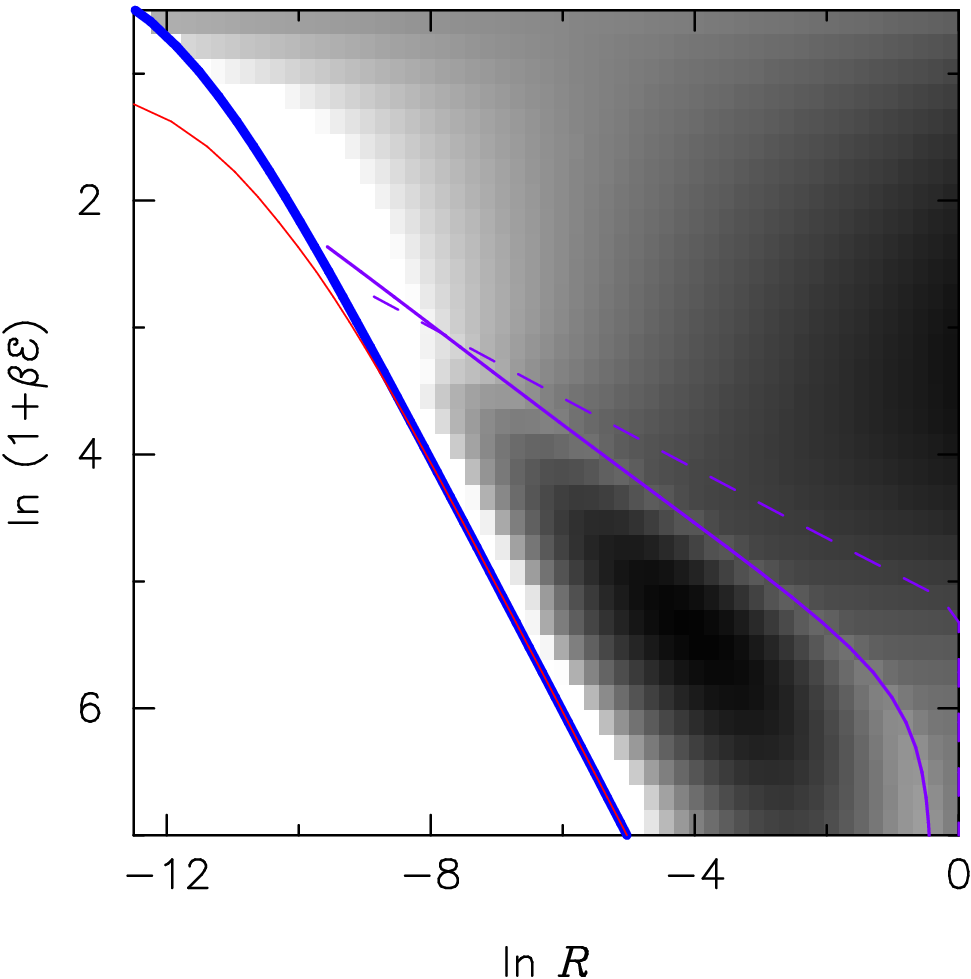}}}
\mbox{
\subfigure{\includegraphics[angle=0.,width=2.25in]{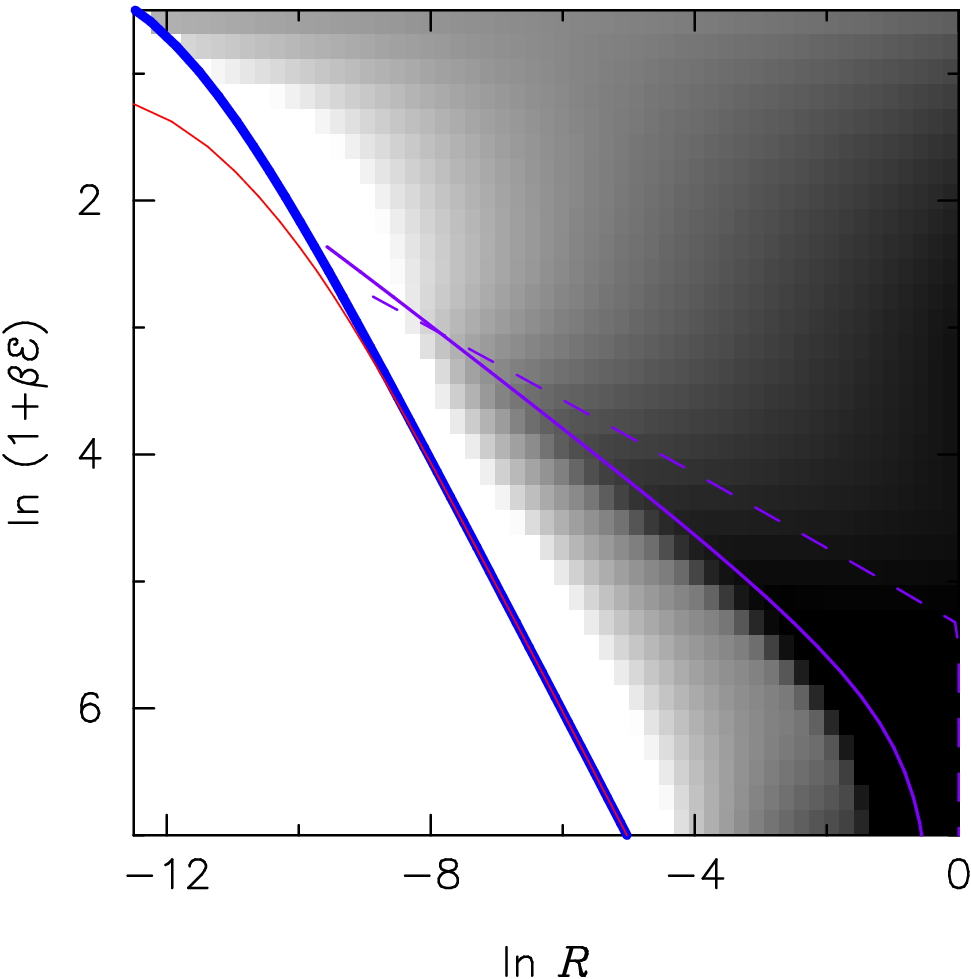}}\quad
\subfigure{\includegraphics[angle=0.,width=2.25in]{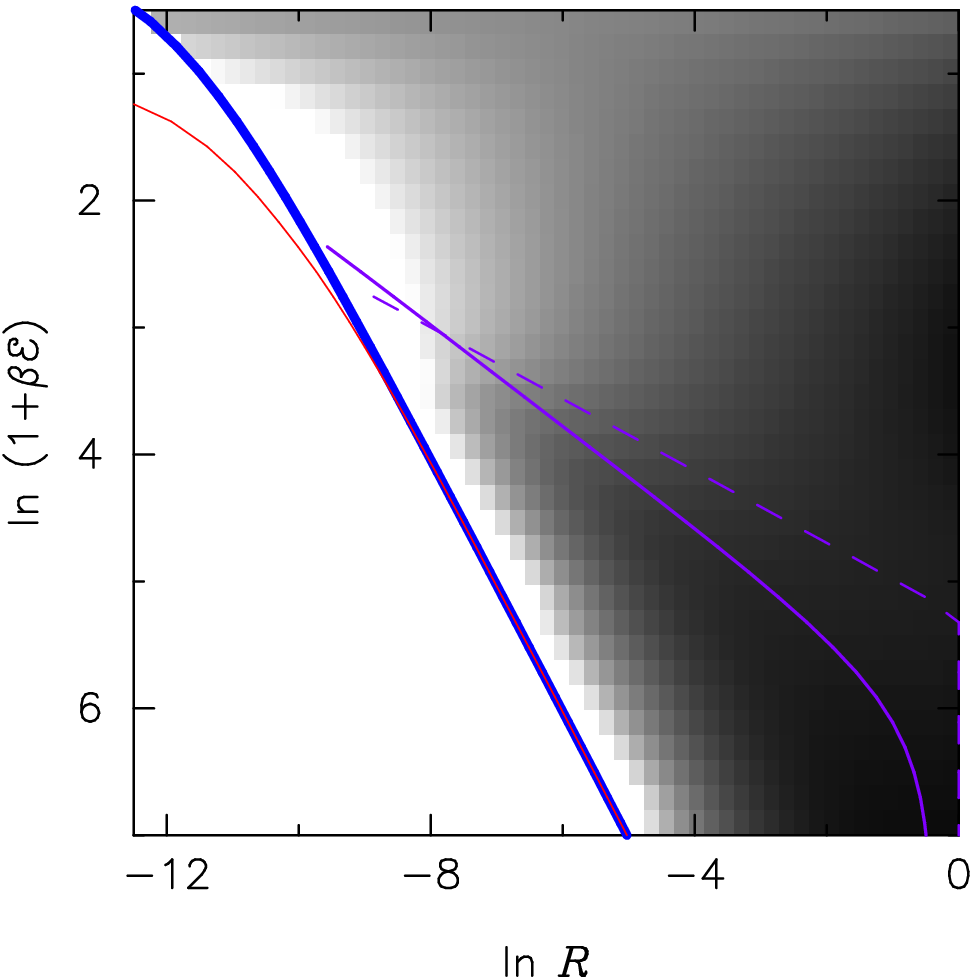}}\quad
\subfigure{\includegraphics[angle=0.,width=2.25in]{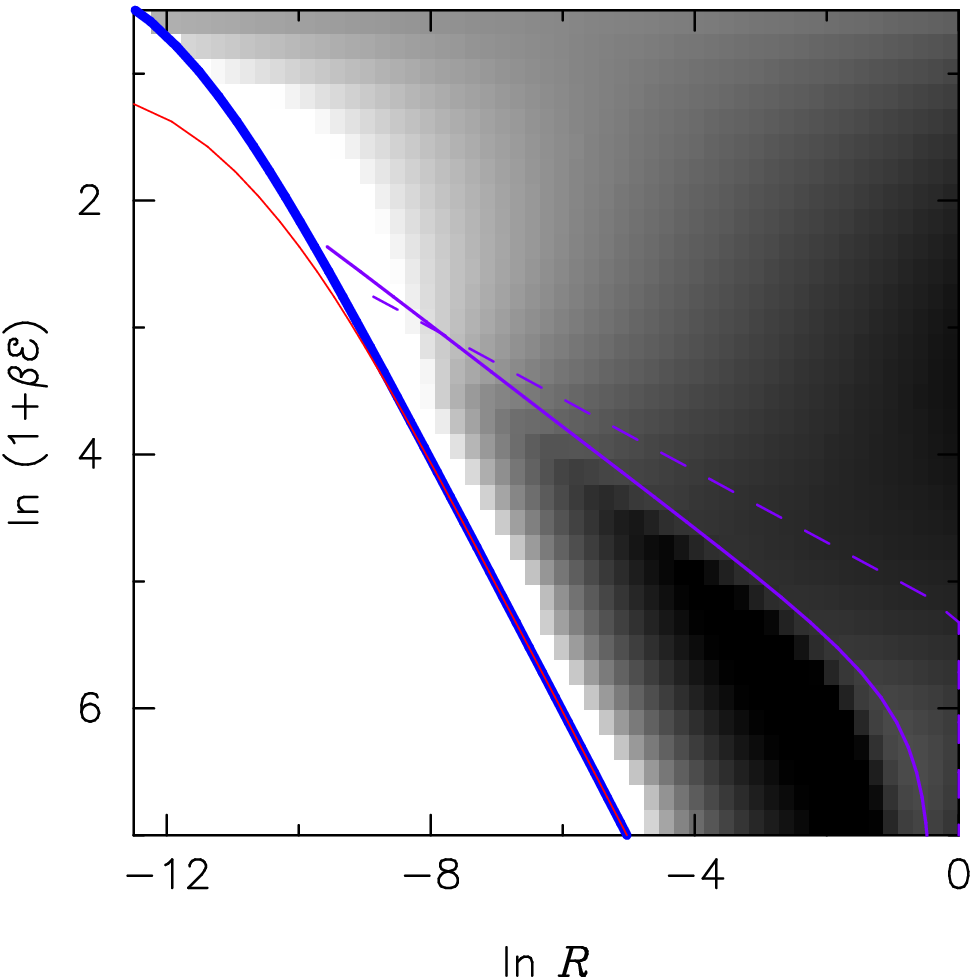}}}
\caption{Grey-scale of $\log f^\star$ in a set of six, steady-state models.
{\bf Top:} models integrated using the power-law forms of the anomalous diffusion coefficients,
with  ${\cal R}_2/{\cal R}_\mathrm{SB}=2$, and with ${\cal R}_1/{\cal R}_2 = 0.8$ (left),
1.0 (middle) and 1.2 (right). 
{\bf Bottom:} models integrated using the exponential forms of the anomalous diffusion coefficients,
with  ${\cal R}_4/{\cal R}_\mathrm{SB}=2$, and with ${\cal R}_3/{\cal R}_4 = 0.9$ (left),
1.0 (middle) and 1.1 (right). 
Other parameters are given in the text.
Curves have the same meaning as in Figure~\ref{Figure:diffR}.
\label{Figure:fofER}}
\end{figure}

Steady-state solutions, $f^*({\cal E}^*, {\cal R}^*)$, are shown in Figure~\ref{Figure:fofER}, again for
models with 
$\rho(1 \mathrm{pc}) \approx 3.5\times 10^6 \msun \mathrm{pc}^{-3}$.
The top(bottom) panels show solutions obtained using the power-law(exponential) 
expressions for the anomalous diffusion coefficients, with  ${\cal R}_2/{\cal R}_\mathrm{SB}=2$
or ${\cal R}_4/{\cal R}_\mathrm{SB}=2$.
What varies, from left to right, is the choice of ${\cal R}_1/{\cal R}_2$ (top) or
 ${\cal R}_3/{\cal R}_4$ (bottom).
 When these ratios are unity (``zero-drift''), the steady-state solutions are characterized
 by $f({\cal R})\sim$ constant near the SB.
 When these ratios are greater or less than one,
 the steady-state solutions behave in roughly the way seen 
 in Figures~\ref{Figure:fofRlastNB} and \ref{Figure:fofRAppend}, becoming either greater or smaller
 in the region below the Schwarzschild barrier, before dropping to zero  at the loss cone boundary.
 Evidently, the form of the steady-state $f$ in this region depends very sensitively on deviations of that ratio
 from unity.
 Depending on the value of that ratio, 
 $f$ in the region below the barrier can either be strongly depleted, or strongly
 enhanced, compared with the ``zero-drift'' solution.
 
\begin{figure}[h!]
\centering
\mbox{\subfigure{\includegraphics[angle=0.,width=3.in]{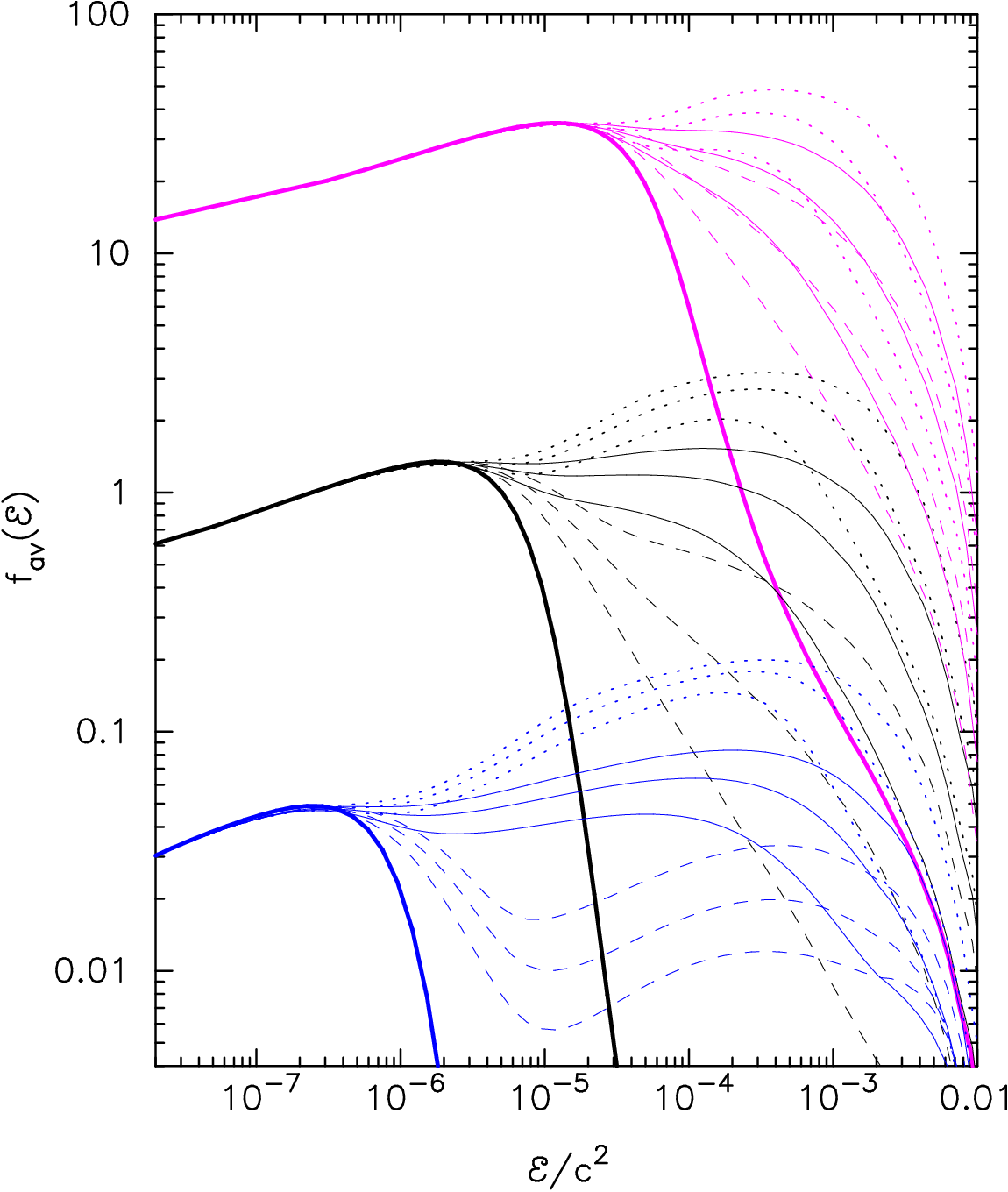}}\quad
\subfigure{\includegraphics[angle=0.,width=3.in]{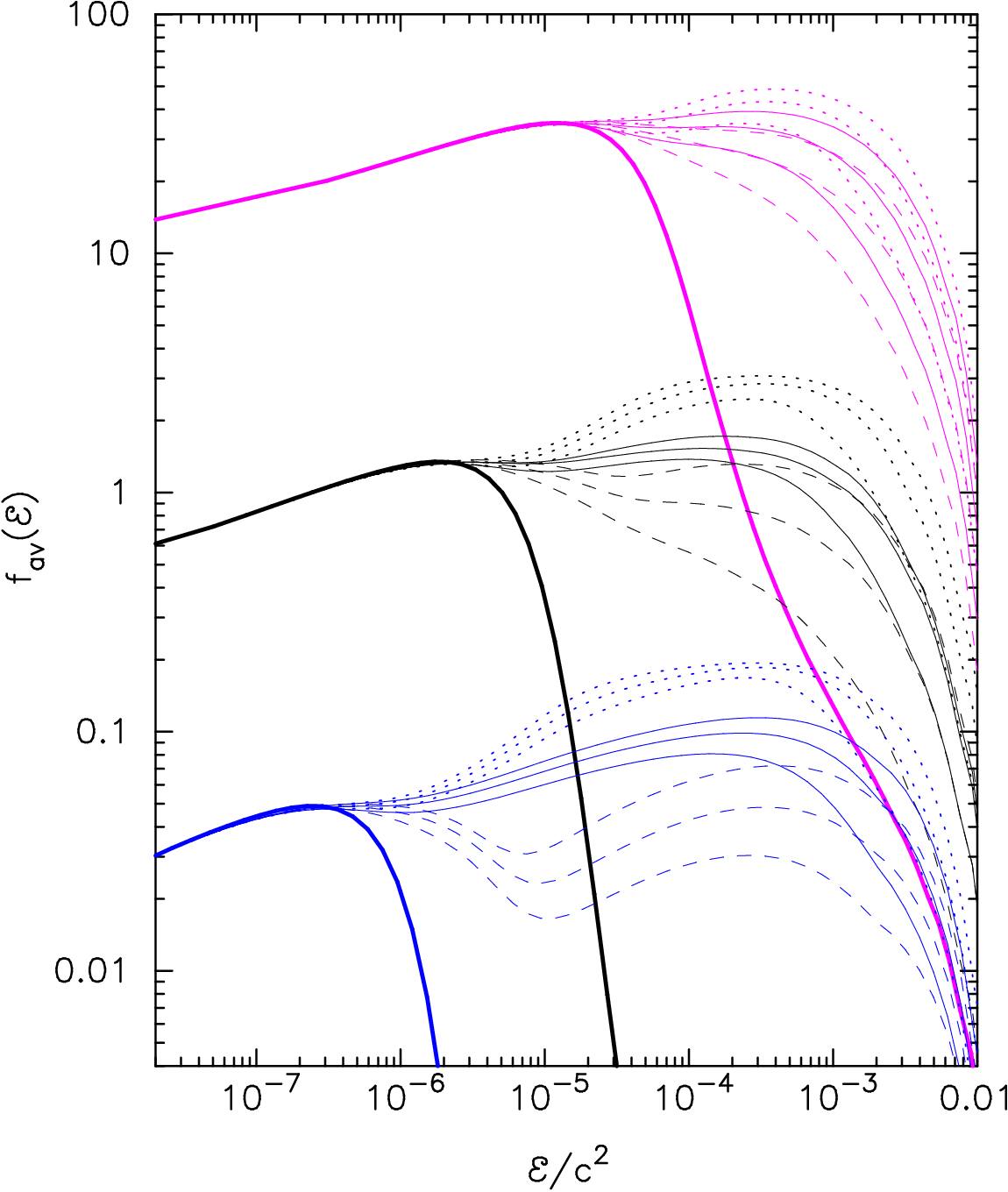} }}
\caption{Angular-momentum-averaged distribution functions, $\overline{f}({\cal E})$,
for all of the steady-state models;
left(right) panel used the power-law(exponential) forms of the anomalous diffusion coefficients.
The three values of the large-radius density normalization, Equation~(\ref{Equation:rhoatonepc}), 
are indicated by the three colors (high density, magenta; intermediate density, black; low density, blue).
In each of the three sets of models, the solid curves have 
${\cal R}_1/{\cal R}_2=1$ (left) or ${\cal R}_3/{\cal R}_4=1$ (right);
the dashed curves have 
${\cal R}_1/{\cal R}_2=1.2$ (left) or ${\cal R}_3/{\cal R}_4=1.1$ (right);
the dotted curves have 
${\cal R}_1/{\cal R}_2=0.8$ (left) or ${\cal R}_3/{\cal R}_4=0.9$ (right).
For each choice of these parameters, there are three curves, with the same line style, corresponding
to the three choices $\{0.5,1.0.2.0\}$ for ${\cal R}_2/{\cal R}_\mathrm{SB}$
or ${\cal R}_4/{\cal R}_\mathrm{SB}$.
The three thick, solid curves in each panel are from integrations that did not account for anomalous
relaxation; these are the same curves plotted in Figure 5 of Paper II.
\label{Figure:fofelast}}
\end{figure}

Angular-momentum-averaged distribution functions, defined as
\beq
\overline{f}({\cal E}) \equiv \int_{{\cal R}_\mathrm{lc}({\cal E})}^1 f\left({\cal E}, {\cal R}\right) d{\cal R} ,
\eeq
are plotted in Figure~\ref{Figure:fofelast}  for all of the steady-state solutions.
Shown for comparison, as the thick solid curves, are $\overline{f}$ for models computed
without anomalous diffusion; these are the same three curves plotted in Figure~5 of Paper II.
As discussed in that paper, inclusion of the resonant diffusion coefficients  has the effect
of sharply truncating the steady-state $\overline{f}({\cal E})$, at binding energies above a certain
value, where the timescale for resonant diffusion (in angular momentum) 
drops below the timescale for classical diffusion (in energy), and stars are carried rapidly into the
\sbh.
The truncation of $\overline{f}$ can still be seen in the new models; but it is  mediated by the presence of
anomalous relaxation.
The reason is the increase in the angular-momentum diffusion time when anomalous relaxation
is accounted for: stars ``pile up'' near the Schwarzschild barrier, until their density is high enough
to drive the requisite flux.
 
\begin{figure}[h!]
\centering
\mbox{\subfigure{\includegraphics[angle=0.,width=3.in]{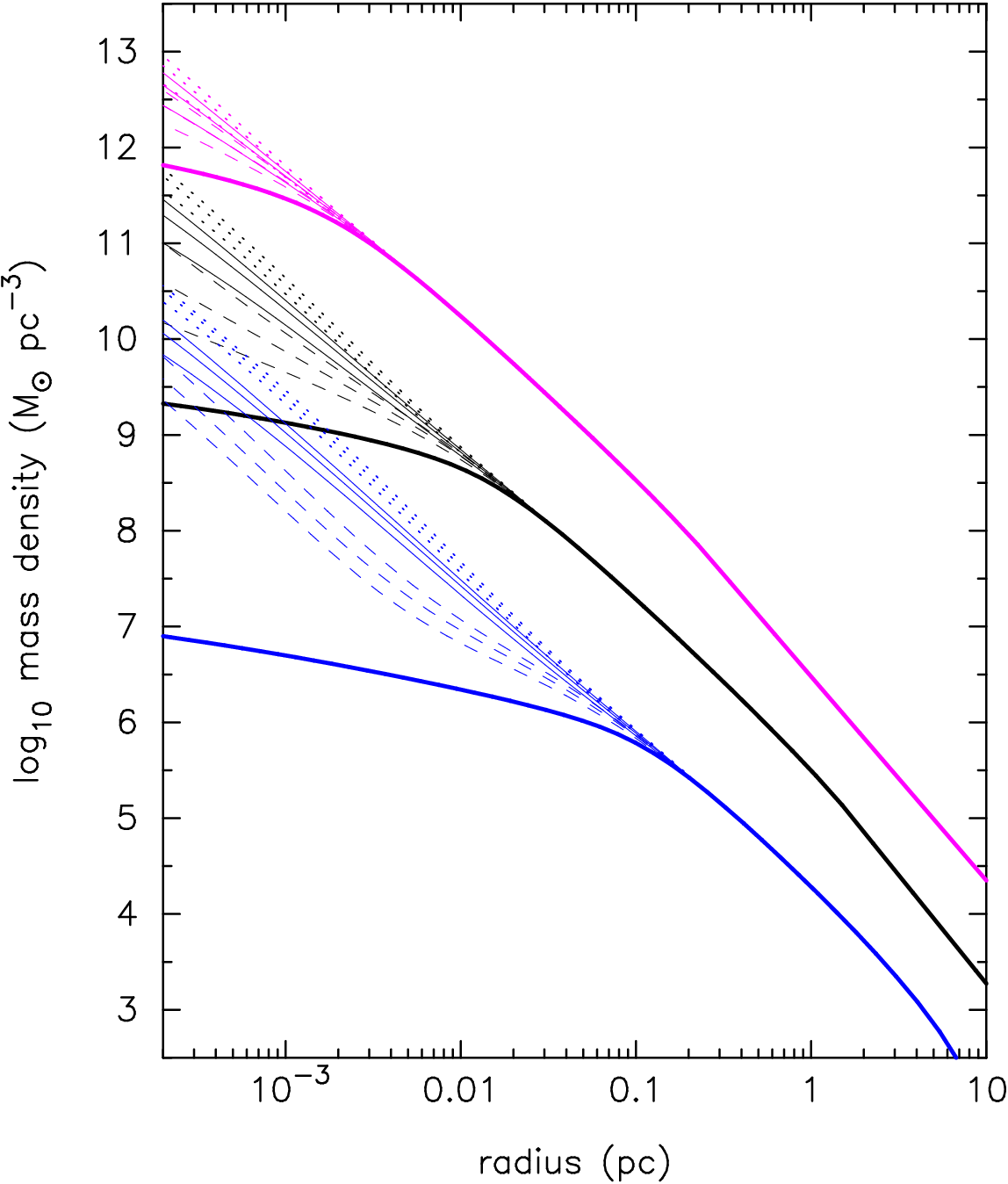}}\quad
\subfigure{\includegraphics[angle=0.,width=3.in]{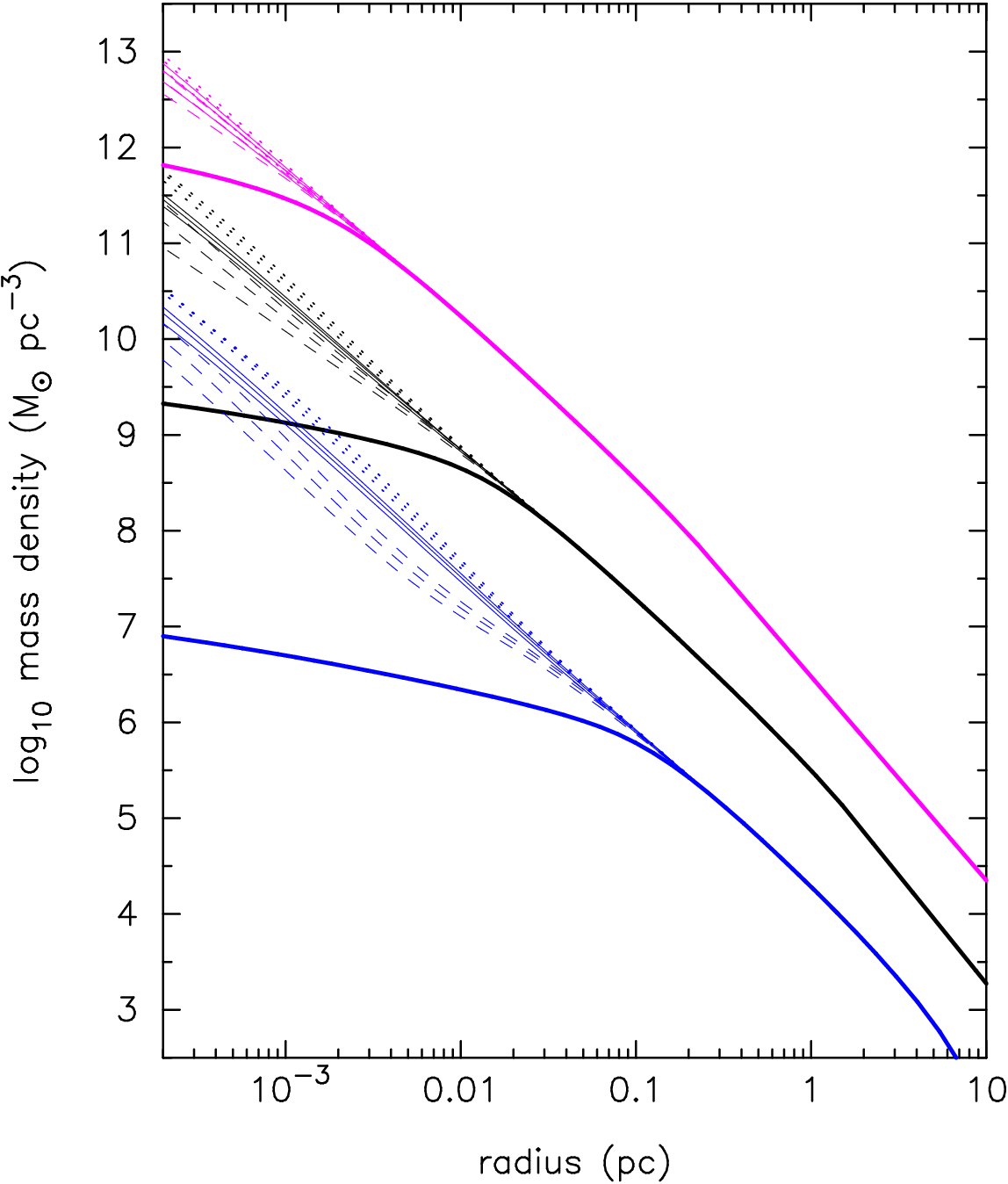} }}
\caption{Mass density as a function of radius for the steady-state models.
Line colors and line styles have the same meanings as in Figure~\ref{Figure:fofelast}.
Scaling to physical units assumes $\mh=4\times 10^6\msun$.
}
\label{Figure:rholast}
\end{figure}

Closely related to $\overline{f}({\cal E})$ is $\rho(r)$, the mass density.
Figure~\ref{Figure:rholast} shows steady-state density profiles for all the integrations.
Also shown  are three density profiles from Figure 4 of Paper II
(no anomalous relaxation), which exhibit cores corresponding 
to the depletion in $\overline{f}({\cal E})$ at large binding energies due to resonant relaxation.
Once again, the lesser depletion in the models that account for anomalous relaxation translates into
cores of lesser prominence.
Indeed in the models with ${\cal R}_1<{\cal R}_2$ or ${\cal R}_3<{\cal R}_4$, 
the steady-state density profiles turn out to be very close to the classical
Bahcall-Wolf cusp at all radii plotted.
This plot confirms a conjecture made in Paper I: namely: that the inhibition in angular-momentum diffusion
associated with the Schwarzschild barrier would reduce the ability of resonant relaxation to form a core.
The generality of this result is discussed in \S\ref{Section:Discussion}.

\begin{figure}[h!]
\centering
\mbox{\subfigure{\includegraphics[angle=0.,width=3.in]{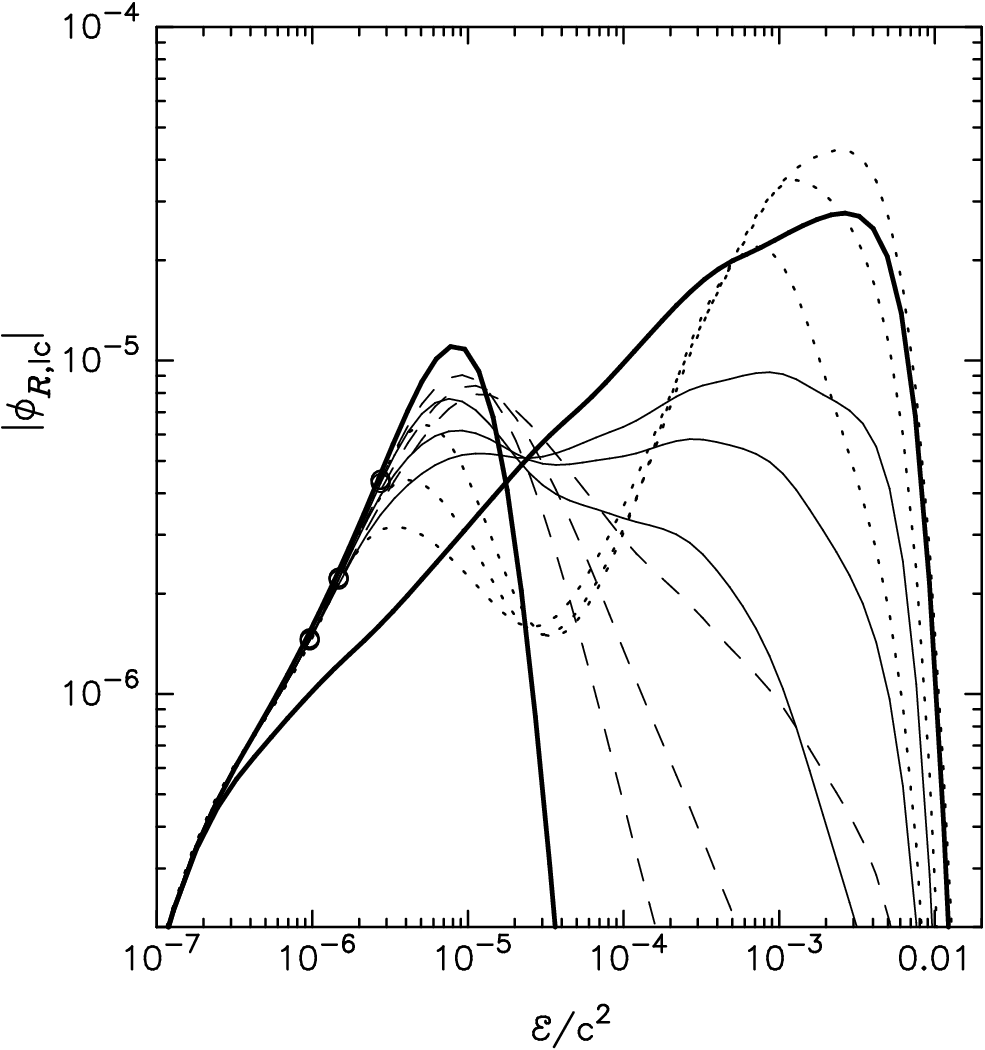}}\quad
\subfigure{\includegraphics[angle=0.,width=3.in]{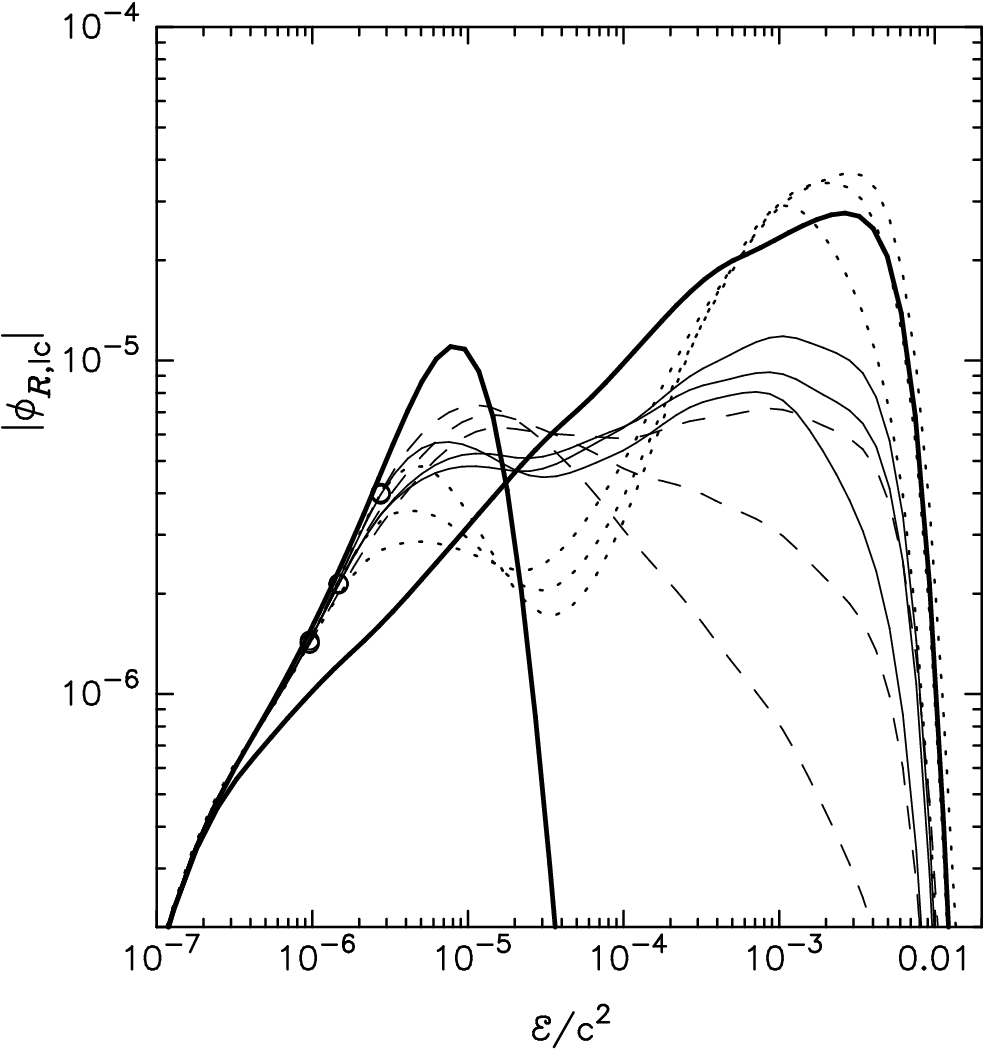} }}
\caption{Dimensionless flux of stars into the loss cone as a function of energy
in the steady-state models;
all of these models used the intermediate of the three values given in 
Equation~(\ref{Equation:rhoatonepc}) for the density at one parsec.
Left(right) panels adopted the power-law(exponential) forms of the anomalous diffusion coefficients.
The two, thick solid curves in each panel show models that included only classical relaxation
(curves that peak near the right)
and only classical plus resonant relaxation (curves that peak near the left).
Line styles have the same meanings as in Figures~\ref{Figure:fofelast} and \ref{Figure:rholast}.
Circles are plotted at values of ${\cal E}$ corresponding to the outer edge of the Schwarzschild barrier.}
\label{Figure:fluxes}
\end{figure}

Even models having similar $\rho(r)$ or $\overline{f}({\cal E})$ can have very different loss rates into the \sbh,
since the latter depends also on the timescale for angular-momentum diffusion.
The flux of stars into the loss cone, Equation~(\ref{Equation:FPFluxConserve}), 
is plotted for the steady-state models as a function of energy in Figure~\ref{Figure:fluxes}.
Shown for comparison are loss rates in steady-state models computed using only classical,
or classical plus resonant, relaxation.
These plots show that the flux of stars into the \sbh\ can depend strongly on the assumed 
forms of the anomalous diffusion coefficients.
There are two, competing effects.
Including anomalous relaxation tends to increase the steady-state density at small radii compared
to $\rho(r)$ computed using resonant relaxation alone, resulting in a larger flux.
Anomalous relaxation also increases the timescales for angular momentum diffusion, which
tends to reduce the flux.

The dependence of the flux on the assumed value of ${\cal R}_1/{\cal R}_2$ or ${\cal R}_3/{\cal R}_4$
is similarly complex.
At low binding energies, Figure~\ref{Figure:fluxes} shows that 
small values of this ratio imply lower fluxes; while at high binding energies, the
reverse is true.
The former result is consistent with the analysis in the Appendix (Figure~\ref{Figure:beta}).
The reason for the latter result can be seen in Figure~\ref{Figure:fofelast}: models
with smaller values of this ratio maintain larger $\overline{f}$ at large binding energies, which
tends to increase the flux.
{\bf Total, or integrated, loss rates for these models can be computed using Equation~(\ref{Equation:DefineFofE}).
The results turn out to be nearly the same -- within a few percent -- for each of the models, roughly
$7\times 10^{-4}$ stars per year. As discussed in Paper I, this is a consequence of the fact that
the total loss rate is dominated by stars at low binding energies, where the effects of resonant
and anomalous relaxation are small.}

\section{Discussion}
\label{Section:Discussion}

\subsection{Steady-state density profiles}
\label{Section:DiscussionDensity}

In Paper II, the formation of cores due to resonant relaxation was discussed.
Equation (38) of that paper gave an estimate of the radius 
of the core so formed:
\beq\label{Equation:RRcore}
\frac{a_\mathrm{core}}{r_m}\approx 0.028 \left(\frac{\ln\Lambda}{15}\right)^{-4/5} .
\eeq
Here, $r_m$ is the gravitational influence radius of the \sbh, defined as the radius of a sphere
containing a mass in stars of $2\mh$.

As described in this paper, at least in the case of the Milky Way, the inclusion of anomalous relaxation
tends to counteract core formation by causing stars to accumulate near the Schwarzschild barrier.

Here we consider the generality of that result.
Suppose that nuclei of galaxies fainter than the Milky Way have \sbhs\ that satisfy the $\mh-\sigma$
relation:
\beq
\frac{\mh}{10^6\msun} \approx 5.76 \left(\frac{\sigma}{100\,\mathrm{km\ s}^{-1}}\right)^{4.86} \nonumber
\eeq
\citep[][Eq.~2.33]{DEGN} and that their nuclear densities are close to the Bahcall-Wolf form,
$\rho(r) \propto r^{-7/4}$.
The latter is not too different from $\rho\propto r^{-2}$, for which $r_m\approx r_h\equiv G\mh/\sigma^2$.
Under these assumptions,
\beq\label{Equation:rmofmh}
r_m \approx 0.88 \left(\frac{\mh}{10^6\msun}\right)^{0.59} \mathrm{pc},\ \ \ \ 
\mh\lesssim 4\times 10^6\msun .
\eeq

We first verify the assumptions that led to Equation~(\ref{Equation:RRcore}).
That equation was derived assuming that $t_\mathrm{coh}=t_{\mathrm{coh,M}}$
(Equation~\ref{Equation:Definetcoh}).
The radius at which $t_\mathrm{coh,M}=t_\mathrm{coh,S}$, for a nucleus with $\rho(r) \propto r^{-7/4}$, is given by Equation~(30) from Paper II:
\begin{eqnarray}
\frac{a_\mathrm{coh}}{r_m} \approx 1.2\times 10^{-3} 
\left(\frac{\mh}{10^6\msun}\right)^{4/9} \left(\frac{r_m}{1\mathrm{pc}}\right)^{-4/9} 
\approx 1.3\times 10^{-3} \left(\frac{\mh}{10^6\msun}\right)^{0.18}
\end{eqnarray}
suggesting that indeed $t_\mathrm{coh} \sim t_\mathrm{coh,M}$ at $a=a_\mathrm{core}$
in the galaxies of interest.

Next we ask how $a_\mathrm{core}$ compares with the radii that define the Schwarzschild barrier.
Equation~(\ref{Equation:SBlimits}) gave approximate limits on the extent of the SB
in a $\rho\propto r^{-7/4}$ nucleus, which we recast here as
\begin{eqnarray}
\frac{a_\mathrm{min}}{r_m} \approx \left(\frac{\mh}{2m_\star}\right)^{4/13}
\left(\frac{r_g}{r_m}\right)^{8/13},
\ \ \ \ 
\frac{a_\mathrm{max}}{r_m} \approx \left(4\Theta\right)^{-4/9}
\left(\frac{\mh}{m_\star}\right)^{4/9}
\left(\frac{r_g}{r_m}\right)^{4/9} . 
\end{eqnarray}
Comparing $a_\mathrm{min}$ and $a_\mathrm{max}$ to $a_\mathrm{core}$:
\begin{subequations}
\begin{eqnarray}
\frac{a_\mathrm{min}}{a_\mathrm{core}} &\approx& 29
 \left(\frac{\mh}{m_\star}\right)^{4/13} \left(\frac{r_g}{r_m}\right)^{8/13} 
\left(\frac{\ln\Lambda}{15}\right)^{4/5} ,\\
\frac{a_\mathrm{max}}{a_\mathrm{core}} &\approx& 5.8 \left(\frac{\Theta}{15}\right)^{-4/9}
 \left(\frac{\mh}{m_\star}\right)^{4/9} \left(\frac{r_g}{r_m}\right)^{4/9} 
\left(\frac{\ln\Lambda}{15}\right)^{4/5} .
\end{eqnarray}
\end{subequations}
Using $ r_g \approx 4.78\times 10^{-8}  \left(\mh/10^6 \msun\right) \mathrm{pc} $,
these become
\begin{subequations}
\begin{eqnarray}
\frac{a_\mathrm{min}}{a_\mathrm{core}} &\approx& 0.064
 \left(\frac{\mh}{10^6\msun}\right)^{12/13}
\left(\frac{m_\star}{\msun}\right)^{-4/13} \left(\frac{r_m}{1\;\mathrm{pc}}\right)^{-8/13} 
\left(\frac{\ln\Lambda}{15}\right)^{4/5} ,\\
\frac{a_\mathrm{max}}{a_\mathrm{core}} &\approx& 1.5 \left(\frac{\Theta}{15}\right)^{-4/9}
 \left(\frac{\mh}{10^6\msun}\right)^{8/9}
\left(\frac{m_\star}{\msun}\right)^{-4/9} \left(\frac{r_m}{1\;\mathrm{pc}}\right)^{-4/9} 
\left(\frac{\ln\Lambda}{15}\right)^{4/5} .
\end{eqnarray}
\end{subequations}

Setting $a_\mathrm{max} < a_\mathrm{core}$ implies that anomalous relaxation is
unlikely to affect the formation of the core due to resonant relaxation.
This condition is:
\begin{eqnarray}\label{Equation:rmcond}
r_m&\gtrsim& 2.5 \left(\frac{\Theta}{15}\right)^{-1}
 \left(\frac{\mh}{10^6\msun}\right)^{2}
\left(\frac{m_\star}{\msun}\right)^{-1} 
\left(\frac{\ln\Lambda}{15}\right)^{9/5} \mathrm{pc}.
\end{eqnarray}
In the case of the Milky Way ($\mh\approx 4\times 10^6\msun$), satisfying this condition
 for $m_\star=\msun$  would require $r_m\gtrsim 40\;\mathrm{pc}$ -- about ten times larger
than the value inferred from stellar kinematics.
This is consistent with Figure~\ref{Figure:rholast}, which showed that anomalous relaxation
inhibits the formation of a core for all reasonable values of $r_m$.
In the case of galaxies with central black holes less massive than the Milky Way's,
 Equations~(\ref{Equation:rmcond}) and~(\ref{Equation:rmofmh}) allow us to write
 the condition $a_\mathrm{max}<a_\mathrm{core}$ as
\beq
\mh \lesssim 5\times 10^5 \left(\frac{\Theta}{15}\right)^{0.48} 
\left(\frac{m_\star}{\msun}\right)^{0.48} \left(\frac{\ln\Lambda}{15}\right)^{-0.86} \msun .
\eeq
Thus, the core formed by resonant relaxation is expected to become progressively more
prominent as $\mh$ is reduced below its value in the Milky Way.
This fact is likely to be important in determining the rate of formation of gravitational-wave sources,
particularly since theoretical estimates often focus on galaxies with $\mh\lesssim 10^6\msun$.
Estimating this rate will be the topic of upcoming papers in this series.

\subsection{Constraining the forms of the anomalous diffusion coefficients}
\label{Section:DiscussionAR}

Until  recently, discussions of gravitational encounters near a \sbh\ have usually been presented
in terms of diffusion {\it timescales}; that is; in terms of second-order diffusion coefficients
like $\langle(\Delta L)^2\rangle$
\citep{RauchTremaine1996,HopmanAlexander2006L,HopmanAlexander2006,GurkanHopman2007,Eilon2009,Madigan2011}.
\citet{Hamers2014} were apparently the first to consider the forms of the first-order diffusion coefficients.
Of course, both first- and second-order diffusion coefficients are essential when computing the
evolution of $f$ via the Fokker-Planck equation.

As shown here through a number of examples, the form of the steady-state $f({\cal E}, {\cal R})$
near the Schwarzschild barrier can depend very sensitively on the relative amplitude of the
first- and second-order coefficients in the anomalous-relaxation regime.
A natural case to consider is that of ``zero drift'', in which the two diffusion coefficients
imply a net flux in angular momentum that is zero when $f({\cal R})=\mathrm{constant}$
(\S\ref{Section:AR_PL}).
While it may be natural -- it is consistent with a ``maximum-entropy'' steady state -- 
this assumption is not compelled by any physical argument of which we are aware.
In stellar dynamics, incorrect conclusions drawn from entropy arguments are legion, and
numerical experiments, when available, would seem to be a better guide.
As discussed in detail in
\S\ref{Section:AR_Constrain} , the highest-quality $N$-body simulations carried out to date of this regime
\citep{MAMW2011} seem to require anomalous diffusion coefficients that differ slightly, though
significantly, from the ``zero-drift'' condition.
A state of ``positive drift" seems to better characterize the existing simulations.

Elucidation of the long-term effects of gravitational encounters in the Schwarzschild regime
near a \sbh\ will ultimately require a better specification of the anomalous diffusion coefficients.
By far the best way to do this -- at least in principle -- is via direct $N$-body integrations,
which impose the fewest approximations.
In practice, integrations of the required accuracy become very time-consuming when $N\gtrsim 10^2$.
A major effort should be devoted to increasing the efficiency of the $N$-body integrators.

\section{Summary}
\label{Section:Summary}

Integrations of the Fokker-Planck equation describing $f(E,L,t)$, the phase-space density
of stars around a supermassive black hole (\sbh) at the center of a galaxy, were carried out using
a numerical algorithm described in two earlier papers \citep{Paper1,Paper2}.
Diffusion coefficients describing classical, resonant and ``anomalous'' relaxation were included;
the latter accounting for the evolution of orbits in the regime below the Schwarzschild barrier (SB)
where the timescale for general relativistic precession is short compared with the coherence time,
invalidating the assumptions that underlie the theory of resonant relaxation \citep{MAMW2011}.
The principal results follow.

\bigskip
\noindent
1. Since a good theoretical understanding of anomalous relaxation is still lacking, two functional forms
were considered for the angular momentum diffusion coefficients in this regime, 
having either a power-law or exponential dependence on ${\cal R}\equiv L^2/L_c^2$.
In either case, a further choice must be made in terms of how to relate the first- and second-order
diffusion coefficients.
It was argued that a natural starting point is a ``zero-drift'' condition that implies no net
flux in angular momentum when $f({\cal R})$ is constant.
Parameterized functional forms for $\langle\Delta{\cal R}\rangle$ and $\langle(\Delta{\cal R})^2\rangle$
were proposed that have the ``zero-drift'' property as a special case.

\bigskip
\noindent
2. Two attempts were made to constrain the forms of the anomalous diffusion coefficients by comparison
with published numerical simulations.
First, as in \citet{Hamers2014}, diffusion coefficients extracted from a large set of test-particle
integrations were compared with the two functional forms.
The power-law form was found to be strongly preferred, at least in the case of the second-order coefficient, 
confirming a result already presented in that paper.
The first-order coefficient was also well fit by the power-law form, 
although data were more noisy and no clear preference
could be established for the ``zero-drift'' conditions.
Second, a set of Fokker-Planck integrations were carried out based on the same initial conditions
that were used in the exact $N$-body integrations of \citet{MAMW2011}.
Two properties of those models were then compared: the dependence of $f$ on ${\cal R}$,
and the capture rate; in both cases, results from the $N$-body integrations were averaged over 
a set of different runs to reduce noise.
Both the power-law and exponential forms for the diffusion coefficients could be made consistent
with these data; it was argued that this was due in part to the effects of classical relaxation,
which always dominates the diffusion rate at sufficiently small ${\cal R}$.
However, a clear preference was established for diffusion coefficients that imply a steady-state
drift toward larger ${\cal R}$, inconsistent with the ``zero-drift'' hypothesis.

\bigskip
\noindent
3. Fokker-Planck integrations were then carried out to find steady-state models having parameters
similar to those of the nuclear star cluster in the Milky Way.
These models were identical to the steady-state models computed in Paper II except for the inclusion of the anomalous diffusion coefficients.
The steady-state $f({\cal E}, {\cal R})$ in regions of phase space below the Schwarzschild barrier
(${\cal R}<{\cal R}_\mathrm{SB}$) was found to be most strongly dependent on the assumed relation
between first- and second-order diffusion coefficients.
Diffusion coefficients satisfying the ``zero-drift'' condition produced steady-state solutions in
which $f$ was nearly constant with respect to ${\cal R}$ below the SB.
Integrations incorporating positive- or negative-drift diffusion coefficients had steady-state
$f$'s that respectively increased or decreased below the SB, before falling to zero at the loss cone
boundary.
In all of these models, departures of the steady-state density, $n(r)$,
from the classical Bahcall-Wolf solution were less pronounced
than in the models of Paper II that did not incorporate anomalous relaxation
(a result that was suggested already in that paper).
The reason is the tendency of stars to accumulate near and
below the SB, thus counteracting the depletion that occurs when only resonant relaxation is accounted for.

\bigskip
\noindent
4. Steady-state loss rates in the presence of anomalous relaxation differ from those in all models
previously published, for two reasons: different steady-state phase-space densities, and different
forms of the angular-momentum diffusion coefficients.
A robust conclusion is that the incorporation of anomalous relaxation implies a lower
capture rate at energies where the SB exists, compared with models that only incorporate 
resonant relaxation, and this is true in spite of the generally higher steady-state densities in the
former models.
However the reduction in the capture rate  depends sensitively on the parameters adopted
for the anomalous diffusion coefficients, being most(least) extreme for the positive-(negative-)
drift cases.

\bigskip
\noindent
5. In galaxies with \sbhs\ less massive than the Milky Way's, core formation by resonant relaxation
is likely to be progressively less affected by anomalous relaxation.



\acknowledgements
A. Hamers kindly provided data from his {\tt TPI} code that were used in constraining
the functional forms of the anomalous diffusion coefficients in \S \ref{Section:AR}.
I also thank him, F. Antonini and E. Vasiliev for  comments that improved the manuscript.
This work was supported by the National Science Foundation under grant no. AST 1211602 
and by the National Aeronautics and Space Administration under grant no. NNX13AG92G.

 \appendix
  \renewcommand{\theequation}{A\arabic{equation}}
  \setcounter{equation}{0} 
  \centerline{Properties of steady-state solutions in the anomalous-relaxation regime} 
   \label{Appendix:SS}
\bigskip

We consider solutions to the time-independent Fokker-Planck equation in ${\cal R}$-space
in the anomalous-relaxation regime.
We assume that the diffusion coefficients are modified versions of the resonant diffusion  coefficients:
\begin{subequations}\label{Equation:DiffCoefsAppendix}
\begin{eqnarray}
\langle \Delta{\cal R}\rangle &=& w_1({\cal E},{\cal R}) \langle\Delta{\cal R}\rangle_\mathrm{RR},
\ \ \ \ 
\langle (\Delta{\cal R})^2\rangle =  w_2({\cal E},{\cal R}) \langle(\Delta{\cal R})^2\rangle_\mathrm{RR} ,
\\
\langle\Delta{\cal R}\rangle_\mathrm{RR} &=&  2A({\cal E})\left(1-2{\cal R}\right) g_1({\cal E}, {\cal R}),
\ \ \ \ 
\langle(\Delta{\cal R})^2\rangle_\mathrm{RR} =  4A({\cal E}){\cal R}\left(1-{\cal R}\right) g_2({\cal E}, {\cal R})
\end{eqnarray}
\end{subequations}
and consider the two functional forms for $\{w_1, w_2\}$ that were considered in \S\ref{Section:AR}: 
a power-law modification, and an exponential modification.
Diffusion in energy is ignored.

\bigskip
\subsection{(1) Power-law}
Identify $w_1$ and $w_2$ with $g_1$ and $g_2$ given respectively by Equations 
(\ref{Equation:Defineg1}) and (\ref{Equation:Defineg2}):
\begin{eqnarray}\label{Equation:g1g2Appendix}
g_1({\cal E}, {\cal R}) &=&  \left\{1+\left[\frac{R_1({\cal E})}{\cal R}\right]^n\right\}^{-2/n} 
+ \frac {2(1-{\cal R})}{1-2{\cal R}} \left(\frac{{\cal R}_1}{{\cal R}}\right)^n
 \left\{1+\left[\frac{R_1({\cal E})}{\cal R}\right]^n\right\}^{-2/n-1} , \nonumber \\
g_2({\cal E},{\cal R}) &=& \left\{1+\left[\frac{R_2({\cal E})}{\cal R}\right]^n\right\}^{-2/n} .
\end{eqnarray}

The flux coefficients, equations~(\ref{Equation:RFluxCoefs}), are
\begin{subequations}\label{Appendix:RFluxCoefs}
\begin{eqnarray}
D_{\cal R}&=&  
-\langle\Delta {\cal R}\rangle + \frac12\frac{\partial}{\partial {\cal R}} \langle(\Delta {\cal R})^2\rangle\\
&=& 2A({\cal E}) \left[1+\left(\frac{{\cal R}_2}{\cal R}\right)^n\right]^{-(1+2/n)}
\left[1-2{\cal R} + \left(3-4{\cal R}\right) \left(\frac{{\cal R}_2}{\cal R}\right)^n\right] \nonumber
\\ &-& 
2A({\cal E}) \left[1+\left(\frac{{\cal R}_1}{\cal R}\right)^n\right]^{-(1+2/n)}
\left[1-2{\cal R} + \left(3-4{\cal R}\right) \left(\frac{{\cal R}_1}{\cal R}\right)^n\right] \\
D_{\cal RR} &=& \frac12\langle(\Delta {\cal R})^2\rangle \\
&=& 2A({\cal E})  {\cal R}\left(1-{\cal R}\right) \left[1+\left(\frac{{\cal R}_2}{\cal R}\right)^n\right]^{-2/n} .
\end{eqnarray}
\end{subequations}
It is easy to verify that in this case,
\beq
D_{\cal R} = -\langle\Delta {\cal R}\rangle + \frac{\langle(\Delta {\cal R})^2\rangle}{2{\cal R}}\left[\frac{1-2{\cal R}}{1-{\cal R}} + \frac{2}{1+\left({\cal R}/{\cal R}_2\right)^n}\right]  .
\eeq
In the limit ${\cal R}\ll \{{\cal R}_1, {\cal R}_2\}$, these expressions become
\begin{eqnarray}
D_{\cal R} \rightarrow 6A({\cal E}) 
\left[ \left(\frac{{\cal R}}{{\cal R}_2}\right)^{2} - 
 \left(\frac{\cal R}{{\cal R}_1}\right)^{2} \right],\ \ \ \ 
D_{\cal RR} \rightarrow 2A({\cal E})  {\cal R} \left(\frac{\cal R}{{\cal R}_2}\right)^{2} .
\end{eqnarray}
The ${\cal R}$-directed flux is
\begin{eqnarray}\label{Appendix:phiofR}
\phi_{\cal R} = - D_{\cal RR}\frac{\partial f}{\partial {\cal R}} - D_{\cal R} f 
\end{eqnarray}
which, in the small-${\cal R}$ limit, becomes
\begin{eqnarray}
\phi_{\cal R} &\rightarrow& -2A({\cal E}) \left(\frac{{\cal R}}{{\cal R}_2}\right)^2
\left(3\lambda f + {\cal R}\frac{\partial f}{\partial {\cal R}}\right) , \ \ \ \ \ 
\lambda \equiv 1 - \left(\frac{{\cal R}_2}{{\cal R}_1}\right)^2 .
\end{eqnarray}

Steady-state solutions can be characterized by either a constant, or a zero, flux.
Setting $\phi_{\cal R}=0$ yields
\beq
f({\cal R}) = f({\cal R}_2) \left(\frac{\cal R}{{\cal R}_2}\right)^{-3\lambda} , \ \ 
{\cal R}\ll \{{\cal R}_1, {\cal R}_2\}.
\eeq
When ${\cal R}_1<{\cal R}_2$, $\lambda <0$ and the solution decreases toward ${\cal R}=0$;
the reverse is true when ${\cal R}_1>{\cal R}_2$.
Setting ${\cal R}_1={\cal R}_2$ yields $\lambda = 0$ and $f({\cal R})=\mathrm{const.}$,
the ``zero-drift'' solution.

A steady-state solution with constant but nonzero flux has the small-${\cal R}$ form
\beq
f({\cal R}) = \left(\frac{\cal R}{{\cal R}_2}\right)^{-2} 
\left[\frac{1}{2-3\lambda}  \frac{\phi_{\cal R}}{2 A} + 
C_\lambda \left(\frac{\cal R}{{\cal R}_2}\right)^{2-3\lambda}\right]
\eeq
for $\lambda \ne 2/3$, with $C_\lambda$ an integration constant.
For $\lambda=2/3$,
\beq
f({\cal R}) = \left(\frac{\cal R}{{\cal R}_2}\right)^{-2} 
\left[-\frac{\phi_{\cal R}}{2 A} \log\left(\frac{\cal R}{{\cal R}_2}\right) + C_\frac23 \right] .
\eeq
We compute $C_\lambda$ by requiring $f$ to fall to zero at ${\cal R}\equiv {\cal R}_0$ (``empty loss cone'').
The results are
\begin{subequations}
\begin{eqnarray}
f({\cal R}) &=& f({\cal R}_2) \left(\frac{\cal R}{{\cal R}_2}\right)^{-2}
\frac{\left({\cal R}/{\cal R}_0\right)^q - 1}
{\left({\cal R}_2/{\cal R}_0\right)^q - 1} , \ \ \ \ q \equiv 2 - 3\lambda , \\
\phi_{\cal R}({\cal E}) &=& -2qA({\cal E}) \frac{f({\cal R}_2)} 
{1 - \left({\cal R}_2/{\cal R}_0\right)^{q}} 
\end{eqnarray}
\end{subequations}
for $\lambda\ne 2/3$, and 
\begin{subequations}
\begin{eqnarray}
f({\cal R}) &=& f({\cal R}_2) \left(\frac{\cal R}{{\cal R}_2}\right)^{-2}
\frac{\log\left({\cal R}/{\cal R}_0\right) }
{\log\left({\cal R}_2/{\cal R}_0\right)} , \\
\phi_{\cal R}({\cal E}) &=& -2A({\cal E}) \frac{f({\cal R}_2)} 
{\log\left({\cal R}_2/{\cal R}_0\right)} 
\end{eqnarray}
\end{subequations}
for $\lambda= 2/3$.

At most energies, ${\cal R}_0\ll {\cal R}_2$. 
Assuming this inequality, $f$ and $\phi_{\cal R}$ have the following forms,
depending on the value of $\lambda$:

\smallskip
\noindent
1. $\lambda > 2/3$, i.e. ${\cal R}_1/{\cal R}_2 >\sqrt{3}$.
Defining $p\equiv 3\lambda - 2 >0$,
\beq
\frac{f({\cal R})}{f({\cal R}_2)} \approx \left(\frac{{\cal R}_2}{\cal R}\right)^2\left[1-\left(\frac{{\cal R}_0}{\cal R}\right)^p\right],\ \ \ \ 
\phi_{\cal R} \approx -2p A({\cal E}) f({\cal R}_2) .
\eeq

\smallskip
\noindent
2. $\lambda < 2/3$, i.e. ${\cal R}_1/{\cal R}_2 <\sqrt{3}$.
In terms of $q\equiv 2-3\lambda >0$,
\beq
\frac{f({\cal R})}{f({\cal R}_2)} \approx \left(\frac{{\cal R}_2}{\cal R}\right)^2 
\left[\left(\frac{\cal R}{{\cal R}_2}\right)^q - \left(\frac{{\cal R}_0}{{\cal R}_2}\right)^q\right],\ \ \ \ 
\phi_{\cal R} \approx -2q A({\cal E}) \left(\frac{{\cal R}_0}{{\cal R}_2}\right)^q f({\cal R}_2) .
\eeq
The ``zero-drift'' case has ${\cal R}_1={\cal R}_2$, $\lambda=0$, $q=2$.

\smallskip
\noindent
3. $\lambda = 2/3$, i.e. ${\cal R}_1/{\cal R}_2 =\sqrt{3}$ :
\beq
\frac{f({\cal R})}{f({\cal R}_2)} \approx \left(\frac{{\cal R}_2}{\cal R}\right)^2
\frac{\log ({\cal R}/{\cal R}_0)}{\log({\cal R}_2/{\cal R}_0)},
\ \ \ \ 
\phi_{\cal R} \approx -\frac{2A({\cal E})f({\cal R}_2)}{\log({\cal R}_2/{\cal R}_0)} .
\eeq

\begin{figure}[h]
\centering
\mbox{\subfigure{\includegraphics[width=3.0in]{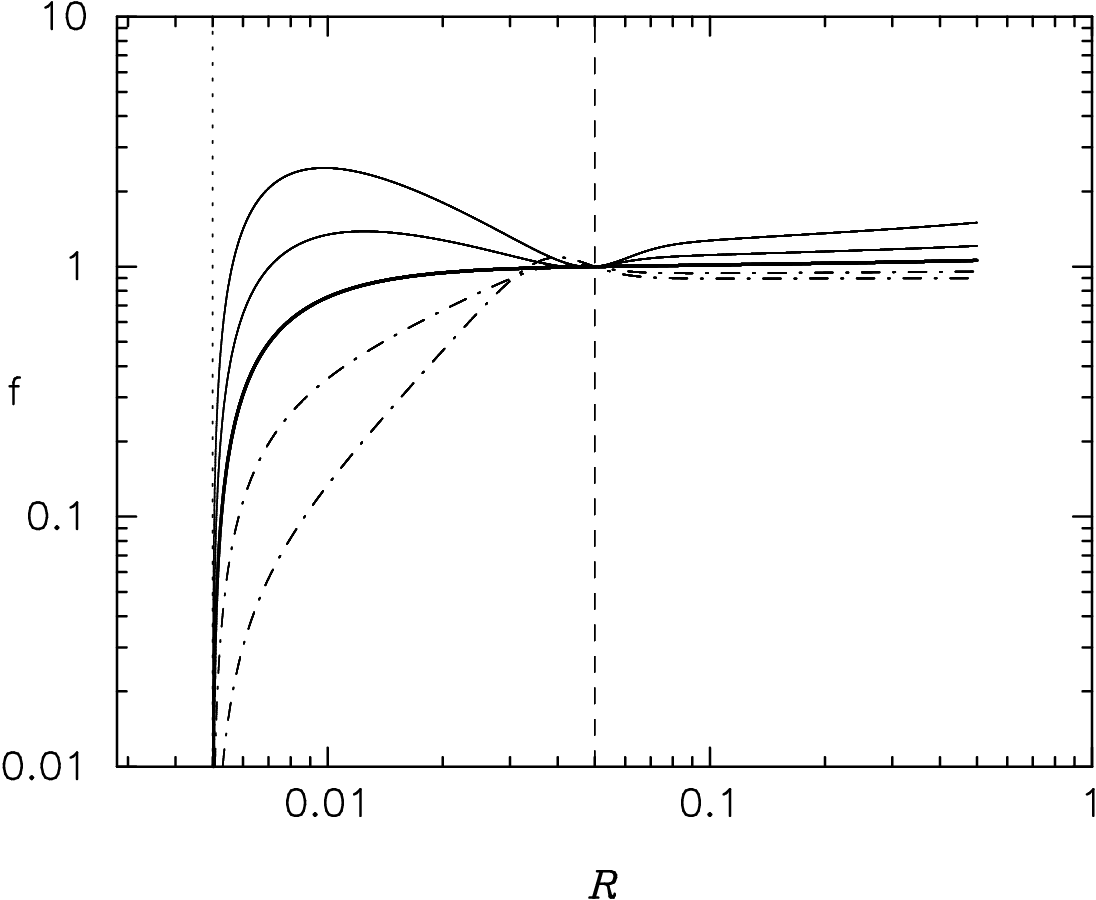}}\quad
\subfigure{\includegraphics[width=3.0in]{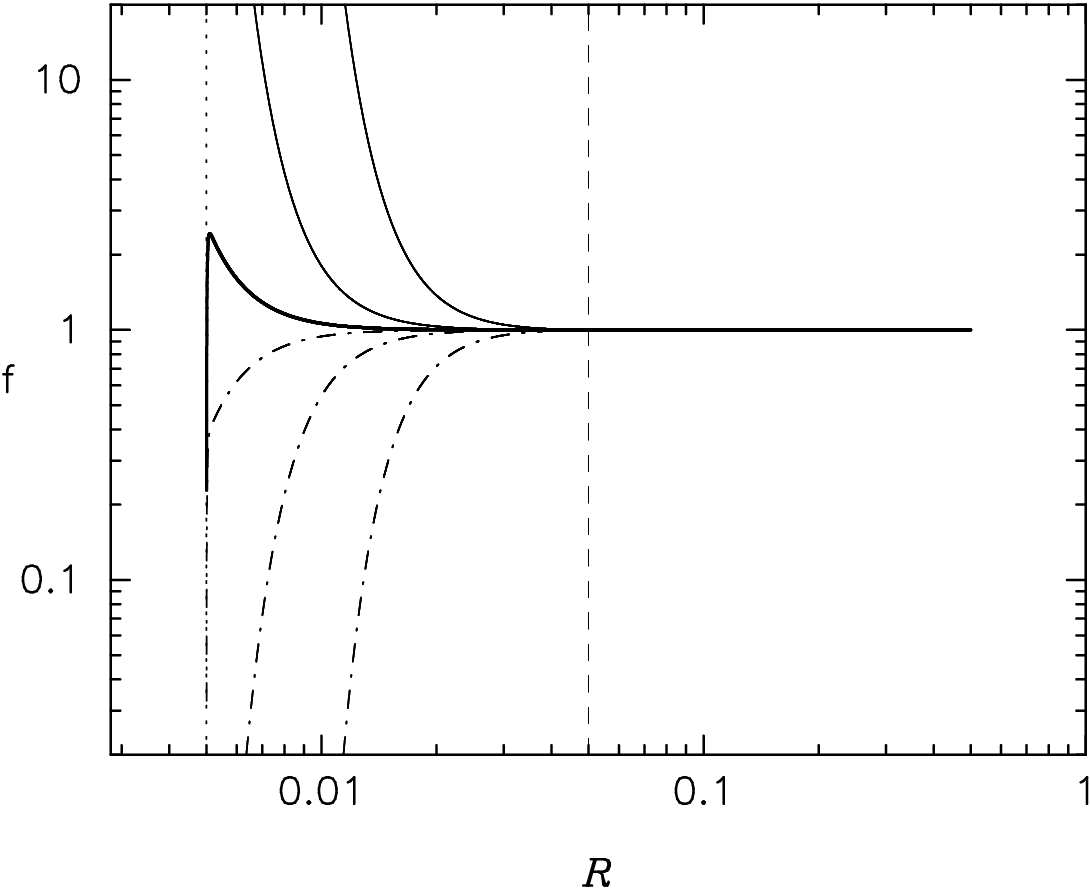} }}
\caption{Steady-state solutions in the anomalous-relaxation regime, under two assumptions about
how the resonant diffusion coefficients are modified.
{\bf Left panel:} Power-law modification, equations (\ref{Equation:g1g2Appendix}), 
with $n=8$, ${\cal R}_2=0.05$
(shown by the vertical dashed line), and
boundary condition $f({\cal R})=0$ at ${\cal R}=0.005$ (show by the vertical dotted line).
${\cal R}_1/{\cal R}_2 = \{0.8,0.9,1,1.1,1.25\}$.
Solid lines: ${\cal R}_1/{\cal R}_2 \ge 1$; dot-dashed lines: ${\cal R}_1/{\cal R}_2 < 1$. 
{\bf Right panel:} Exponential modification, equations (\ref{Equation:h1h2Appendix}), with ${\cal R}_4=0.05$ and
${\cal R}_3/{\cal R}_4 = \{0.99,0.999,0.9999,1.0001,1.001,1.01\}$.
Solid lines: ${\cal R}_3/{\cal R}_4 > 1$; dot-dashed lines: ${\cal R}_3/{\cal R}_4 < 1$. 
\label{Figure:fofRAppend}}
\end{figure}

\begin{figure}[h]
\centering
\mbox{\subfigure{\includegraphics[width=3.0in]{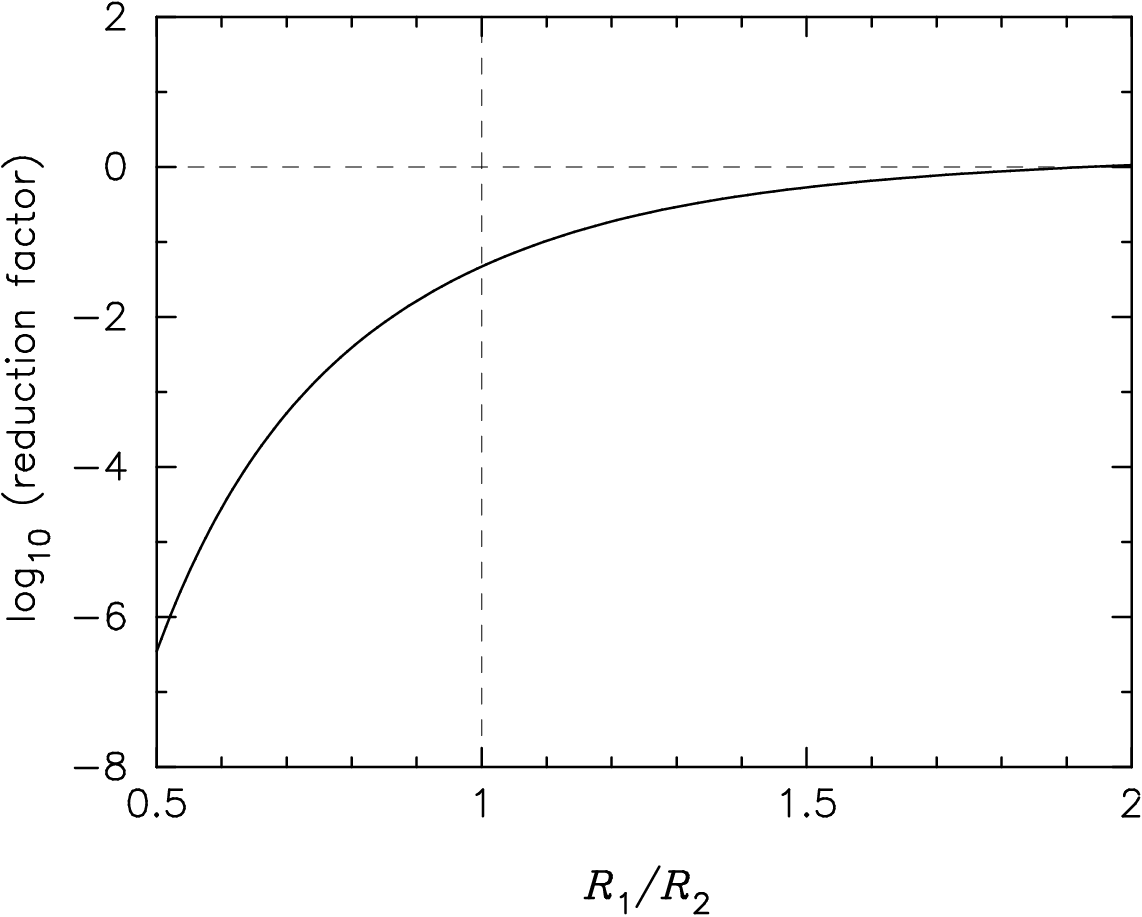}}\quad
\subfigure{\includegraphics[width=3.0in]{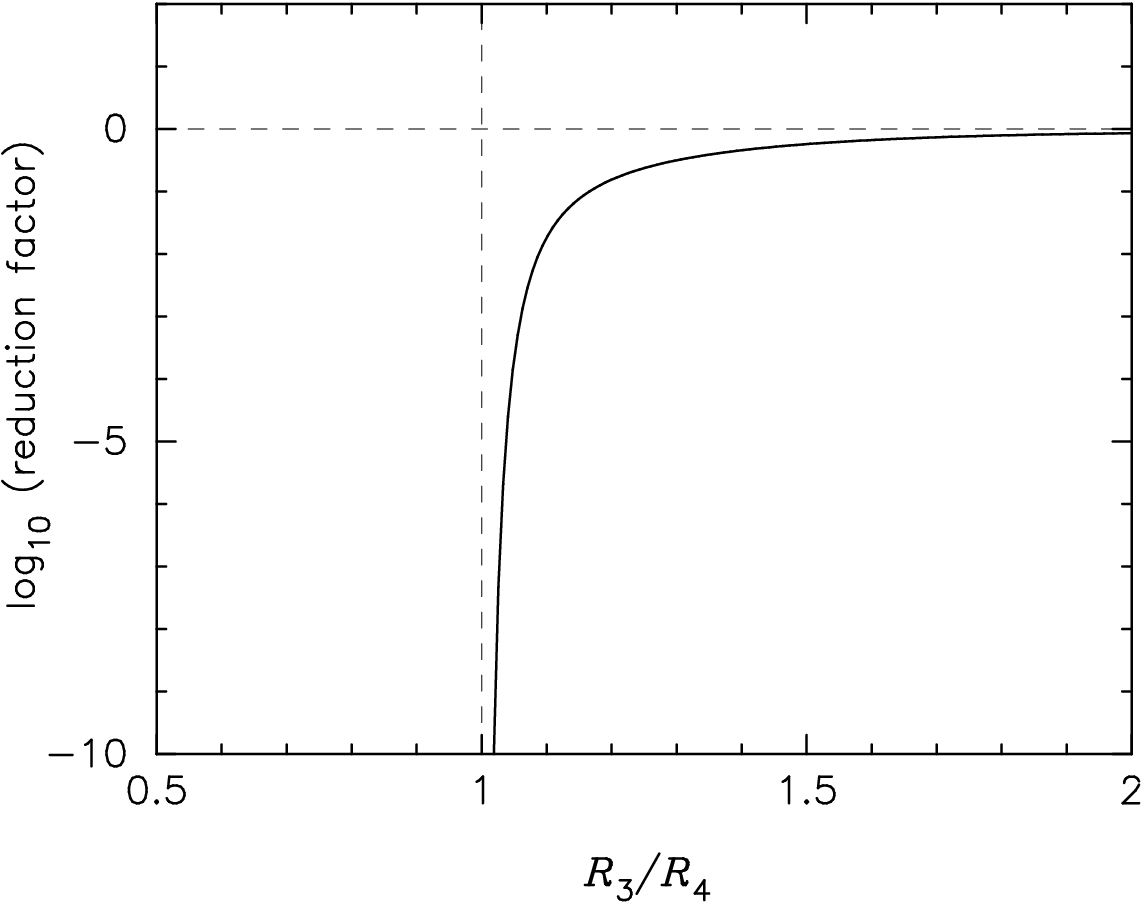} }}
\caption{Reduction in the steady-state flux due to anomalous diffusion.
{\bf Left panel:} power-law modification; {\bf right panel:} exponential modification.
Both curves assume $f({\cal R})=0$ at ${\cal R}=0.005$, as in Figure~\ref{Figure:fofRAppend}.
\label{Figure:beta}}
\end{figure}

It is interesting to compare the expressions for the flux to those that would obtain in the 
absence of anomalous relaxation.
Repeating the analysis with $g_1=g_2=1$, we find:
\begin{eqnarray}
D_{\cal R} &=& 0, \ \ \ \ 
D_{\cal RR} = 2A({\cal E}) {\cal R}\left(1-{\cal R}\right), \ \ \ \ 
\phi_{\cal R} = -2A({\cal E}) {\cal R} \left( 1-{\cal R}\right) \frac{\partial f}{\partial {\cal R}} ,
\end{eqnarray}
so that the steady state is characterized by
\beq
2A({\cal E}) {\cal R}\left(1-{\cal R}\right) \frac{\partial f}{\partial{\cal R}} + \phi_{\cal R} = 0 .
\eeq
The solution is
\beq
\frac{f({\cal R})}{f({\cal R}_2)} = 
\frac{\log\left(\frac{\cal R}{1-{\cal R}}\right) - \log\left(\frac{{\cal R}_0}{1-{\cal R}_0}\right)}
{\log\left(\frac{{\cal R}_2}{1-{\cal R}}\right) - \log\left(\frac{{\cal R}_0}{1-{\cal R}_0}\right)}
\eeq
with flux
\beq
\phi_{\cal R} = -2A({\cal E}) \frac{f({\cal R}_2)}
{\log\left(\frac{{\cal R}_2}{1-{\cal R}_2} \right) - \log \left(\frac{{\cal R}_0}{1-{\cal R}_0} \right)}
\eeq
Again supposing that ${\cal R}_0\ll {\cal R}_2$, these expressions become:
\begin{eqnarray}
\frac{f({\cal R})}{f({\cal R}_2)} \approx 
\frac{\log{\cal R} - \log{\cal R}_0} {\log{\cal R}_2 - \log{\cal R}_0} ,
\ \ \ \ \ 
\phi_{\cal R} \approx  -\frac{2A({\cal E}) f({\cal R}_2)}
{\log\left( {\cal R}_2 / {\cal R}_0 \right)} .
\end{eqnarray}
Define a ``reduction factor,'' $\eta$, as the ratio of this flux to the flux that would obtain in the
presence of anomalous relaxation, assuming the same value for $f({\cal R}_2)$.
Then
\[
 \eta \approx
  \begin{cases}
   p \log ({\cal R}_2/{\cal R}_0) & \text{if } {\cal R}_1/{\cal R}_2 > \sqrt{3} \\
    q ({\cal R}_0/{\cal R}_2)^q \log ({\cal R}_2/{\cal R}_0)    & \text{if }    {\cal R}_1/{\cal R}_2 < \sqrt{3} \\
   1       & \text{if }    {\cal R}_1/{\cal R}_2 = \sqrt{3} .
     \end{cases}
\]
Figures~\ref{Figure:fofRAppend} and \ref{Figure:beta} plot more accurate expressions for $f({\cal R})$
and $\eta$, computed using equations~(\ref{Appendix:RFluxCoefs}) and (\ref{Appendix:phiofR}), 
without assuming the smallness of ${\cal R}$ or ${\cal R}_0$.

The ``zero-drift'' solution has 
${\cal R}_1={\cal R}_2$, $\lambda=0$ and $q=2$.
For these values,
\beq
\eta = 2 \left(\frac{{\cal R}_0}{{\cal R}_2}\right)^2 \log\left(\frac{{\cal R}_2}{{\cal R}_0}\right).
\eeq
For ${\cal R}_0/{\cal R}_2 = \{0.5,0.1,0.01,0.001 \}$,
$\eta \approx \{0.35, 0.046, 9.2\times 10^{-4}, 1.4\times 10^{-5} \}$.

The value of ${\cal R}_1/{\cal R}_2$ favored in the numerical experiments was $\sim 0.8$,
implying $\lambda\approx -0.56$ and $q\approx 3.7$.
For these values,
\beq
\eta \approx 3.7 
\left(\frac{{\cal R}_0}{{\cal R}_2}\right)^{3.7} \log\left(\frac{{\cal R}_2}{{\cal R}_0}\right).
\eeq
For ${\cal R}_0/{\cal R}_2 = \{0.5,0.1,0.01,0.001 \}$,
$\eta \approx \{0.20, 1.7\times 10^{-3}, 6.8\times 10^{-7}, 2.0\times 10^{-10}\}$.

\bigskip\bigskip
\subsection{(2) Exponential}
Next we identify $\{w_1,w_2\}$ with $\{h_1,h_2\}$ given by equations (\ref{Equation:Defineh1})
and (\ref{Equation:Defineh2}):
\begin{eqnarray}\label{Equation:h1h2Appendix}
h_1({\cal E}, {\cal R}) &=& \left[1 + \frac{2(1-{\cal R})}{1-2{\cal R}}\left(\frac{{\cal R}_3}{{\cal R}}\right)^2\right]
\exp\left(-\frac{{\cal R}_3^2}{{\cal R}^2}\right) , \nonumber \\
h_2({\cal E},{\cal R}) &=& \exp\left(-\frac{{\cal R}_4^2}{{\cal R}^2}\right) .
\end{eqnarray}
The flux coefficients are
\begin{subequations}
\begin{eqnarray}
D_{\cal R}&=& 2A({\cal E}) \left[1-2{\cal R} + 2\left(1-{\cal R}\right)\left(\frac{{\cal R}_4}{{\cal R}}\right)^2\right]
\exp\left(-\frac{{\cal R}_4^2}{{\cal R}^2}\right) 
\nonumber \\ 
&-& 2A({\cal E}) \left[1-2{\cal R} + 2\left(1-{\cal R}\right)\left(\frac{{\cal R}_3}{{\cal R}}\right)^2\right]
\exp\left(-\frac{{\cal R}_3^2}{{\cal R}^2}\right) ,
\\
D_{\cal RR} &=& 2A({\cal E})  {\cal R}\left(1-{\cal R}\right)  \exp\left(-\frac{{\cal R}_4^2}{{\cal R}^2}\right) 
\end{eqnarray}
\end{subequations}
and
\beq
D_{\cal R} = -\langle\Delta {\cal R}\rangle + \frac{\langle(\Delta {\cal R})^2\rangle}{2{\cal R}}
\left(\frac{1-2{\cal R}}{1-{\cal R}} + 2\frac{{\cal R}_4^2}{{\cal R}^2}\right)  .
\eeq
In the limit ${\cal R}\ll \{{\cal R}_3, {\cal R}_4\}$, these expressions become
\begin{eqnarray}
D_{\cal R} \rightarrow 4A({\cal E}) \left(\frac{{\cal R}_4}{\cal R}\right)^2
\exp\left(-\frac{{\cal R}_4^2}{{\cal R}^2}\right)
\left[1 - \frac{{\cal R}_3^2}{{\cal R}_4^2}\exp{\left(-\frac{{\cal R}_3^2 - {\cal R}_4^2}{{\cal R}^2}\right)}\right] ,
\ \ \ \ 
D_{\cal RR} \rightarrow 2A({\cal E}){\cal R}\exp{\left(-\frac{{\cal R}_4^2}{{\cal R}^2}\right)}  .
\end{eqnarray}
In the same limit, the ${\cal R}$-directed flux is
\begin{eqnarray}
\phi_{\cal R} &\rightarrow& -2A({\cal E}) \exp\left(-\frac{{\cal R}_4^2}{{\cal R}^2}\right)
\bigg\{2\left(\frac{{\cal R}_4}{{\cal R}}\right)^2
\left[1 - \frac{e^{-\delta{\cal R}_3^2/{\cal R}^2}}{1-\delta}\right] f + 
{\cal R}\frac{\partial f}{\partial {\cal R}}\bigg\} , \ \ \ \ \ 
\delta \equiv 1 - \left(\frac{{\cal R}_4}{{\cal R}_3}\right)^2 .
\end{eqnarray}

Setting $\phi_{\cal R}=0$ yields
\beq
\frac{\log f({\cal R})}{\log f({\cal R}_4)} = 
 \frac{{\cal R}_4^2}{{\cal R}^2} - 1 + \frac{{\cal R}_3^2}{{\cal R}_3^2 - {\cal R}_4^2}
\left[\exp\left(\frac{{\cal R}_4^2 - {\cal R}_3^2}{{\cal R}^2}\right) -
\exp\left(\frac{{\cal R}_4^2 - {\cal R}_3^2}{{\cal R}_4^2}\right)\right] , 
\ \ 
{\cal R}\ll \{{\cal R}_3, {\cal R}_4\}.
\eeq
For ${\cal R}_3 < {\cal R}_4$, the dominant terms imply
\beq
f({\cal R}) \rightarrow f({\cal R}_4) \exp\left[\delta^{-1} \exp\left(\alpha \frac{{\cal R}_4^2}{{\cal R}^2}\right)\right]
\eeq
where $\delta = 1-{\cal R}_4^2/{\cal R}_3^3 <0$, $\alpha = 1-{\cal R}_3^2/{\cal R}_4^2 >0$, hence
$f$ drops very rapidly to zero below ${\cal R}={\cal R}_4$.
Whereas for ${\cal R}_3 > {\cal R}_4$,
\beq
f({\cal R}) \rightarrow f({\cal R}_4) e^{{\cal R}_4^2/{\cal R}^2},
\eeq
a rapid rise toward ${\cal R}=0$.
This is qualitatively the same behavior as in the power-law case when $\phi_{\cal R}=0$.

In the case of a constant but nonzero flux, only the ``zero-drift'' solution
(with ${\cal R}_3={\cal R}_4$) can be expressed in terms of simple functions:
\begin{subequations}
\begin{eqnarray}
f({\cal R}) &\approx& -\frac{\phi_{\cal R}}{4A}\left[E_i\left(\frac{{\cal R}_4^2}{{\cal R}_0^2}\right)  -
E_i\left(\frac{{\cal R}_4^2}{{\cal R}^2}\right)\right],\ \ \ \ {\cal R}\ll {\cal R}_3, {\cal R}_4 \\
\frac{f({\cal R})}{f({\cal R}_4)} &\approx &
\frac{E_i\left({\cal R}_4^2/{\cal R}_0^2\right) -  E_i\left({\cal R}_4^2/{\cal R}^2\right)}
{E_i\left({\cal R}_4^2/{\cal R}_0^2\right) -  E_i\left(1\right)} , \\
-\frac{\phi_{\cal R}}{4A} &\approx& f({\cal R}_4) \left[E_i\left(\frac{{\cal R}_4^2}{{\cal R}_0^2}\right) -
E_i\left(1\right) \right]^{-1} .
\end{eqnarray}
\end{subequations}
In these expressions, $E_i$ is the exponential function.
The reduction factor defined above becomes, in this case,
\beq
\eta = 2\frac{{\cal R}_4}{{\cal R}_0} 
\log \left(\frac{{\cal R}_4}{{\cal R}_0}\right)
e^{-{\cal R}_4/{\cal R}_0} 
\eeq

For ${\cal R}_0/{\cal R}_4 = \{0.5,0.1,0.01\}$,
$\eta \approx \{0.38, 2.1\times 10^{-3}, 3.4\times 10^{-41}\}$.

\clearpage


\begin{thebibliography}{}

\bibitem[Antonini \& Merritt(2012)]{AntoniniMerritt2012} 
Antonini, F., \& Merritt, D.\ 2012, \apj, 745, 83 

\bibitem[Antonini \& Merritt(2013)]{AntoniniMerritt2013} 
Antonini, F., \& Merritt, D.\ 2013, \apjl, 763, LL10 

\bibitem[Bahcall \& Wolf(1976)]{BahcallWolf1976}
Bahcall, J.~N., \& Wolf, S. 1976, \apj, 209, 214

\bibitem[Bahcall \& Wolf(1977)]{BahcallWolf1977} 
Bahcall, J.~N., \& Wolf, R.~A.\ 1977, \apj, 216, 883

\bibitem[Bar-Or \& Alexander(2014)]{BarorAlexander2014}
Bar-Or, B., \& Alexander, T.\ 2014, Classical and Quantum Gravity, 31, 244003 

\bibitem[Bartko et al.(2010)]{Bartko2010}
Bartko, H., et al. \ 2010, \apj, 708, 834 

\bibitem[Begelman et al.(1980)]{BBR1980} 
Begelman, M.~C., Blandford, R.~D., \& Rees, M.~J.\ 1980, \nat, 287, 307 

\bibitem[Brem et al.(2014)]{Brem2014} 
Brem, P., Amaro-Seoane, P., \& Sopuerta, C.~F.\ 2014, \mnras, 437, 1259 

\bibitem[Buchholz et al.(2009)]{Buchholz2009} 
Buchholz, R.~M., Sch{\"o}del, R., \& Eckart, A.\ 2009, \aap, 499, 483 

\bibitem[Chandrasekhar(1942)]{Chandrasekhar1942} 
Chandrasekhar, S.\ 1942, The Principles of Stellar Dynamics.
Chicago, The University of Chicago press.  

\bibitem[Chatzopoulos et al.(2015)]{Chatz2015} Chatzopoulos, S., 
Fritz, T.~K., Gerhard, O., et al.\ 2015, \mnras, 447, 952 

\bibitem[Cohn \& Kulsrud(1978)]{CohnKulsrud1978} 
Cohn, H.\& Kulsrud, R. 1978,
\apj, 226, 1087

\bibitem[Do et al.(2009)]{Do2009} 
Do, T., Ghez, A.~M., Morris,  M.~R., Lu, J.~R., Matthews, K., Yelda, S.,  \& Larkin, J.\ 2009, 
\apj, 703, 1323 

\bibitem[Do et al.(2013)]{Do2013} 
Do, T., Martinez, G.~D., Yelda, S., et al.\ 2013, \apjl, 779, L6 

\bibitem[Eilon et al.(2009)]{Eilon2009} 
Eilon, E., Kupi, G.,  \& Alexander, T.\ 2009, \apj, 698, 641 

\bibitem[Freitag et al.(2006)]{Freitag2006} Freitag, M., 
Amaro-Seoane, P., \& Kalogera, V.\ 2006, \apj, 649, 91 

\bibitem[G{\"u}rkan \& Hopman(2007)]{GurkanHopman2007} 
G{\"u}rkan, M.~A., \& Hopman, C.\ 2007, \mnras, 379, 1083 

\bibitem[Hamers, Portegies Zwart \& Merritt(2014)]{Hamers2014}
Hamers, A., Portegies Zwart, S. \& Merritt, D. 2014,
MNRAS, in press

\bibitem[H{\'e}non(1961)]{Henon1961} 
H{\'e}non, M.\ 1961, Annales d'Astrophysique, 24, 369

\bibitem[Hopman(2009)]{Hopman2009} 
Hopman, C.\ 2009, Classical and  Quantum Gravity, 26, 094028 

\bibitem[Hopman \& Alexander(2006a)]{HopmanAlexander2006L} 
Hopman, C., \& Alexander, T.\ 2006, \apjl, 645, L133 

\bibitem[Hopman \& Alexander(2006b)]{HopmanAlexander2006} 
Hopman, C., \& Alexander, T.\ 2006, \apjl, 645, L133 

\bibitem[Lee(1969)]{Lee1969} 
Lee, E.~P.\ 1969, \apj, 155, 687 


\bibitem[Madigan et al.(2011)]{Madigan2011} 
Madigan, A.-M., Hopman, C., \& Levin, Y.\ 2011, \apj, 738, 99 

\bibitem[Magorrian \& Tremaine(1998)]{MagorrianTremaine1998}
Magorrian, J. \& Tremaine, S.\ 1998, MNRAS, 309, 447.

\bibitem[Merritt(2009)]{Merritt2009} Merritt, D.\ 2009, \apj, 694, 959 

\bibitem[Merritt(2010)]{Merritt2010} 
Merritt, D.\ 2010, \apj, 718, 739
 
\bibitem[Merritt(2013)]{DEGN}
Merritt, D. 2013, Dynamics and Evolution of Galactic Nuclei
(Princeton: Princeton University Press).

\bibitem[Merritt(2015a)]{Paper1}
Merritt, D. 2015a, ApJ, 804:52 (Paper I)

\bibitem[Merritt(2015b)]{Paper2}
Merritt, D. 2015b, ApJ, 804:128 (Paper II)

\bibitem[Merritt(2015c)]{Paper4}
Merritt, D. 2015c, ApJ, in press (Paper IV)

\bibitem[Merritt et al.(2010)]{MAMW2010} 
Merritt, D., Alexander, T., Mikkola, S., \& Will, C.~M.\ 2010, \prd, 81, 062002 

\bibitem[Merritt et al.(2011)]{MAMW2011} 
Merritt, D., Alexander, T., Mikkola, S., \& Will, C.~M.\ 2011, \prd, 84, 044024 

\bibitem[Merritt et al.(2015)]{MAV2015}
Merritt, D., Antonini, F., \& Vasiliev, E. 2015, submitted to {\it The Astrophysical Journal} 

\bibitem[Merritt et al.(2006)]{Merritt2006} Merritt, D., 
Storchi-Bergmann, T., Robinson, A., et al.\ 2006, \mnras, 367, 1746 

\bibitem[Merritt \& Szell(2006)]{MerrittSzell2006} 
Merritt, D., \& Szell, A.\ 2006, \apj, 648, 890 

\bibitem[Merritt \& Wang(2005)]{MerrittWang2005} 
Merritt, D., \& Wang, J.\ 2005, \apjl, 621, L101 

\bibitem[Milosavljevi{\'c} \& Merritt(2001)]{MM2001} 
Milosavljevi{\'c}, M., \& Merritt, D.\ 2001, \apj, 563, 34 

\bibitem[Milosavljevi{\'c} \& Merritt(2003)]{MM2003} 
Milosavljevi{\'c}, M., \& Merritt, D.\ 2003, \apj, 596, 860 

\bibitem[Merritt \& Vasiliev(2012)]{MerrittVasiliev2012} 
Merritt, D., \& Vasiliev, E.\ 2012, \prd, 86, 102002 

\bibitem[Rauch \& Tremaine(1996)]{RauchTremaine1996} 
Rauch, K.~P., \& Tremaine, S.\ 1996, New Astron., 1, 149 

\bibitem[Rosenbluth et al.(1957)]{Rosenbluth1957} Rosenbluth, M.~N., 
MacDonald, W.~M., \& Judd, D.~L.\ 1957, Physical Review, 107, 1 

\bibitem[Sch{\"o}del(2011)]{Schoedel2011} 
Sch{\"o}del, R.\ 2011,  Highlights of Spanish Astrophysics VI, 36 

\bibitem[Sch{\"o}del et al.(2009)]{Schoedel2009} 
Sch{\"o}del, R., Merritt, D., \& Eckart, A.\ 2009, \aap, 502, 91 

\bibitem[Sigurdsson \& Rees(1997)]{SigurdssonRees1997} 
Sigurdsson, S., \& Rees, M.~J.\ 1997, \mnras, 284, 318 

\end{thebibliography}
\end{document}